\documentclass[11pt,a4paper]{article}
\usepackage{graphics,color,cite}
\usepackage{amsmath}
\usepackage{latexsym,amsmath,amssymb,amsfonts,amsthm,amsbsy,enumerate}
\usepackage{amsmath,amssymb}
\usepackage[affil-it]{authblk}

\usepackage[a4paper, total={6.5in, 8in}]{geometry}

\newcommand{\nn}{\nonumber}

\begin{document}
\title{\textbf{The dissipative Bose-Hubbard model} \\ Methods and examples}

\author[1]{G. Kordas \thanks{gekordas@phys.uoa.gr}}

\author[2,3]{D. Witthaut}

\author[4]{P. Buonsante}

\author[5,6]{A. Vezzani}

\author[6,7]{R. Burioni}

\author[1]{A.~I. Karanikas}

\author[6,7]{S. Wimberger}

\affil[1]{University of Athens, Physics Department, Nuclear \& Particle
Physics Section, Panepistimiopolis, Ilissia 15771, Athens, Greece}
\affil[2]{Forschungszentrum J\"{u}lich, Institute for Energy and Climate
Research (IEK-STE), 52428 J\"{ul}ich, Germany} 
\affil[3]{Institute for Theoretical Physics, University of Cologne, 50937 K\"{o}ln, Germany}
\affil[4]{QSTAR, INO-CNR and LENS, Largo Fermi 2, I-50125 Firenze, Italy}
\affil[5]{S3, CNR Istituto di Nanoscienze, Via Campi 213A, 41125 Modena, Italy}
\affil[6]{Dipartimento di Fisica e Scienze della Terra, Universit\`{a} di Parma, Via G.P. Usberti 7/a, 43124 Parma, Italy}
\affil[7]{INFN, Sezione di Milano Bicocca, Gruppo Collegato di Parma, Italy}
\maketitle
\abstract{Open many-body quantum systems have attracted renewed interest in the context of quantum information science and quantum transport with biological clusters and ultracold atomic gases. The physical relevance in many-particle bosonic systems lies in the realization of counter-intuitive transport phenomena and the stochastic preparation of highly stable and entangled many-body states due to engineered dissipation. We review a variety of approaches to describe an open system of interacting ultracold bosons which can be modeled by a tight-binding Hubbard approximation. Going along with the presentation of theoretical and numerical techniques, we present a series of results in diverse setups, based on a master equation description of the dissipative dynamics of ultracold bosons in a one-dimensional lattice. Next to by now standard numerical methods such as the exact unravelling of the master equation by quantum jumps for small systems and beyond mean-field expansions for larger ones, we present a coherent-state path integral formalism based on Feynman-Vernon theory applied to a many-body context.
}
\section{Introduction}
\label{intro}
Since the first realization of Bose-Einstein condensates in 1995 \cite{BEC,ketterle02}, ultracold atoms have provided a vast research field for investigating the quantum nature of matter with an amazing experimental precision. For a few years now, single particles can be imaged, e.g. in lattice confinements \cite{albi05,gross10,bakr09,miranda15,sherson10,weiten11,ott2015} and in the continuum \cite{gerike08}. Such experiments allow one in situ measurements of strongly correlated many-body quantum systems. The Hubbard model introduced originally for electrons in 1963 by John Hubbard \cite{Hubbard} describes many-particle effects in a lattice. For ultracold atoms, periodic lattice structures can be realized in one, two or three spatial dimensions using optical lattices \cite{morsch06,Weidemueller}. The Fermi- as well as the Bose-Hubbard model for ultracold fermions or bosons, respectively, is nowadays realized in the laboratories essentially without unwanted coupling to environments \cite{lewen07,bloch08}. All these possibilities permit one the study of originally idealized models from solid-state physics. Many-body non-equilibrium dynamics \cite{Bloch2015,Schmiedmayer2015} and transport problems \cite{Esslinger2012,Nist2014,ott2015}, for instance, can be experimentally (quantum) simulated \cite{Bloch2012}. 

In this review, we explicitly focus on dissipative many-body dynamics of bosons. Our work horse will be the Bose-Hubbard model \cite{fisher89,jaks98,jaks05}, which in particular, represents a paradigm for bosonic many-body quantum systems. It is used to describe, e.g., chains of coupled photonic cavities \cite{Tomadin2012}, superconducting Josephson junctions \cite{brud05}, and, of course, strongly interacting ultracold bosons. We treat a Bose-Hubbard system in the presence of localized particle loss and noise. Modern experiments indeed allow one to induce site-resolved loss, either by optical means \cite{weiten11} or by using a strongly focused electron beam \cite{gerike08,ott2015}. 

From the field of quantum optics we know for quite some time that dissipation and decoherence in general may be used to control single- and many-body quantum systems \cite{carmichael,Gard04,Gard09,WM2010,daley14}. Such ideas were applied to many-body massive systems around 2008 by various groups \cite{witt08,kraus08,diel08,syassen08,cirac2009}. Controlled dissipation can indeed drive a many-body quantum system towards interesting states. If such states are stable stationary solutions of the quantum master equation, they are typically called dark states \cite{darkstate,yang99}. In the context of quantum information science, the robust preparation and the stability of highly entangled states, e.g. within subspaces protected against decoherence, is obviously of great practical interest. Moreover, there is the hypothesis around that engineered dissipation and noise might be responsible for the efficient transport of excitations in bimolecular clusters, a  highly complex many-body system on the verge of behaving classical and (maybe) quantum like, see e.g. \cite{Plenio2013}.

In the following, we will briefly present the major experimental techniques to realize controlled loss in a gas of ultracold atoms moving in a lattice. Then we shall introduce the Bose-Hubbard Hamiltonian. In sec. \ref{sec:0} the dissipative Bose-Hubbard problem is defined. Secs. \ref{sec:1} to \ref{sec:5} collect a series of exact and approximate numerical methods which can be used to deal with a dissipative Bose-Hubbard system described by a master equation. Along the way, advantages and disadvantages of these techniques are discussed. Sec. \ref{sec:7} opens the door towards a systematic description of a many-body system in the presence of decoherence based on the formalism of coherent-state path integrals. Our conclusions are presented in Sec. \ref{sec:8}. Finally, we would like to note that our choice of material is necessarily selective given the overwhelming amount of literature on dissipation in quantum systems and ultracold atoms in general. We nevertheless hope that this review will give a helpful overview for experts and beginners in the subject at the same time, as well as for the interested readers from related fields.

\subsection{Dissipative processes in experiments}
\label{intro:1}
Dissipative processes are always present in real life experiments. So experimentalists put great efforts to control these sources of noise. For example, the authors in~\cite{sabin15} present a gravitational wave detector that is robust to thermal noise and depletion. Elastic collisions of the atoms of a BEC with the atoms of the background gas or the absorption and spontaneous emission of photons from the lattice beam, so-called phase noise processes, lead to the loss of coherence of the BEC. Furthermore, there are present inelastic collisions between atoms~\cite{syassen08,gerton99} and three-body recombination of atoms which correspond to decay into the continuum of unbound states~\cite{gerton99,kraemer06,eismann15,weber03}, leading to atom losses from the trap.

Recent experimental advances has allowed a new level of local manipulation of individual lattice sites. Single-site access can be implemented optically by increasing the lattice period~\cite{albi05,gross10}, by implementing a high-resolution optical imaging system~\cite{bakr09,miranda15} or by a strongly focused laser beam~\cite{sherson10,weiten11}. An even higher resolution can be achieved by a focused electron  beam~\cite{gerike08,gerik10,santra15,ott2015,ott2013}. In the latter case a focused electron beam locally produces ions in the atomic cloud, due to electron-atom interaction. The produced ions are extracted and collected with the aid of ion optics and an ion detector.  This detection procedure allows the reconstruction of the atoms' position where the ionization occurred.

Since, in this article, we are going to discuss Markovian environments, let us discuss the electron microscopy case in more detail. In these experiments, the vacuum chamber is equipped with an electron microscope. After the preparation of a Bose-Einstein condensate the electron beam impact on it. The electron-atom collisions can ionize the atoms which subsequently are removed from the trap and detected by the ion detector. The ionized atoms are about $40\%$ of all the scattered atoms. These ions, while leaving from the trap, can collide with other atoms producing more losses. We furthermore have about $55\%$ inelastic scattering events and $5\%$ elastic, which cannot produce a detectable signal.
\begin{figure}%\sidecaption
%\resizebox{0.4\hsize}{!}{
\centering
\resizebox{0.75\columnwidth}{!}{
\includegraphics{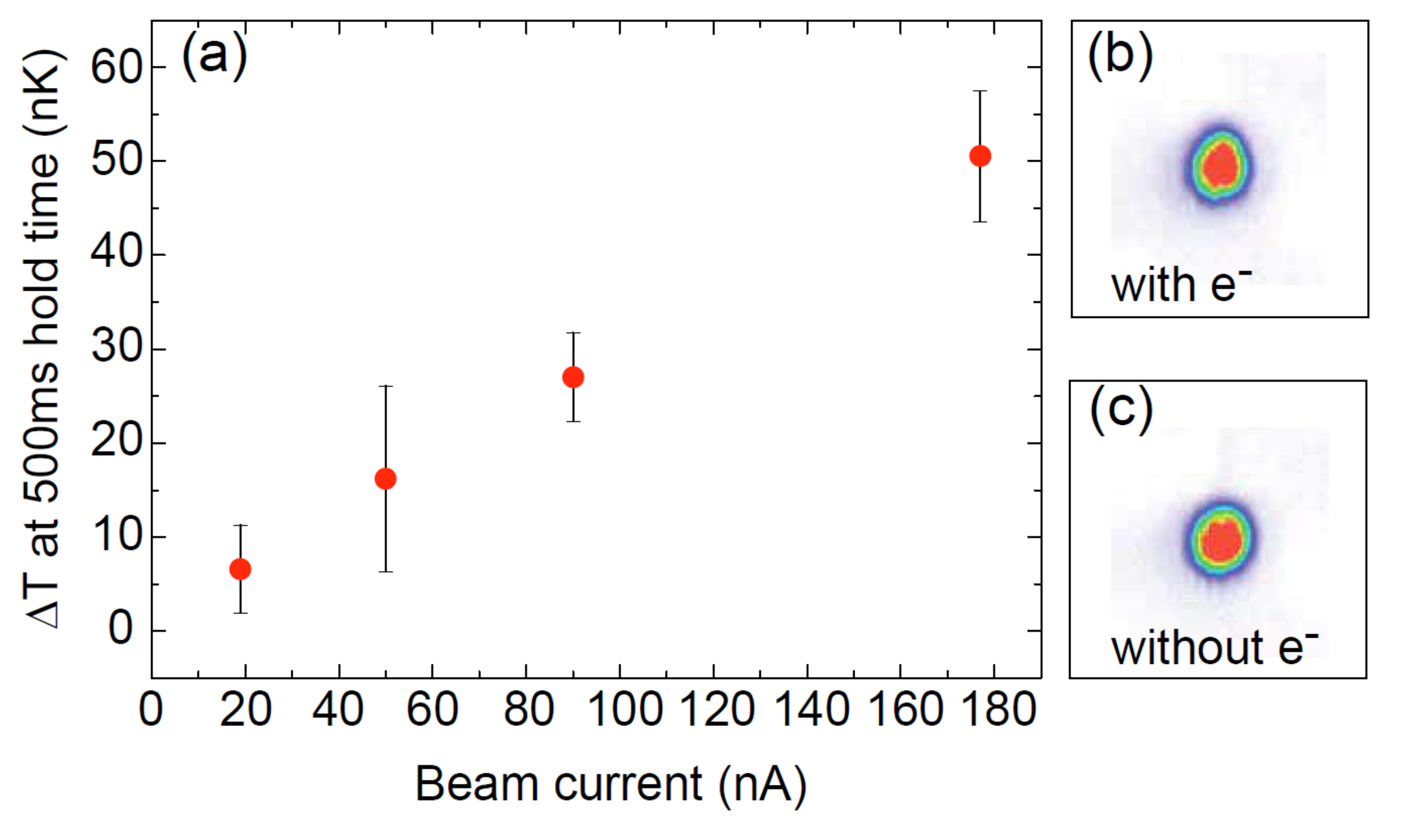}}
\caption{\label{ott} (a) The temperature difference between an illuminated and a non-illuminated atom cloud after a holding time of 500 ms as a function of the electron beam current. (b) Absorption image of an illuminated BEC and (c) of a BEC that has not been illuminated. Reprinted from~\cite{gerik10} with kind permission from Herwig Ott.
}
\end{figure}

As we described it above, we can say that this system is a \emph{special} case of an open quantum system. It is an open system in which we can accurately control the dissipation process. We can remove single atoms from the position we want with the desired rate. This is already interesting since usually dissipation is connected with an uncontrollable environment. Furthermore, as we are going to argue, the ``side effects'' of the electron impact on the cloud are negligible. Figs.~\ref{ott} (a)-(c) report  experimental data, which show the effect of heating of the atomic cloud due to the impact of the electrons. Fig.~\ref{ott} (a) shows the temperature gained by the atomic cloud for different electron beam currents. As one can see, for an electron beam current of 20 nA the temperature of the sample increases only by 6 nK as compared to an initial temperature of typically around 100 nk. So in a very good approximation, the only effect of the electron beam on the Bose-Einstein condensate is the loss of atoms. In other words, the back action from the lost atoms on the remaining system is rather small, allowing us helpful approximations based on the Master equation approach. This is further illustrated in Figs.~\ref{ott} (b,c) where we have the absorption images of an illuminated atomic cloud, Fig.~\ref{ott} (b), and a cloud that has not been illuminated, Fig.~\ref{ott} (c). The only significant difference between the two images is that the atom number has been reduced by $7\%$. As we are going to discuss in Sec.~\ref{sec:0}, the situation is exactly that of a quantum system coupled to a Markovian environment.
\subsection{The Bose-Hubbard model}
\label{intro:2}
\begin{figure}%\sidecaption
\resizebox{0.9\hsize}{!}{
\centering
%\resizebox{0.6\columnwidth}{!}{
\includegraphics{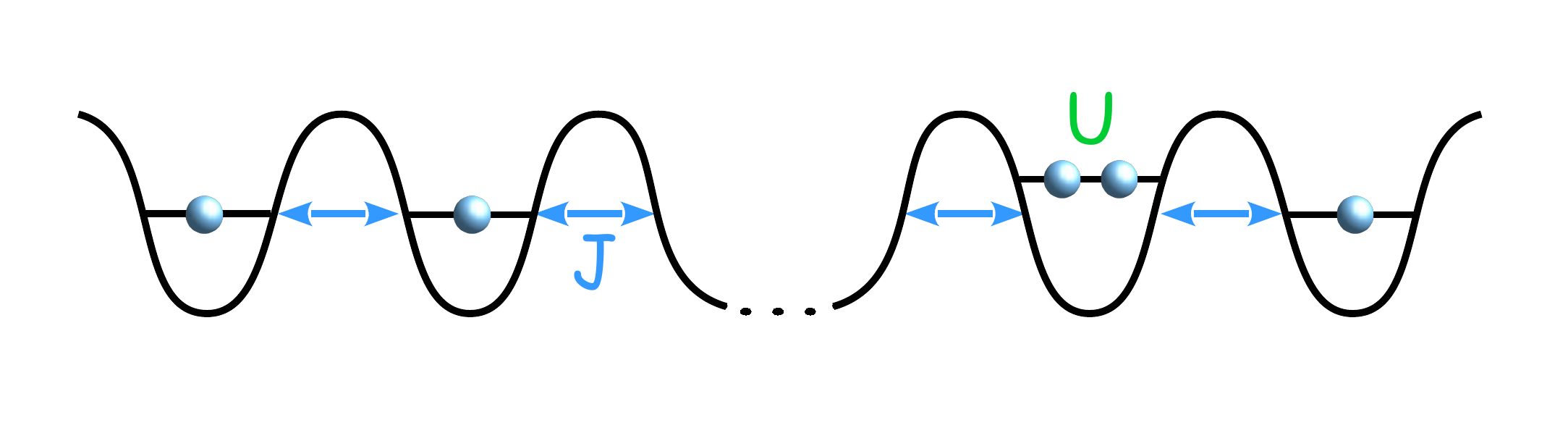}}
\caption{\label{fig:sk0}
Schematic of a Bose-Hubbard chain.
}
\end{figure}
The Bose-Hubbard (BH) model has been extensively used to describe the dynamics of a dilute bose gas trapped in a deep optical lattice~\cite{lewen07,jaks98,jaks05}. One of its great success was the prediction of the quantum phase transition from superfluid to Mott insulator~\cite{fisher89,jaks98} which was experimentally confirmed~\cite{greiner02}. Except from ultracold atoms in optical lattices, the BH model has been used to describe the physics of a variety of interacting quantum systems, ranging from superconducting Josephson junction arrays~\cite{brud05} to interacting polaritons in hybrid optical systems~\cite{hart08,leib10} and circuit QED~\cite{nae14}.

In order to derive the BH model~\cite{jaks98} one begins from the general many-body Hamiltonian
\begin{eqnarray}\label{eq:ham2nd}
\nonumber \hat{H} &=& \int d\vec{r}~\hat{\Psi}^\dag(\vec{r}) \left[ -\frac{\hbar^2}{2m}\nabla^2 +
 V_{\rm lat.}(\vec{r}) + V_{\rm trap.}(\vec{r})\right]\hat{\Psi}(\vec{r}) +\\
&& +\frac{U_0}{2}\int d\vec{r}~\hat{\Psi}^\dag(\vec{r}) \hat{\Psi}^\dag(\vec{r}) \hat{\Psi}(\vec{r}) \hat{\Psi}(\vec{r}),
\end{eqnarray}
where the operators $\hat{\Psi}^\dag(\vec{r})$ and $\hat{\Psi}(\vec{r})$ create and annihilate a boson at the position $\vec{r}$, respectively. 
Other than the periodic potential, in Hamiltonian (\ref{eq:ham2nd}) we have taken into account also an overall trapping potential $V_{\rm trap.}(\vec{r})$. In (\ref{eq:ham2nd}) we have assumed that the gas is dilute and at low temperatures so the interaction term can be approximated by a contact interaction in the coordinate space, the $U_0$ term. For a deep lattice the energy gap between the bands is very large if we compare it with the chemical potential, the interaction and kinetic energy. This ensures that all the particles will stay in the ground band
\begin{equation}\label{eq:annih}
 \hat{\Psi}(\vec{r}) = \sum_{j=1}^M w_0(\vec{r}-\vec{r}_j) \hat{\alpha}_j,
\end{equation}
where $\hat{\alpha}_j$ annihilates a boson at site $j$ and $w_0$ is the Wannier function of the ground bloch band. Now if we insert Eq. (\ref{eq:annih}) into the Hamiltonian (\ref{eq:ham2nd}) we obtain the BH Hamiltonian
\begin{equation}\label{eq:BH}
\hat{H}_{\rm BH} = \sum_{j=1}^M \varepsilon_j \hat{\alpha}^\dag_j \hat{\alpha}_j - J\sum_{j=1}^{M-1} (\hat{\alpha}^\dag_{j+1} \hat{\alpha}_j + \hat{\alpha}^\dag_j \hat{\alpha}_{j+1}) + \frac{U}{2}\sum_{j=1}^M \hat{\alpha}^\dag_j \hat{\alpha}^\dag_j \hat{\alpha}_j\hat{\alpha}_j,
\end{equation}
where in (\ref{eq:BH}) we have defined the following parameters
\begin{eqnarray}
 \varepsilon_j &=& \int d\vec{r}~V_{\rm trap.}(\vec{r}_{\phantom{.}}) |w_0(\vec{r}-\vec{r}_j)|^2,\\
 J &=& \int d\vec{r}~w^*_0(\vec{r}-\vec{r}_j)\left[-\frac{\hbar^2}{2m}\nabla^2 +
 V_{\rm lat.}(\vec{r}_{\phantom{.}})\right] w_0(\vec{r}-\vec{r}_{i=j\pm1}),\\
 U &=& U_0\int d\vec{r}~|w_0(\vec{r}-\vec{r}_j)|^4.
\end{eqnarray}
From the above expressions we can easily understand the physical meaning of the BH parameters: $\varepsilon_j$ is the on-site potential, $J$ is connected to the kinetic energy, so it parametrizes the tunneling between adjacent sites and $U$ is the on-site interaction strength. As most of the previous studies so far, we restrict here to discuss a one-dimensional model, already represented by Eq. \eqref{eq:BH}.

In the model Eq. \eqref{eq:BH} a pure BEC state is the ground state for zero interparticle interactions, $U=0$, and has the form
\begin{equation}\label{eq:BEC}
|\Psi^{(N)}\rangle = \frac{1}{\sqrt{N!}} \left(\sum_{j=1}^M\psi_j \hat{\alpha}^\dag_j \right)^N |0\rangle,
\end{equation}
where the coefficients obey the normalization condition
\begin{equation}
\sum_{j=1}^M |\psi_j|^2 = 1.
\end{equation}
\section{The dissipative Bose-Hubbard model}
\label{sec:0}
A dissipative or open quantum system is a system coupled to another, frequently much larger, classical or quantum system to which we have no access. Here we assume that the system is a BH lattice connected to a Markovian environment. In the latter the dynamics is given by a master equation in Lindblad form~\cite{bre06}
\begin{equation}\label{eq:DBH}
\frac{d\hat{\rho}}{dt} = -i[\hat{H}_{\rm BH},\hat{\rho}] - \frac{1}{2} \sum_\ell \left( \hat{L}^\dag_\ell \hat{L}_\ell \hat{\rho} + \hat{\rho}\hat{L}^\dag_\ell\hat{L}_\ell - 2\hat{L}_\ell \hat{\rho}\hat{L}^\dag_\ell\right),
\end{equation}
where $\hat{L}_\ell$ are called {\it Lindblad} or {\it jump} operators and are in general non-Hermitian. The form of the Lindblad operators depends on the specific coupling of the quantum system with its environment. We must underline that in the general derivation of the equation (\ref{eq:DBH}) one applies a Born-Markov approximation. This means that the typical time during which the internal correlations in the environment exists is much shorter than the the characteristic time of the system-environment interaction. A more general formalism which allows the handle of non-Markovian environments is the so-called Feynman-Vernon influence functional. We will introduce the latter for open BH lattices in Sec.~\ref{sec:5}.

Master equation (\ref{eq:DBH}) can describe a great variety of experimentally relevant dissipation processes. For example, phase noise is described by a Lindblad operator of the form $\hat{L}_\ell = \sqrt{\kappa}\hat{\alpha}^\dag_\ell \hat{\alpha}_\ell$, where $\kappa$ is the rate of phase noise~\cite{kordas12,kordas13,ang97,witt08,trim08_1,witt09,witt11,pichler10,pole12,kollath2015}. The specific Lindblad operator conserves the total particle number, but destroys the coherence globally in the system. Another process we can describe is a three-body recombination of particles which corresponds to decay into the continuum of unbound states, $\hat{L}_\ell =\sqrt{\gamma_3}(\hat{\alpha}_\ell)^3$~\cite{daley09,moerdijk96}. In fact it was shown~\cite{daley09} that three body losses give rise to effective hard-core three-body interactions and, for attractive interactions, it may lead to a dimer superfluid phase.

Inelastic collisions between the atoms of the condensate can induce two-body losses from the system~\cite{moerdijk96,li08}. This dissipative process can be described by a Lindblad operator of the form: $\hat{L}_\ell =\sqrt{\gamma_2}(\hat{\alpha}_\ell)^2$~\cite{garc,kiffner11}. Starting from the master equation (\ref{eq:DBH}) the authors in~\cite{garc} showed that strong two-body losses can create a Tonks-Girardeau gas. The term is used to describe the fact that 1D hard-core bosons can be mapped exactly to that of an identical fermionic model. So in the spirit of Tonk and Girardeau strong repulsion in an 1D system is equivalent to a Pauli exclusion principle. The same can be achieved with strong inelastic collisions in an 1D lattice. The particles act as hard-core bosons, with dissipation playing the role of strong repulsion and one can show that the system can be effectively described by a reduced loss rate $\gamma_{\rm eff}\propto J^2 / \gamma_2$, which is much slower than the original loss rate, $\gamma_2$, or the tunnelling strength $J$. The behaviour described above was confirmed also experimentally~\cite{syassen08}.

Although dissipation is frequently an unwanted experimental reality, there were theoretical suggestions that it can be used to control the many-body dynamics. In this context, we speak of ``quantum reservoir engineering''.  The system which is described by a master equation (\ref{eq:DBH}) will go to a dynamical steady state, which is in general a mixed state. However, under certain conditions, the steady state may be a pure state, $|DS\rangle$, a so-called dark state. The conditions for the system to converge to a unique dark state are: (a) the dark state must be an eigenstate of the Lindblad operators with zero eigenvalues, $\hat{L}_\ell|DS\rangle = 0,~\forall\ell$, which holds if (b) $\hat{H}_{\rm BH}|DS\rangle = \varepsilon|DS\rangle$. So, by designing the Lindblad operators one can prepare pure many-body states or non-equilibrium quantum phases. In~\cite{diel08,kraus08,Diehl_PRL_105_015702}, laser driven atoms are coupled to a phonon bath provided by a second BEC in order to drive the system to a pure BEC state. This process was described by a master equation of the form (\ref{eq:DBH}) with Lindblad operators $\hat{L}_\ell \equiv \hat{c}_{i,j} = \sqrt{\kappa} (\hat{\alpha}^\dag_i + \hat{\alpha}^\dag_j)(\hat{\alpha}_i - \hat{\alpha}_j)$. The interplay between interactions and dissipation leads to a nonequilibrium phase diagram in which there is a dynamical phase transition from a condensed to a homogeneous thermal steady state~\cite{Diehl_PRL_105_015702}. Although the thermal state is always dynamically stable, for weak hopping $J$ there is a region, in the phase diagram, in which the condensed state is unstable. The instability turns out to originate from  the interplay of short-time quantum and long wavelength classical fluctuations. In the same spirit, the authors in~\cite{wata12,caba14} use dissipation in order to prepare phase- and number-squeezed states. A double-well filled with ultracold atoms is immersed in a background BEC. The latter BEC consists of atoms of different species than the atoms in the double-well and act as a quantum reservoir. It is shown that squeezing develops on a time scale proportional to $1/N$, where $N$ is the number of particles in the double well, making the time scale for the creation of squeezed states very short.

Localized single-particle losses ($\hat{L}_\ell =\sqrt{\gamma_\ell}\hat{\alpha}_\ell$) are another case in which dissipative processes can help to the dynamical control of the quantum system. This fundamental dissipation process has attracted a lot of theoretical attention~\cite{Shchesnovich10,witt08,trim08_1,witt09,graefe08,franzoi11,
franzoi07,barme11,ng09,henn13,witt11,trim11,wimb13,liv06,kordas12,
kordas13,pudlik13,kepe12,penna13,Shchesnovich12,Shchesnovich12_1}, not only because of its simplicity, but also due to the interesting phenomena that one can study. For example, as it was shown in~\cite{witt08,trim08_1,witt09,graefe08} this dissipation mechanism can create attractive fixed points in the classical phase space of the system. As a consequence a wide range of initial conditions converge to the state corresponding to the classical fixed point. This mechanism suggests that localized losses can be used to control the many-body dynamics. Indeed, it has been shown that localized single particle losses can be used to create localized non-linear structures. In the presence of strong interactions, discrete breathers are produced in lattices with boundary dissipation~\cite{franzoi11,franzoi07,ng09,henn13,witt11,trim11,wimb13,liv06}, while coherent dark solitons can be engineered if we additionally use phase imprinting~\cite{witt11,trim11}, see subsection~\ref{subsec:3.1}.
Furthermore, it has been shown~\cite{kordas12,kordas13} that unstable initial conditions in the presence of strong interactions and boundary losses lead to the emergence of quantum superpositions of discrete breathers which are localized in different lattice sites. We discuss the latter case in subsection~\ref{subsec:1.1}.

Another example in which coherent and dissipative dynamics cooperate to drive the system to a desirable direction is distance-selective pair particle loss~\cite{ates12,cui14}. In this case particles are expelled from the system only if they occupy lattice sites of a certain distance. For a BH chain this mechanism can be modeled by a master equation with Lindblad operators: $\hat{L}_\ell = \sqrt{\gamma_2} \hat{\alpha}_\ell \hat{\alpha}_{\ell+d}$, that is losses occur nonlocally with $d$ be the critical distance and $\gamma_2$ the pair loss rate. For hard-core bosons in an 1D lattice~\cite{ates12}, the conjuction of this dissipation process  and the tunneling leads to the formation of coherent long-lived complexes of two or more particles, as far the loss rate exceeds the tunneling rate. The authors in~\cite{cui14} included an external driving to a BH chain, so that the particles could move in and out of the chain via the boundary sites. For $U<2J$ the nonlocal pair loss helps the localization and slowdown the transport of the particles. The system relaxes to a strongly bunched steady state. If $U>2J$, the pair loss enhances the localization, while the external driving tends to destroy it. In both cases the particles relax to a steady state which is spatially bunched.
\section{The quantum jump method}
\label{sec:1}
The quantum jump or Monte Carlo wave-function method~\cite{mol93,dum92,plenio98,daley14} is a useful numerical method which allows one an exact unravelling of master equations of the form given in Eq. (\ref{eq:DBH}). The advantage of this method is that instead of evolving the initial density matrix $\hat{\rho}(0)=|\Psi(0)\rangle\langle \Psi(0)|$, we evolve the state vector $|\Psi(0)\rangle$ using a stochastic procedure and averaging over the possible outcomes. The method can be easily generalized to the case where the initial state is a mixed state.

The method consists of a continuous evolution of $|\Psi(0)\rangle$, using an effective non-Hermitian Hamiltonian, which is interrupted by stochastic quantum jumps. Specifically, if we know the wave-function at time $t$ then we obtain the wave-function at time $t+\delta t$ using the following stochastic procedure. We define the quantity
\begin{equation}
 \delta p = \sum_\ell \delta p_\ell,
\end{equation}
where
\begin{equation}
 \delta p_\ell = \langle \Psi(t)|\hat{L}_\ell^\dag \hat{L}_\ell|\Psi(t) \rangle \delta t
\end{equation}
is the probability of the process that is described by the operator $\hat{L}_\ell$ to occur, with $\delta t$ such that $\delta p \ll 1$. We further need a random number $\epsilon \in [0,1]$. 

If $\epsilon > \delta p$ the state $|\Psi(t)\rangle$ is evolved continuously with the effective non-Hermitian Hamiltonian
\begin{equation}
 \hat{H}_{\rm eff.} = \hat{H}_{\rm BH} - \frac{i}{2}\sum_\ell \hat{L}_\ell^\dag \hat{L}_\ell
\end{equation}
giving the state
\begin{equation}\label{eq:qj6}
 |\Psi(t+\delta t)\rangle = \frac{1}{\sqrt{1-\delta p}}\left( 1 - i\hat{H}_{\rm eff.}\delta t\right)|\Psi(t)\rangle,
\end{equation}
where the pre-factor in the right-hand side of (\ref{eq:qj6}) ensures the normalization of the wave-function, since the effective Hamiltonian, $\hat{H}_{\rm eff}$, is non-Hermitian. 

If $\epsilon \leq \delta p$ the state performs a jump and we choose the specific jump operator, $\hat{L}_\ell$, according to the probability law $P_\ell = \delta p_\ell / \delta p$,
\begin{equation}
 |\Psi(t+\delta t)\rangle = \frac{\hat{L}_\ell |\Psi(t)\rangle}{\sqrt{\delta p_\ell / \delta t}},
\end{equation}
with a probability $P_\ell$.
\begin{figure}
%\resizebox{0.4\hsize}{!}{
\centering
\resizebox{0.6\columnwidth}{!}{
\includegraphics{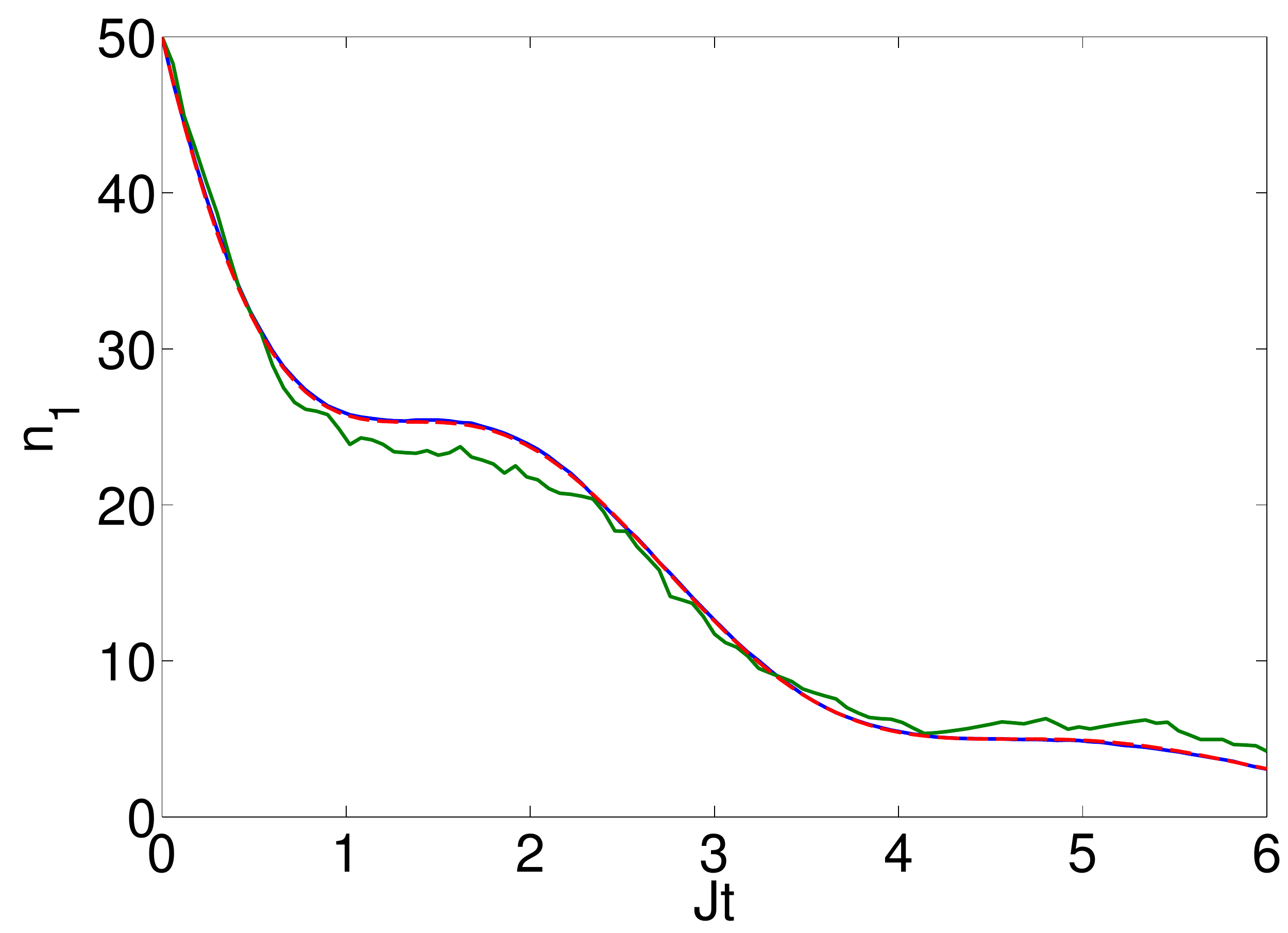}}
\caption{\label{fig:QJvsAN}
A comparison of a quantum jump simulation for 2 (solid green line) and 200 (solid blue line) trajectories and the analytical results (dashed red lines) for a double well. Shown is the time evolution of the population in the first well $n_1$. The parameters are: $U=0$, $\gamma_1=J$ and $\gamma_2=0$. The initial states is a pure BEC with $n_1(0)=n_2(0)=50$ and a phase difference of $\pi$ between wells.
}
\end{figure}

We claim that we will obtain the full density matrix if we perform the above procedure for a sufficient number of times and then average over these realization outcomes, $|\Psi^{(k)}(t)\rangle$,
\begin{equation}\label{eq:qj1}
 \overline{\hat{\rho}(t)} = \frac{1}{N_{\rm t}} \sum_{k=1}^{N_{\rm t}}|\Psi^{(k)}(t)\rangle \langle \Psi^{(k)}(t)| \mathop{\xrightarrow{\hspace*{0.7cm}}}\limits_{N_{\rm t}\gg 1} \hat{\rho}(t),
\end{equation}
provided that $\hat{\rho}(t=0)= \overline{\hat{\rho}(t=0)}$. The number of trajectories, $N_{\rm t}$, should be large enough such that the averaged density matrix, $\overline{\hat{\rho}(t)}$, converges.
Indeed, we have
\begin{eqnarray}
\nonumber \hat{\rho}(t) &\rightarrow& \overline{\hat{\rho}(t+\delta t)} = (1-\delta p) |\Psi_{\rm no~jump}\rangle \langle \Psi_{\rm no~jump}| + \delta p |\Psi_{\rm jump}\rangle \langle \Psi_{\rm jump}|\\
\nonumber &=& (1-\delta p)\frac{|\Psi(t+\delta t)\rangle \langle\Psi(t+\delta t)}{\sqrt{1-\delta p}\sqrt{1-\delta p}} + \delta p \sum_\ell P_\ell \frac{\hat{L}_\ell|\Psi(t)\rangle\langle\Psi(t)|\hat{L}_\ell^\dag}{\sqrt{\delta p_\ell/\delta t}\sqrt{\delta p_\ell/\delta t}}\\
\nonumber &=& (1 - i\hat{H}_{\rm eff.})\hat{\rho}(t))(1 - i\hat{H}_{\rm eff.}) + \sum_\ell \hat{L}_\ell \hat{\rho}(t) \hat{L}_\ell^\dag \\
&=& \hat{\rho}(t) + i [\hat{\rho}(t),\hat{H}_{\rm BH}] \delta t
- \frac{1}{2} \sum_\ell \left( \hat{L}^\dag_\ell \hat{L}_\ell \hat{\rho}(t) + \hat{\rho}(t)\hat{L}^\dag_\ell\hat{L}_\ell - 2\hat{L}_\ell \hat{\rho}(t)\hat{L}^\dag_\ell\right) \delta t
\end{eqnarray}
and if we average the above equation over all trajectories
\begin{equation}
\frac{d\overline{\hat{\rho}(t)}}{dt} = i [\overline{\hat{\rho}(t)},\hat{H}_{\rm BH}]
- \frac{1}{2} \sum_\ell \left( \hat{L}^\dag_\ell \hat{L}_\ell \overline{\hat{\rho}(t)} + \overline{\hat{\rho}(t)}\hat{L}^\dag_\ell\hat{L}_\ell - 2\hat{L}_\ell \overline{\hat{\rho}(t)}\hat{L}^\dag_\ell\right),
\end{equation}
we find that the averaged density matrix evolves according to the master equation (\ref{eq:DBH}).

If the initial state is a mixed state, then we must decompose this state into a statistical mixture of pure states, $\hat{\rho}(0)=\sum_j p_j |\phi_j\rangle\langle\phi_j|$. We then choose randomly the initial state for the algorithm among the $|\phi_j\rangle$ states with probability $p_j$.

The calculation of an arbitrary observable $\hat{O}$ is straightforward. One takes the mean value of the quantum average $\langle \Psi^{(k)}(t)|\hat{O}|\Psi^{(k)}(t)\rangle$
over the various outcomes $|\Psi^{(k)}(t)\rangle$ of the above process
\begin{equation}
 \langle O\rangle(t) \mathop{\xrightarrow{\hspace*{0.7cm}}}\limits_{N_{\rm t}\gg 1} \frac{1}{N_{\rm t}} \sum_{k=1}^{N_{\rm t}} \langle \Psi^{(k)}(t)|\hat{O}|\Psi^{(k)}(t)\rangle,
\end{equation}
where $N_{\rm t}$ is the number of trajectories.

In Fig.~\ref{fig:QJvsAN} we have solved master equation (\ref{eq:DBH}) for a BEC trapped in a double well without interactions and single-particle losses in the first site, $\hat{L}_\ell=\hat{\alpha}_1$, using the quantum jump method and compare the results with the analytical one for the population in the first site
\begin{equation}
 n_1(t) = n_{{\rm tot}}(0) {\rm e}^{-\frac{\gamma_1}{2}t}\left[ \frac{J^2}{2\omega^2} - \frac{\gamma_1^2}{32\omega^2}\cos(2\omega t)
-\frac{\gamma_1}{8\omega}\sin(2\omega t)\right],
\end{equation}
with the frequency $\omega = \sqrt{J^2 - \gamma_1^2/16}$. One can see that there is an excellent agreement between the analytical result and the quantum jump method, in this case already for just 200 trajectories.
\subsection{Quantum superposition of discrete breathers}
\label{subsec:1.1}
\begin{figure}%\sidecaption
%\resizebox{0.9\hsize}{!}{
\centering
\resizebox{0.6\columnwidth}{!}{
\includegraphics{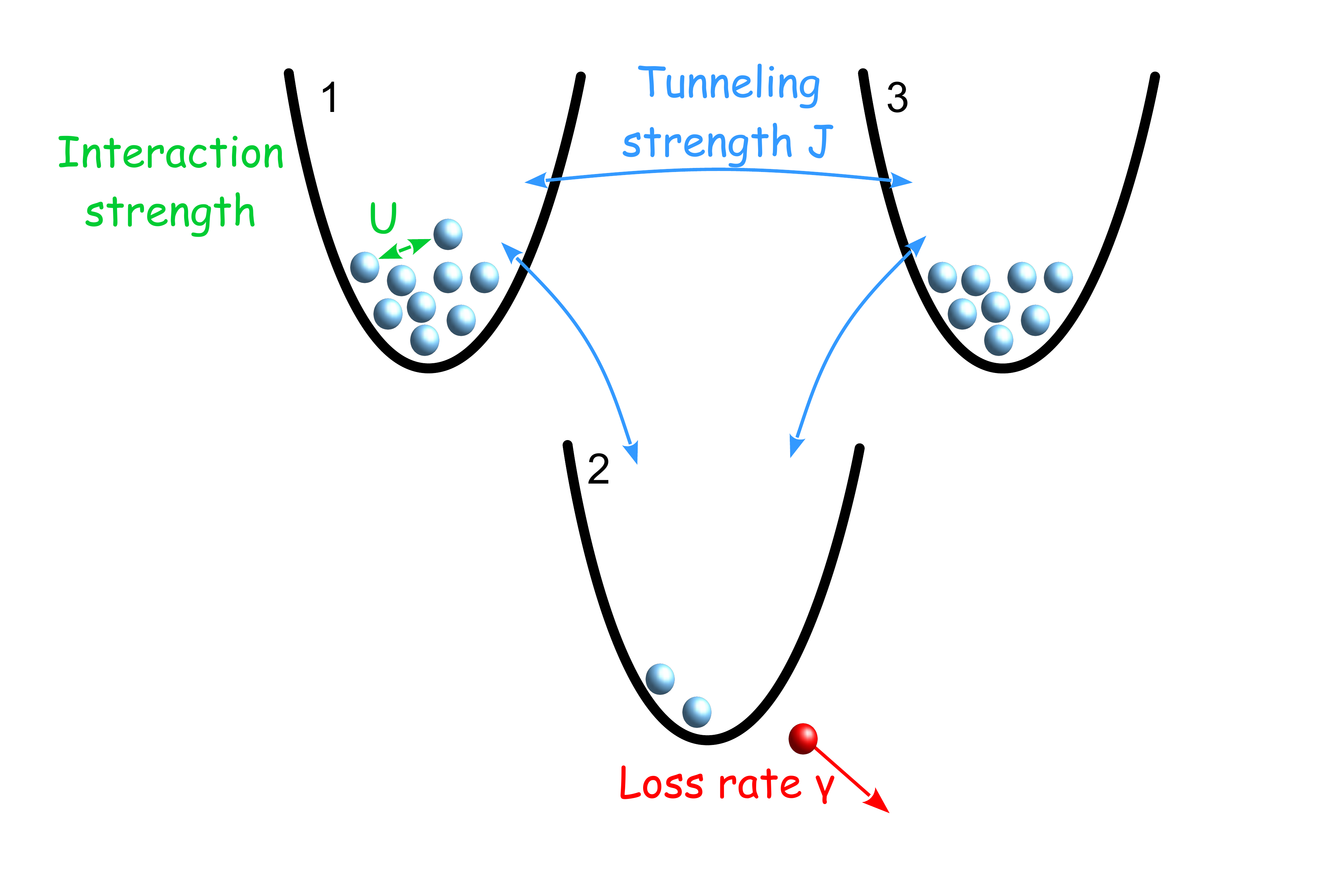}}
\caption{\label{fig:sk1}
Schematic of a BH triple-well trap subject to localized single particle losses.
}
\end{figure}
Discrete breathers are spatially localized, time-periodic, stable excitations in perfectly periodic discrete systems, which arise from the non-linearity and discreteness of the system~\cite{franzoi11,camp04,flach08,Henn10}. These non-linear structures are predicted on the level of the discrete non-linear Schr\"{o}ndinger equation, which is valid only in the classical mean-field limit. In the presence of dissipation these excited states can become attractively stable. This has been shown in the mean-field limit~\cite{ng09,henn13} and beyond this approximation~\cite{witt11,trim11}. Here we will discuss how controlled localized losses can be used to generate quantum superpositions of discrete breathers~\cite{kordas12,kordas13}. In other words, dissipation will help us to create non-linear mean-field objects that present quantum properties like entanglement. In order to study a purely quantum property we need an exact method for numerical many-body simulations, such as the quantum jump technique. Here we will follow the authors in~\cite{kordas12,kordas13}, and study the dynamical generation of  many-body entanglement in a triple-well trap subjected to localized particle losses in the middle site, see Fig.~\ref{fig:sk1}.

\begin{figure}
\centering
\resizebox{0.8\hsize}{!}{
%\centering
%\resizebox{0.75\columnwidth}{!}{
\includegraphics{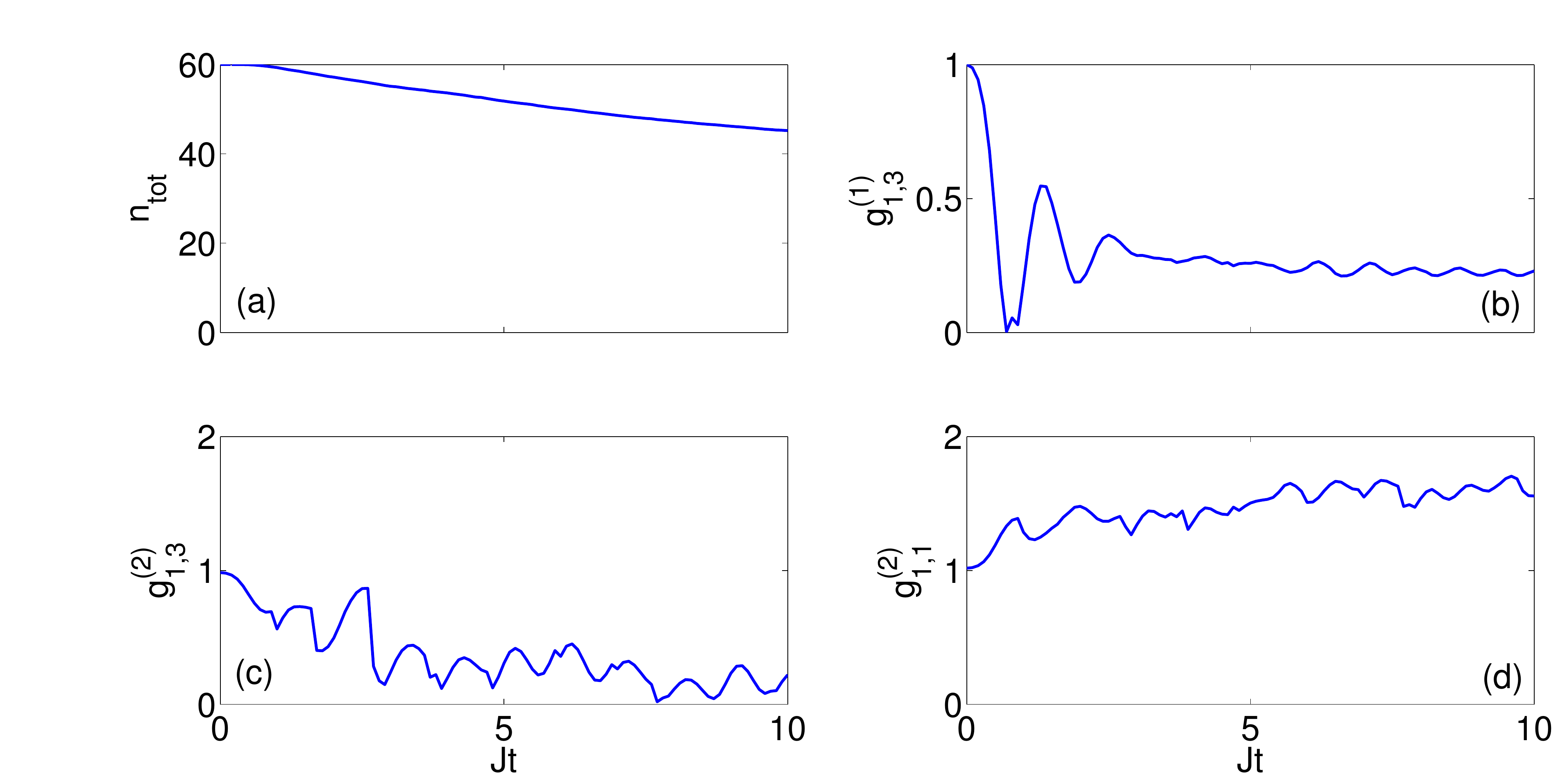}}
\caption{\label{fig:DB1}
Dynamics of (a) the total particle number, $n_{\rm tot}$, (b) the phase coherence, $g^{(1)}_{1,3}$, between sites 1 and 3 (c) the number correlations, $g^{(2)}_{1,3}$, between sites 1 and 3, and (d) the number fluctuations in the first site, $g^{(2)}_{1,1}$. The parameters are $U=0.2J$, $\gamma_2 = 0.5J$ and the initial state is antisymmetric with $n_1(0)=n_3(0)=30$ and $n_2(0) = 0$. For the simulations we have used the quantum jump method averaging over 200 trajectories.
}
\end{figure}

In Fig.~\ref{fig:DB1} we have simulated the dynamics of an initially pure BEC with an anti-symmetric wavefunction
\begin{equation}\label{eq:antiS}
(\psi_1,\psi_2,\psi_3) = \frac{1}{\sqrt{2}}(1,0,-1)
%   |\Psi\rangle = \frac{1}{2^N \, \sqrt{N!}} 
%       (\hat{\alpha}_1^\dagger - \hat{\alpha}_3^\dagger  )^N |0\rangle
\end{equation}
in the presence of strong interparticle interactions, $U=0.1J$. The evolution of the total particle number, Fig.~\ref{fig:DB1} (a), shows that the decay of the anti-symmetric state (\ref{eq:antiS}) has a very slow rate. This is a consequence of the fact that this state is a stationary state of the master equation for the non-interacting system, i.e. for $U=0$, and then there is no decay. The physical reason for this is the exactly destructive interference of atoms tunneling from site 1 and 3 to the leaky site 2.

In order to characterize the many-body state we analyse the normalized first- and second-order correlation functions
\begin{equation}\label{eq:g1}
g^{(1)}_{j,k} = \frac{|\langle\hat{\alpha}_{j}^\dag \hat{\alpha}_k\rangle|}{\sqrt{\langle\hat{n}_j\rangle\langle\hat{n}_k\rangle}}
\end{equation}
and
\begin{equation}\label{eq:g2}
g^{(2)}_{j,k} = \frac{\langle\hat{n}_{j}^\dag \hat{n}_k\rangle}{\langle\hat{n}_j\rangle\langle\hat{n}_k\rangle},
\end{equation}
respectively. The phase coherence of the state is characterized by the first-order correlation function (\ref{eq:g1}), while the density fluctuations ($j=k$) and correlations ($j\neq k$) are given by the second-order correlation function (\ref{eq:g2}). In Fig.~\ref{fig:DB1} (b) we can see that there is a fast decrease of the phase coherence between the sites 1 and 3, indicating the destruction of the BEC. Furthermore, the number correlations, Fig.~\ref{fig:DB1} (c), shows strong anti-correlations. This result shows that the atoms bunch at one of the non-lossy sites, while in the other non-lossy site very few atoms remain. From Figs.~\ref{fig:DB1} (a-d) it is apparent that the system reaches a meta-stable steady state, which decay very slowly. Thus, for sufficient large number of atoms, this state can survive longer than the experimentally relevant time-scales.

\begin{figure}%\sidecaption
\resizebox{0.9\hsize}{!}{
%\centering
%\resizebox{0.75\columnwidth}{!}{
\includegraphics{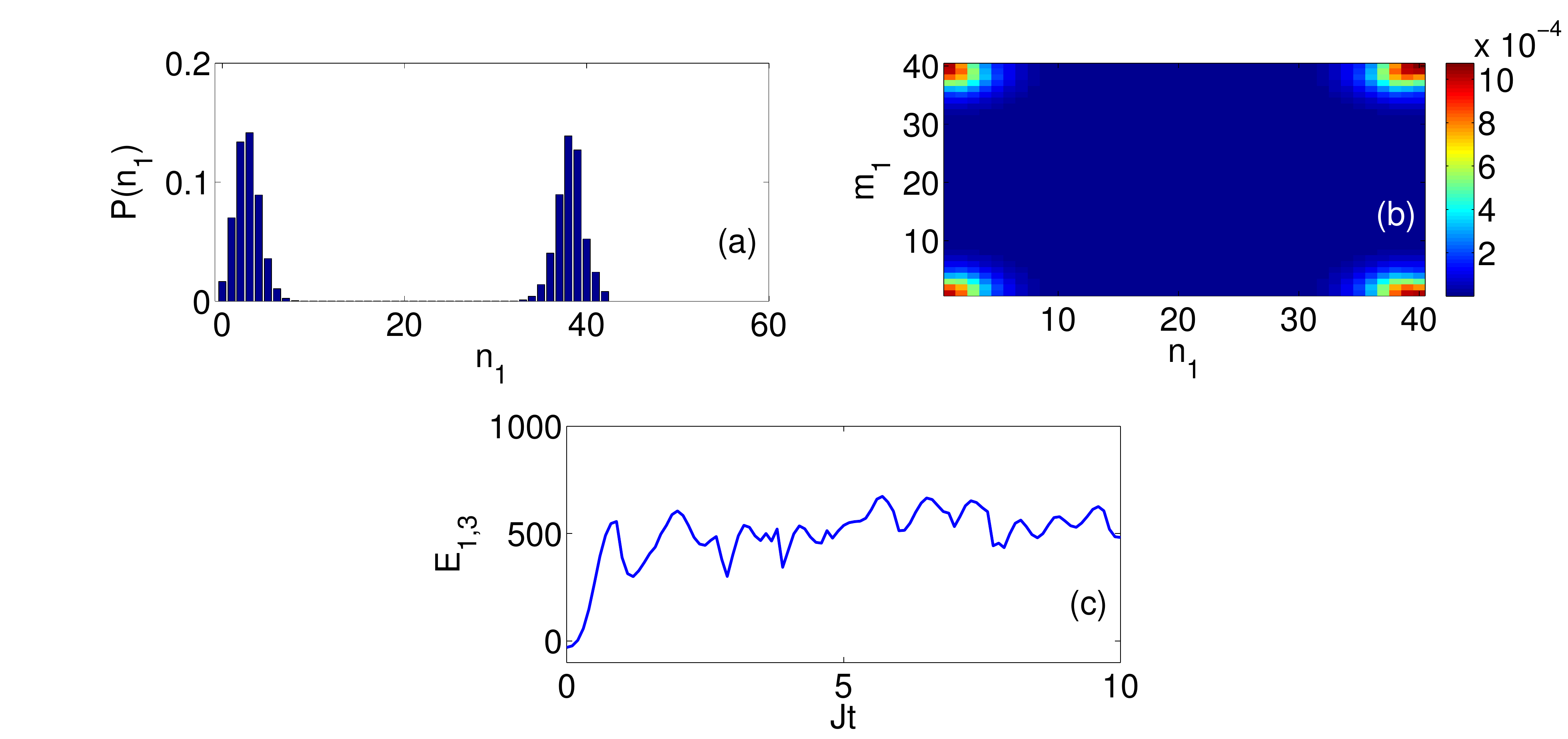}}
\caption{\label{fig:DB2}
(a) Full counting statistics in the first well and (b) density matrix $|\rho(n_1,n_2,n_3;n_1',n_2',n_3')|$ for fixed particle number $n=40$ and $n_2=n_2'=0$, at $t=10J^{-1}$. (c) The evolution of the entanglement parameter $E_{1,3}$. The simulation parameters are the same as in Fig.~\ref{fig:DB1}.
}
\end{figure}

In order to further enlighten the many-body state, in Fig.~\ref{fig:DB2} (a) we have plotted the full counting statistics of the particle number in site 1 at time $t=10J^{-1}$, when the meta-stable steady state has been reached. As one can see the probability distribution $P(n_1)$ becomes bimodal. Almost all atoms are bunched either in the first site or in the third. In other words in site 1 or 3 a discrete breather is formed and the total state is described by a quantum superposition of two breathers one localized in the first and the second to the third lattice site. We call this many-body state as a ``breather state". 

In the breather state the two discrete breathers are fully coherent. To show this we first note that the full many-body state, $\hat{\rho}$, can be written as an incoherent sum of states with different particle number $n$, since the particle loss proceeds via incoherent jumps only. So, we can write
\begin{equation}
\hat{\rho} = \sum_n p_n \hat{\rho}^{(n)}.
\end{equation}
In Fig.~\ref{fig:DB2} (b) we depict the density matrix $\hat{\rho}^{(n=40)}(Jt=10)$ for a subset of indices, fixing $n_2=m_2=0$, after the formation of the breather state. This figure shows that there is full coherence between the contributions with small and large particle number $n_1\lesssim 40$ and $n_1\gtrsim 0$. One can write the breather state as a superposition of states of the form
\begin{equation}\label{eq:int}
|n_1,n_2,n-n_1-n_2\rangle +e^{i\phi} |n-n_1-n_2,n_2,n_1\rangle,
\end{equation}
indicating that these states can be used for precision interferometry~\cite{kordas12,kordas13}. This is true since, although the total particle number forming the state (\ref{eq:int}) varies statistically, the coherence of the sites 1 and 3 is guaranteed.

The previous discussion indicates that the particles of a breather state are strongly entangled, that is if some particles are measured in one site, then the rest of the particles will be projected onto the same site with very large probability. In order to detect this entanglement, the authors of~\cite{kordas12,kordas13} introduced an entanglement parameter, especially useful for quantum jump simulations:
\begin{equation}\label{eqn:ent_para}
E_{a,b} = \langle (\hat{n}_a - \hat{n}_b)^2\rangle - \langle \hat{n}_a - \hat{n}_b\rangle^2 - \langle \hat{n}_a + \hat{n}_b\rangle - \frac{1}{2N_{\rm t}^2} \sum_{k,\ell=1}^{N_{\rm t}} [\langle \hat{n}_a - \hat{n}_b\rangle_k - \langle \hat{n}_a - \hat{n}_b\rangle_\ell]^2,
\end{equation}
for the sites $a$ and $b$, with $ \langle ...\rangle_{k}$ the expectation value in the pure state $|\Psi^{(k)}\rangle$ which is automatically provided by the quantum jump method, see Eq. (\ref{eq:qj1}). It can be shown~\cite{kordas12,kordas13} that $E_{a,b}$ is negative only for separable states, so if $E_{a,b}>0$ then we have entanglement between the sites $a$ and $b$. Since this entanglement criterion turns out to be useful for our BH type of models, we review a rigorous proof of the criterion in appendix~\ref{sec:ent}. In Fig.~\ref{fig:DB2} (c) we have plotted the evolution of the entanglement parameter (\ref{eqn:ent_para}). The entanglement parameter assumes a large positive value, $E_{1,3}\approx 500$, which clearly reveals the presence of many-body entanglement.
\section{The mean-field approximation}
\label{sec:2}
Exact numerical solutions of the master equation (\ref{eq:DBH}) are possible only for small systems, thus approximate methods are needed in order to study systems, in particular, of larger sizes as typically realized in the experiment. For a lattice filled with BEC the mean-field (MF) approximation has proven very accurate, at least for closed systems. This approximation is valid in the limit where $N\rightarrow\infty$ while $UN={\rm const.}$, and the state of the system remains always in a pure BEC state as given by Eq. (\ref{eq:BEC}). For closed systems, the dynamics of $\psi_j$ is given by the discrete non-linear Schr\"{o}ndinger (DNLS) equation~\cite{kevrekidis09,kevrekidis04,brazhnyi04,morsch06}.
We must note that except the DNLS equation other MF methods are available, see for example~\cite{buon04}.
For dissipative systems, however, the above assumptions are not necessarily fulfilled. For example, for particle loss processes the total particle number can significantly be reduced during the time evolution, or phase noise can destroy the global coherence of the system. In consequence, such open systems may not be well described by a pure state, see Eq. (\ref{eq:BEC}).

We will derive the MF approximation using the evolution equations for the single-particle density matrix (SPDM), $\sigma_{j,k}\equiv \langle \hat{\alpha}^\dag_j \hat{\alpha}_k \rangle = {\rm Tr}(\hat{\alpha}^\dag_j \hat{\alpha}_k\hat{\rho})$. Due to the interparticle interaction term these evolution equations do not form a closed system of equations, building that way the so-called Bogoliubov-Born-Green-Kirkwood-Yvon (BBGKY) hierarchy, known from statistical physics \cite{Reichl2004}:
\begin{eqnarray}\nonumber
 i\frac{d}{dt}\langle \hat{\alpha}_i^\dagger \hat{\alpha}_j\rangle &=& f_1\left(\langle \hat{\alpha}_{i'}^\dagger \hat{\alpha}_{j'}\rangle,\langle\hat{\alpha}_{i'}^\dagger \hat{\alpha}_{j'} \hat{\alpha}_{k'}^\dagger \hat{\alpha}_{l'}\rangle\right)\\
 i\frac{d}{dt}\langle \hat{\alpha}_i^\dagger \hat{\alpha}_j \hat{\alpha}_k^\dagger \hat{\alpha}_l\rangle &=& f_2\left(\langle \hat{\alpha}_{i'}^\dagger \hat{\alpha}_{j'} \hat{\alpha}_{k'}^\dagger \hat{\alpha}_{l'}\rangle,\langle\hat{\alpha}_{i'}^\dagger \hat{\alpha}_{j'} \hat{\alpha}_{k'}^\dagger \hat{\alpha}_{l'} \hat{\alpha}_{m'}^\dagger \hat{\alpha}_{n'}\rangle\right)\label{eq:hierarchy}\\
 \nonumber i\frac{d}{dt}\langle \hat{\alpha}_i^\dagger \hat{\alpha}_j \hat{\alpha}_k^\dagger \hat{\alpha}_l \hat{\alpha}_{m}^\dagger \hat{\alpha}_{n}\rangle &=& f_3\left(\langle \hat{\alpha}_{i'}^\dagger \hat{\alpha}_{j'} \hat{\alpha}_{k'}^\dagger \hat{\alpha}_{l'}\hat{\alpha}_{m'}^\dagger \hat{\alpha}_{n'} \rangle,\langle\hat{\alpha}_{i'}^\dagger \hat{\alpha}_{j'} \hat{\alpha}_{k'}^\dagger \hat{\alpha}_{l'} \hat{\alpha}_{m'}^\dagger \hat{\alpha}_{n'}\hat{\alpha}_{e'}^\dagger \hat{\alpha}_{g'}\rangle\right)\\
 \nonumber &\vdots&
\end{eqnarray}
$f_1$, $f_2$, $f_3$, $\ldots$  are functions which increase in complexity with the order. Taking into account the assumptions of the MF approximation we can write the four-point correlation functions in the form
\begin{equation}
 \langle \hat{\alpha}_j^\dag \hat{\alpha}_k \hat{\alpha}_\ell^\dag \hat{\alpha}_m \rangle \approx \langle \hat{\alpha}_j^\dag 
 \hat{\alpha}_k \rangle \langle \hat{\alpha}_\ell^\dag \hat{\alpha}_m \rangle,
\end{equation}
so now we can truncate the hierarchy (\ref{eq:hierarchy}) and keep the first set of equations that includes only the two-point correlation functions.

For example, in the case of the localized single-particle losses and phase noise the MF approximation gives the following equations of motion
\begin{eqnarray}\label{eq:MF}\nonumber
 i\frac{d}{dt} \sigma_{j,k} &=& (\varepsilon_k - \varepsilon_j)\sigma_{j,k} - J(\sigma_{j,k+1} + \sigma_{j,k-1} - \sigma_{j+1,k} - \sigma_{j-1,k})\\
\nonumber && + U(\sigma_{k,k}\sigma_{j,k} - \sigma_{j,j}\sigma_{j,k})\\
&&-i\frac{\gamma_j + \gamma_k}{2}\sigma_{j,k} - i\kappa (1-\delta_{j,k})\sigma_{j,k}.
\label{eq:MF1}
\end{eqnarray}
It is worth commenting on the last two terms of Eq. (\ref{eq:MF1}), which describe the dissipation processes. As one can easily see, the phase noise term (the last term) destroys all the coherences between the sites, $\langle \hat{\alpha}_j^\dag \hat{\alpha}_k \rangle$, with the same rate $\kappa$. So it is a decoherence process. On the other hand, the single-particle loss term does not reduce only the particle number in the leaky site $\langle\hat{\alpha}_j^\dag \hat{\alpha}_j\rangle$, but also destroys all the coherences between the leaky site and the other sites. This allows us to use the localized single particle loss as a tool to control and fine tune the coherences in the system. 
\begin{figure}%\sidecaption
%\resizebox{0.4\hsize}{!}{
\centering
\resizebox{0.4\columnwidth}{!}{
\includegraphics{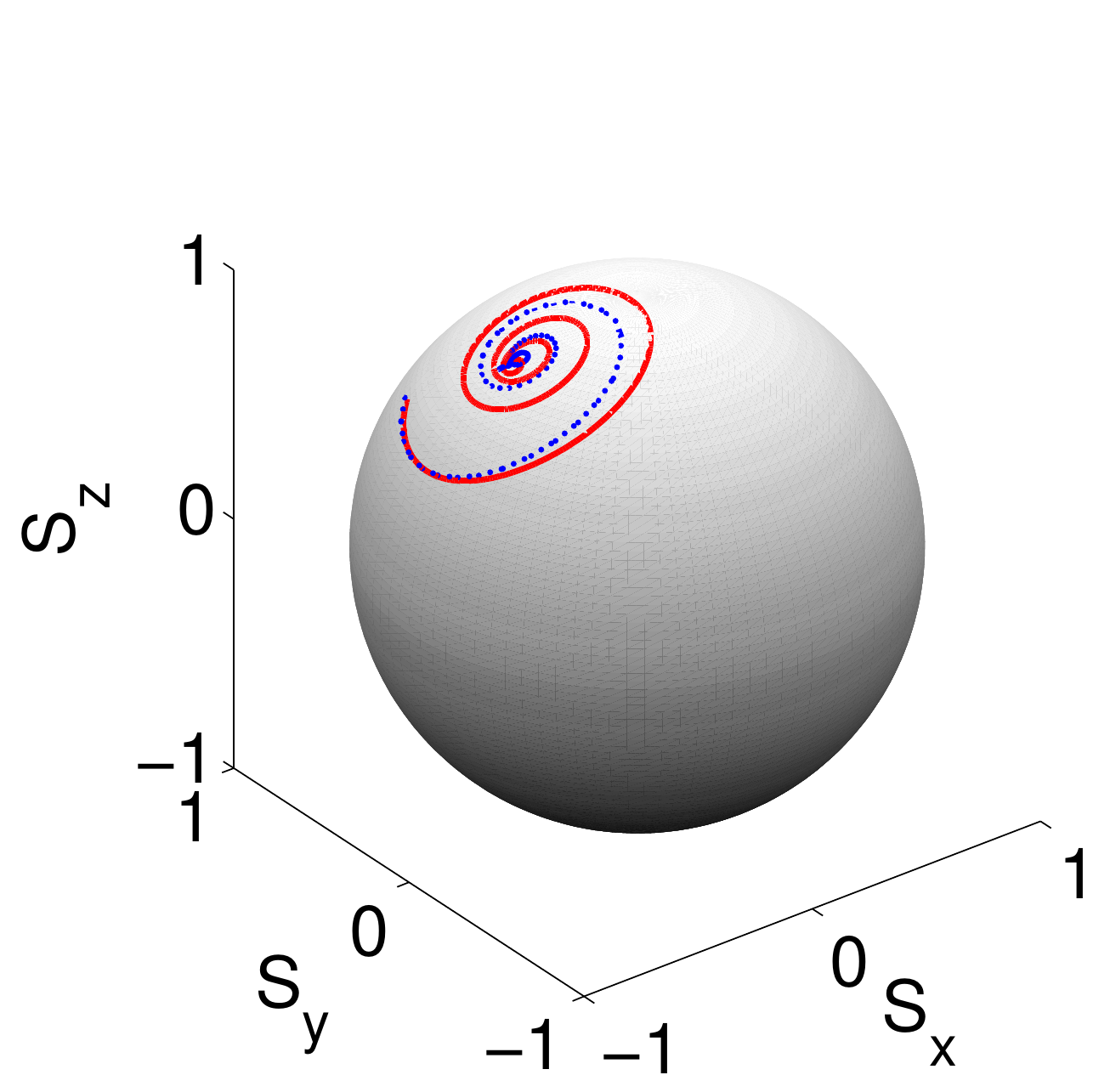}}
\caption{\label{fig:QJvsMF}
Comparison between the mean-field approximation (red solid line) and the quantum jump method (blue dotted line), averaging over 200 trajectories, on the Bloch sphere. The parameters are $Un(0)=6J$, $\gamma_1 = 1J$ and $\gamma_2=0$. The initial state has $\psi_1=\sqrt{0.2}$ and $\psi_2=\sqrt{0.8}e^{1.2\pi i}$ with total particle number is $n(0)=100$.
}
\end{figure}
If we neglect the phase noise term in (\ref{eq:MF1}), $\kappa=0$, then these equations are equivalent to the
non-Hermitian discrete nonlinear Schr\"{o}dinger equation
\begin{equation}\label{eq:DNLS}
 i\frac{d}{dt}\psi_k = (\varepsilon_k - \varepsilon_j)\psi_k -J (\psi_{k+1} + \psi_{k-1}) + U|\psi_k|^2\psi_k - i \frac{\gamma_k}{2}\psi_k
\end{equation}
by the identification $\sigma_{j,k} = \psi_j^*\psi_k$. In this way one comes back to the non-Hermitian DNLS which was heuristically used in older works on dissipative bosonic systems~\cite{liv06,ng09,braz09,Main2014,Graefe2010,QMasterEqMF}. With a gain term, of the same strength as the losses, Eq. (\ref{eq:DNLS}) has been used to study $\mathcal{PT}$-symmetric lattices~\cite{graefe08_2,kono12,pelino13,li13,rodri13}. Such a $\mathcal{PT}$-symmetric equations can be derived as the mean-field limit of a many-body master equation with loss and gain Liouvilians ($\hat{L}_\ell = \hat{\alpha}_\ell^\dag$). However, the strength of the particle loss and gain should be appropriately balanced such that it resembles
the behavior of a discrete $\mathcal{PT}$-symmetric DNLS~\cite{dast14}.

In order to test the approximation we will discuss the case of a Bose-Hubbard dimer with one leaky site, the first, with strong interactions between the bosons and without phase noise. In Fig.~\ref{fig:QJvsMF} we compare the MF approximation with quantum jump results. We observe deviations from the quantum jump results for the dynamics of the Bloch vector, 
\begin{equation}
S\equiv s/n_{\rm tot} \equiv (\langle\hat{\alpha}_1^\dag \hat{\alpha}_2 + \hat{\alpha}_2^\dag \hat{\alpha}_1\rangle, i\langle\hat{\alpha}_1^\dag \hat{\alpha}_2 - \hat{\alpha}_2^\dag \hat{\alpha}_1\rangle, \langle\hat{\alpha}_2^\dag \hat{\alpha}_2 - \hat{\alpha}_1^\dag \hat{\alpha}_1\rangle)/\langle \hat{\alpha}_1^\dag \hat{\alpha}_1 +  \hat{\alpha}_2^\dag \hat{\alpha}_2\rangle.
\end{equation}
The mean-field approximation overestimates the oscillations
of the components of $S(t)$, since it cannot take into account the decoherence of the BEC. However, we must note that the MF approximation captures qualitatively the behaviour of the system.

\subsection{Stochastic resonance in a leaky double well}
\label{subsec:2.1}
The dynamics of the coherence of a BEC in a double-well which is subject to phase noise and single particle losses has been studied in a series of works at the level of the MF approximation and by the exact quantum jump method~\cite{witt08,trim08_1,witt09}. It was shown that a finite particle-loss rate can lead to a maximum of the coherence between the two condensate modes. This phenomenon, that is the response of a system to an external driving can be facilitated in the presence of an appropriate amount of noise, is known in the field of nonlinear dynamical systems as stochastic resonance (SR)~\cite{benzi81,ReviewsSR}.

\begin{figure}
%\resizebox{0.4\hsize}{!}{
\centering
\resizebox{0.9\columnwidth}{!}{
\includegraphics{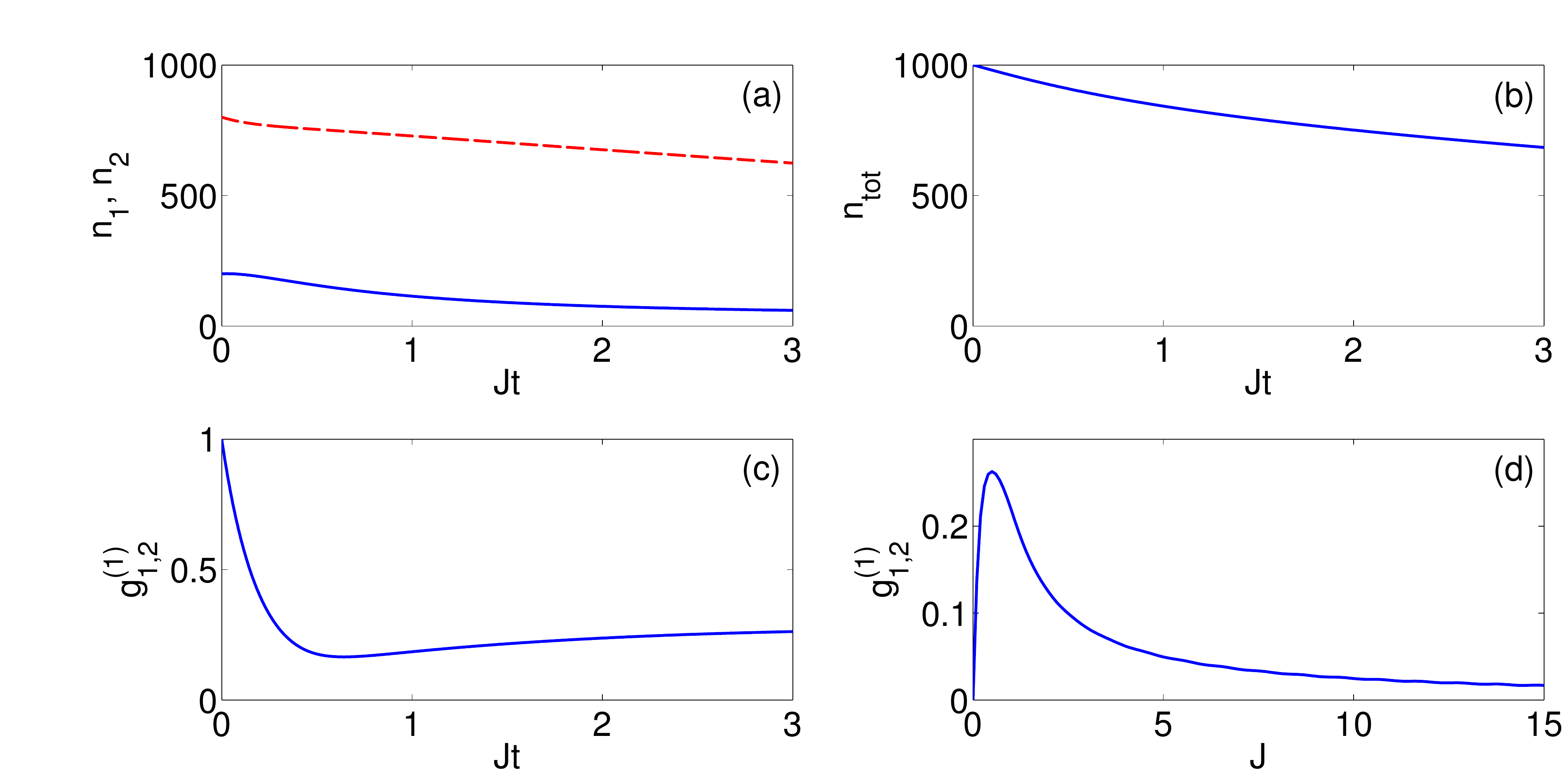}}
\caption{\label{fig:SR}
Relaxation to the quasi-steady state for $\gamma_1=4J$, $\kappa=10J$, $\varepsilon_1=\varepsilon_2=0$ and $Un(0)=4J$. The evolution (a) of the population in the first (blue solid line) and second (red dashed line) site, (b) of the total population, (c) of the first order coherence function $g^{(1)}_{1,2}$, and (d) the development of the stochastic resonance maximum of the first order coherence at fixed propagation time. The parameters in (a-c) corresponds to the maximum of the curve in (d). For all calculations the initial state is $\psi_1=\sqrt{0.2}$, $\psi_2=\sqrt{0.8}e^{1.2\pi i}$ and $n(0)=1000$.
}
\end{figure}

In the MF approximation the evolution of such a system is given by the equations (\ref{eq:MF}). In Figs.~\ref{fig:SR} (a-c) we have plotted the evolution of the particle numbers, $n_1$, $n_2$ and $n_{\rm tot}$, and the first-order coherence function for weak interactions and strong phase noise. We observe that the decay of the particles is slow and most importantly the coherence function, Fig.~\ref{fig:SR}, saturates, $g^{(1)}_{1,2}\approx0.25$, thus the system reaches a metastable steady state. Fig.~\ref{fig:SR} (d) depicts the phase coherence function at fixed propagation time, after the steady-state has been reached, as a function of the tunneling strength. One observes a SR-like maximum of the phase coherence for a finite value of the tunneling strength. This behaviour is counter-intuitive since an increase of the coupling between the two sites reduces their phase coherence. For a non-interacting system one can show that the stochastic resonance maximum occurs when the particle loss and phase noise match the time scales of the intrinsic dynamics~\cite{witt09}, i.e. when $\gamma_1, \kappa \sim 4J$, which defines the single-particle bandwidth of the tight binding model Eq. \eqref{eq:BH}.
\section{The Bogoliubov backreaction method}
\label{sec:3}
The mean-field approximation assumes always a pure BEC, thus many-body effects, like the depletion of the BEC, are inaccessible by this approximation. So if one is interested in quantum effects, beyond-MF methods are necessary. A simple beyond MF approximation can be derived from the BBGKY hierarchy (\ref{eq:hierarchy}). We take into account also the evolution equations for the four-point correlation functions, or equivalently the variances
\begin{equation}\label{eq:var}
\Delta_{jk\ell m} = \langle\hat{\alpha}_j^\dag \hat{\alpha}_k \hat{\alpha}_\ell^\dag \hat{\alpha}_m\rangle - \langle\hat{\alpha}_j^\dag \hat{\alpha}_k\rangle \langle\hat{\alpha}_\ell^\dag \hat{\alpha}_m\rangle
\end{equation}
explicitly and find an appropriate truncation for the six-point correlation function. This method, the so-called Bogoliubov back-reaction (BBR) method, was first presented for closed systems~\cite{vard01,ang01,Tikho07} and gives accurate predictions for large, but finite particle numbers, and for systems close to a pure BEC state.

Let us start from the evolution equations for the four-point correlation functions
\begin{eqnarray}
 \nn \frac{d}{dt}\langle \hat{\alpha}_j^\dag \hat{\alpha}_k \hat{\alpha}_\ell^\dag \hat{\alpha}_m\rangle &=& 
{\rm Tr} \{\hat{\alpha}_j^\dag \hat{\alpha}_k \hat{\alpha}_\ell^\dag \hat{\alpha}_m [\hat{H},\hat{\rho}]\}\\
\nn &=& (\varepsilon_k + \varepsilon_m - \varepsilon_j - \varepsilon_\ell)\langle \hat{\alpha}_j^\dag \hat{\alpha}_k \hat{\alpha}_\ell^\dag \hat{\alpha}_m\rangle\\
\nn && -J (\langle \hat{\alpha}_j^\dag \hat{\alpha}_k \hat{\alpha}_\ell^\dag \hat{\alpha}_{m+1}\rangle + \langle \hat{\alpha}_j^\dag \hat{\alpha}_k \hat{\alpha}_\ell^\dag \hat{\alpha}_{m-1}\rangle +
\langle \hat{\alpha}_j^\dag \hat{\alpha}_{k+1} \hat{\alpha}_\ell^\dag \hat{\alpha}_m\rangle \\
\nn&&+\langle \hat{\alpha}_j^\dag \hat{\alpha}_{k-1} \hat{\alpha}_\ell^\dag \hat{\alpha}_m\rangle - \langle \hat{\alpha}_{j+1}^\dag \hat{\alpha}_k \hat{\alpha}_\ell^\dag \hat{\alpha}_m\rangle
-\langle \hat{\alpha}_{j-1}^\dag \hat{\alpha}_k \hat{\alpha}_\ell^\dag \hat{\alpha}_m\rangle\\
\nn &&-\langle \hat{\alpha}_j^\dag \hat{\alpha}_k \hat{\alpha}_{\ell+1}^\dag \hat{\alpha}_m\rangle - \langle \hat{\alpha}_j^\dag \hat{\alpha}_k \hat{\alpha}_{\ell-1}^\dag \hat{\alpha}_m\rangle)\\
\nn && +U( \langle \hat{\alpha}_j^\dag \hat{\alpha}_k \hat{n}_k \hat{\alpha}_\ell^\dag \hat{\alpha}_m\rangle +\langle \hat{\alpha}_j^\dag \hat{\alpha}_k \hat{\alpha}_\ell^\dag \hat{\alpha}_m\hat{n}_m\rangle\\
&& - \langle \hat{n}_j\hat{\alpha}_j^\dag \hat{\alpha}_k \hat{\alpha}_\ell^\dag \hat{\alpha}_m\rangle - \langle \hat{\alpha}_j^\dag \hat{\alpha}_k \hat{n}_\ell\hat{\alpha}_\ell^\dag \hat{\alpha}_m\rangle)
\end{eqnarray}
The interaction term involves a six-point correlation function, so to obtain a closed set of equations we use the following truncation~\cite{Tikho07}
\begin{eqnarray}
 \langle \hat{\alpha}_j^\dag \hat{\alpha}_k \hat{\alpha}_\ell^\dag \hat{\alpha}_m \hat{\alpha}_r^\dag \hat{\alpha}_s\rangle &\approx&
\langle \hat{\alpha}_j^\dag \hat{\alpha}_k \hat{\alpha}_\ell^\dag \hat{\alpha}_m\rangle \langle\hat{\alpha}_r^\dag \hat{\alpha}_s\rangle
+\langle \hat{\alpha}_j^\dag \hat{\alpha}_k\hat{\alpha}_r^\dag \hat{\alpha}_s\rangle\langle \hat{\alpha}_\ell^\dag \hat{\alpha}_m \rangle \label{eq:BBR2}\\
\nn &&  +\langle\hat{\alpha}_\ell^\dag \hat{\alpha}_m \hat{\alpha}_r^\dag \hat{\alpha}_s\rangle \langle \hat{\alpha}_j^\dag \hat{\alpha}_k\rangle - 2 \langle \hat{\alpha}_j^\dag \hat{\alpha}_k\rangle \langle \hat{\alpha}_\ell^\dag \hat{\alpha}_m\rangle \langle\hat{\alpha}_r^\dag \hat{\alpha}_s\rangle.
\label{eq:truBBR}
\end{eqnarray}
For a pure BEC, described by the state (\ref{eq:BEC}), the relative error induced by this truncation vanishes as $1/N^2$ with increasing particle number. Indeed, the six-point correlation function scales as $N^3$, while the error introduced by the truncation (\ref{eq:BBR2}) increases linearly with $N$. Thus, the BBR method provides a better description of the many-body dynamics than the simple MF approximation, since it includes higher order correlation functions at least approximately.
Using the truncation (\ref{eq:BBR2}), the evolution of the variances is given by
\begin{eqnarray}
  \nn i\frac{d}{dt}\Delta_{jk\ell m} &=& -J [\Delta_{j,k,\ell,m+1} + \Delta_{j,k,\ell,m-1} + \Delta_{j,k+1,\ell,m} + \Delta_{j,k-1,\ell,m}\\
  \nn && -\Delta_{j,k,\ell+1,m} - \Delta_{j,k,\ell-1,m} - \Delta_{j+1,k,\ell,m} -\Delta_{j-1,k,\ell,m}]\\
  \nn && + U [\Delta_{kk\ell m}\sigma_{jk} - \Delta_{jj\ell m}\sigma_{jk} + \Delta_{jkm m}\sigma_{\ell m} - \Delta_{jk\ell \ell}\sigma_{\ell m}\\
  && + \Delta_{jk\ell m}(\sigma_{kk}+\sigma_{mm} - \sigma_{\ell\ell} - \sigma_{jj})].
\end{eqnarray}

The BBR method works only close to a pure BEC state. It breaks down if we have significant depletion of the condensate mode~\cite{ang01}. This hierarchy truncation is a systematic perturbative approximation, but it is state dependent. That is, the perturbative parameter is small only for a special class of states, the coherent states defined in Eq.~(\ref{eq:BEC}). Let us discuss, for example, the case of a BEC trapped in a double well. If $\lambda_0$ is the largest eigenvalue of the SPDM, $\sigma_{j,k}$, then for a state close to pure BEC we have $e=1-\lambda_0/n_{\rm tot}\ll 1$. Thus, if the evolution starts from a pure BEC state one can use $e$ as a perturbative parameter for the evolution. In this context, the MF approximation is zeroth order in $e$ since the state is always a pure BEC. One order higher we have the BBR approximation. We write
\begin{equation}
 \hat{\alpha}_j^\dag \hat{\alpha}_k = \sigma_{j,k} + \hat{\delta}\sigma_{j,k},
\end{equation}
where the c-numbers, $\sigma_{j,k}$, is of order $N$ and all the matrix elements $\hat{\delta}\sigma_{j,k}$ remain smaller than $\mathcal{O}(Ne^{1/2})$ in the whole evolution. Then the normalized variances $\Delta_{jk\ell m}N^{-2}$ will be of order $e$.

In~\cite{ang01} the authors showed that if one starts from the unstable fixed point, of the classical phase space of the BH dimer, the BBR method predicts the right quantum break time. After this time, however, the BBR results deviates significantly from the exact results. So the BBR method, for closed systems, is a useful tool to provide quantities which are not accessible by the MF approximation and can accurately predict the deviations from the MF dynamics. As we are going to see next, the BBR method, for certain kinds of dissipation, in particular, localized loss is typically quite accurate. This originates from the fact that such a loss, provided it is sufficiently strong, effectively damps excitations to higher orders of the correlation functions.

The BBR approximation can be easily generalized for open systems~\cite{witt11,trim11}. Let us demonstrate the method using the example of
localized particle loss and phase noise. In this case, the dynamics of the four-point correlation function is given by
\begin{eqnarray}
 \nn \frac{d}{dt} \langle \hat{\alpha}_j^\dag \hat{\alpha}_k \hat{\alpha}_\ell^\dag \hat{\alpha}_m\rangle &=& {\rm Tr}\{ \hat{\alpha}_j^\dag \hat{\alpha}_k \hat{\alpha}_\ell^\dag \hat{\alpha}_m (\mathcal{L}_{\rm loss}+\mathcal{L}_{\rm noise})\hat{\rho}\}\\
  \nn &=& -\frac{\gamma_j + \gamma_k + \gamma_\ell + \gamma_m}{2} \langle \hat{\alpha}_j^\dag \hat{\alpha}_k \hat{\alpha}_\ell^\dag \hat{\alpha}_m\rangle + \gamma_k \langle \hat{\alpha}_j^\dag \hat{\alpha}_m\rangle \delta_{k\ell}\\
  \nn &&-\kappa (2 + \delta_{km} + \delta_{j\ell} - \delta_{jk} - \delta_{jm} -\delta_{k\ell} - \delta_{\ell m})\\
  && \times \langle \hat{\alpha}_j^\dag \hat{\alpha}_k \hat{\alpha}_\ell^\dag \hat{\alpha}_m\rangle,
\end{eqnarray}
and in terms of the variances we have
\begin{eqnarray}
 \nn \frac{d}{dt} \Delta_{jk\ell m} &=& -\frac{\gamma_j + \gamma_k + \gamma_\ell + \gamma_m}{2} \Delta_{jk\ell m} + \gamma_k \sigma_{jm} \delta_{k\ell}\\
  \nn && -\kappa(\delta_{km} + \delta_{j\ell} -\delta_{jm} - 2\delta_{k\ell})(\Delta_{jk\ell m} + \sigma_{jk}\sigma_{\ell m})\\
  && - \kappa(2 - \delta_{jk} - \delta_{\ell m})\Delta_{jk\ell m}.
\end{eqnarray}

\begin{figure}
%\resizebox{0.4\hsize}{!}{
\centering
\resizebox{0.7\columnwidth}{!}{
\includegraphics{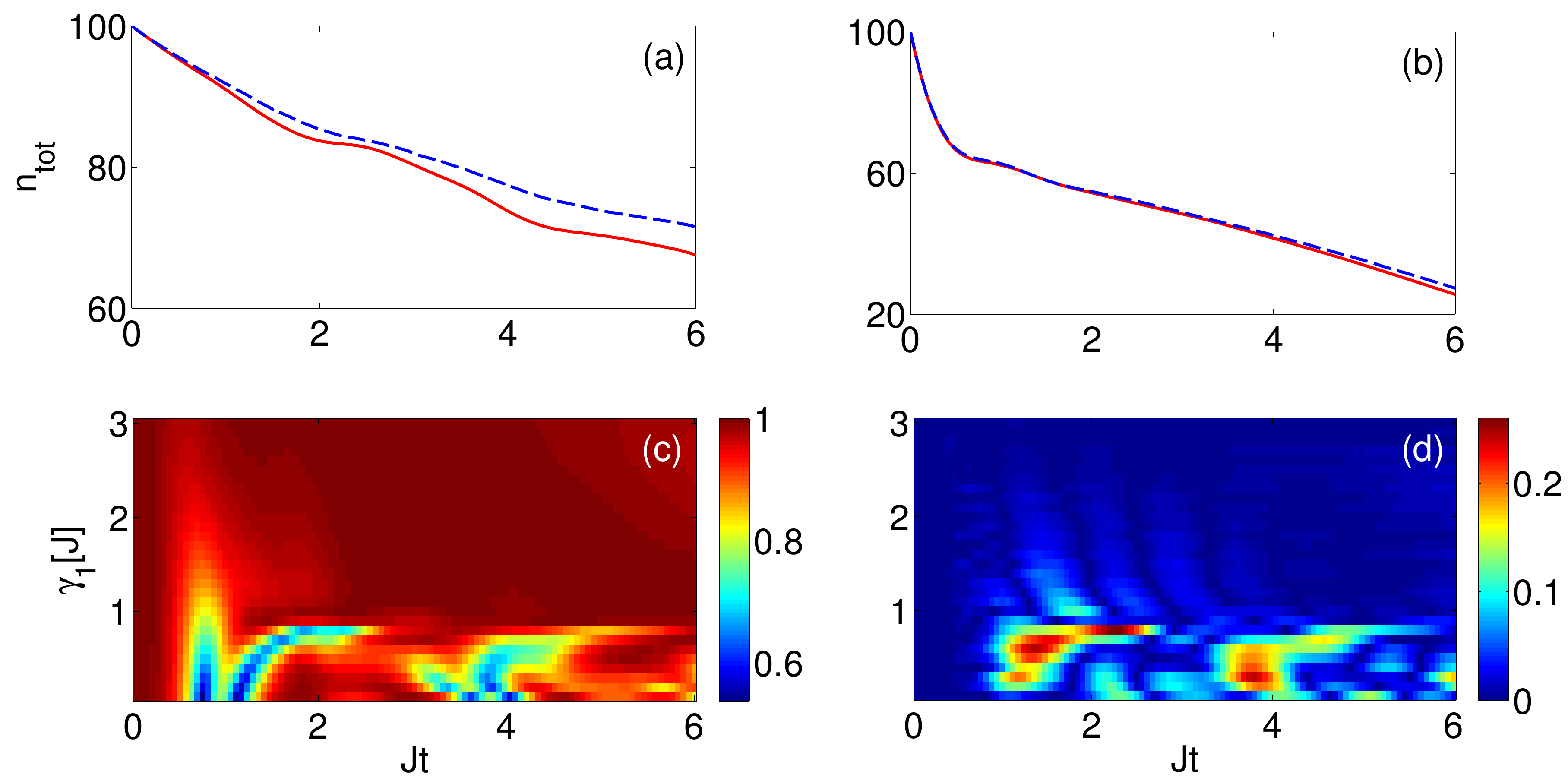}}
\caption{\label{fig:BBRvsQJ}
Numerical test of the BBR approximation for a double well with single particle loss in the first well. (a,b) Dynamics of the total particle number $n_{\rm tot}$ for two different values of the loss rate, $\gamma_1 = 0.2J$ and $\gamma_1 = 2.5J$, respectively.
Red solid line: the BBR approximation, blue dashed line: the quantum jump method, averaging over 200 trajectories. (c) Evolution of the condensate fraction, $\lambda_0/n_{\rm tot}$ as a function of of the loss rate $\gamma_1$. (d) Trace distance between the rescaled SPDMs, $\sigma(t)/n_{\rm tot}(t)$ obtained by the quantum jumps method and the BBR approximation.  The macroscopic interaction strength is $Un(0)=9J$ and the initial state has $\psi_1=\sqrt{0.5}$ and $\psi_2=\sqrt{0.5}e^{i\pi}$ with total particle number $n(0)=100$.
}
\end{figure}

Let us now discuss the validity of the BBR method in the presence of single particle losses. In Fig.~\ref{fig:BBRvsQJ} we have two examples of the dynamics of a BEC in a leaky double-well, comparing the BBR method with numerical exact results of the quantum jump method. As an initial state we have used a pure BEC with equal population and a phase difference of $\pi$ between the wells, that is the unstable fixed point of the dissipationless MF dynamics. As we have discussed, for the closed double well, the BBR method predicts the quantum break time and then it starts to fail. For strong dissipation, Fig.~\ref{fig:BBRvsQJ} (b), the BBR method predicts accurately the evolution of the total population $n_{\rm tot}=\langle \hat{n}_1 + \hat{n}_2\rangle$. Significant differences are observed for weak dissipation, Fig.~\ref{fig:BBRvsQJ} (a). In other words, in the presence of single particle loss the BBR method improves its performance. This is further illustrated in Fig.~\ref{fig:BBRvsQJ} (d), where we have plotted the trace distance of the exact rescaled SPDM and the rescaled SPDM obtained by the BBR method,
\begin{equation}
 D \equiv \frac{1}{2}{\rm Tr}\left\{ \left|\frac{\sigma_{\rm BBR}}{n_{\rm BBR}} - \frac{\sigma_{\rm QJ}}{n_{\rm QJ}}\right| \right\}
\label{eq:diast}
\end{equation}
as a function of time and for different dissipation rates $\gamma_1$. For sufficiently strong losses, we observe that $D$ almost vanishes for all times. Thus, in this regime, the BBR method gives accurate predictions of the quantum dynamics. A hint why this happens is given in Fig.~\ref{fig:BBRvsQJ} (c), where we have the condensate fraction as a function of time for different values of $\gamma_1$. As one can see the BEC tends to remain pure for strong losses for all times. This means that the performance of the BBR method is expected to be accurate -- as an expansion around a pure BEC state.

\begin{figure}
%\resizebox{0.4\hsize}{!}{
\centering
\resizebox{0.9\columnwidth}{!}{
\includegraphics{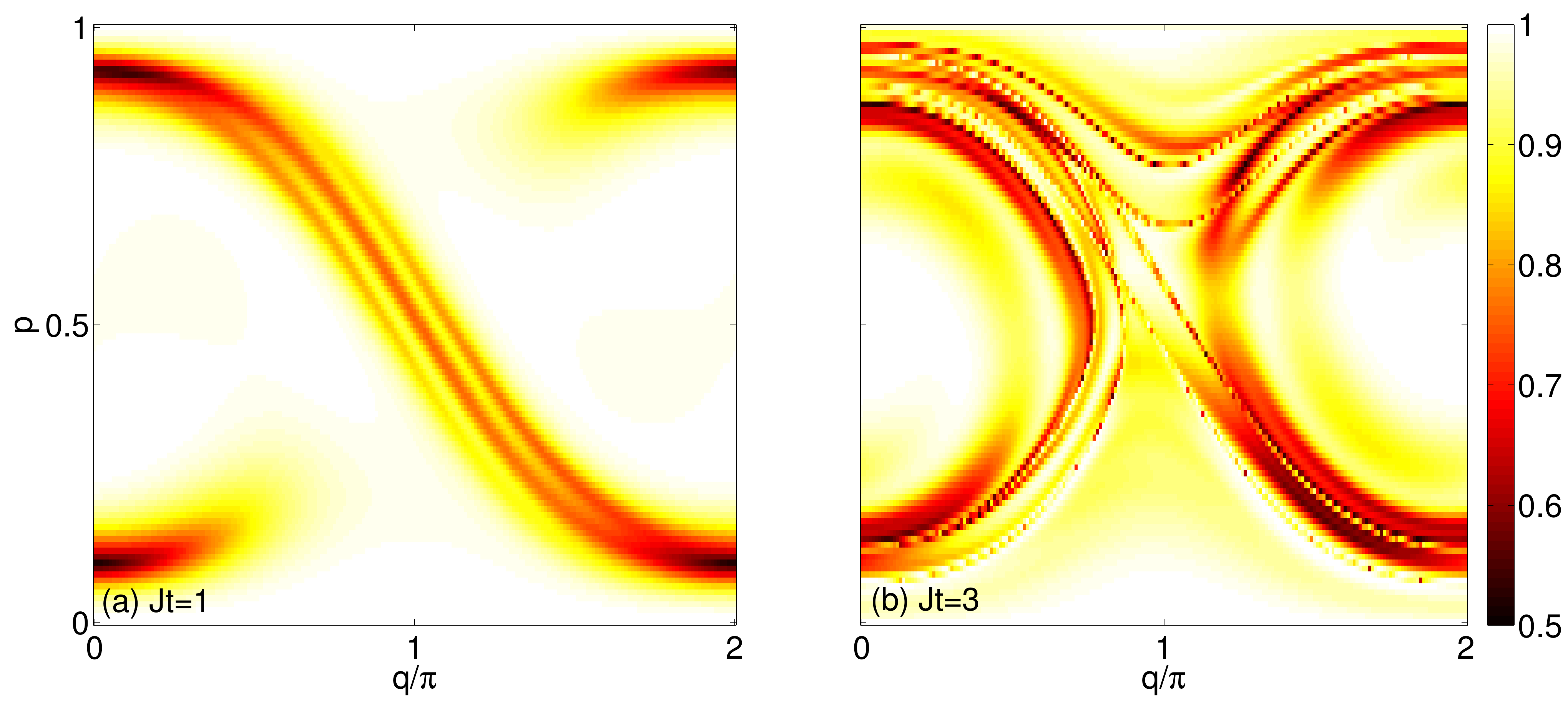}}
\caption{\label{fig:PvsQ}
The condensate fraction $\lambda_0/n_{tot}$ at different time instants scanning the initial state's parameters $p\equiv |\psi_1|^2$ and $q\equiv{\rm\arg}\psi_2-{\rm\arg}\psi_1$.
The parameters are $Un(0)=9J$, $\gamma_1=0.2J$, $\kappa=0$, $n(0)=100$, and $Jt=1$ in (a) and $Jt=3$ in (b).
}
\end{figure}

The BBR method is a state dependent perturbative approximation, so it is important to understand how the initial state affects the predictions of the method. In Fig.~\ref{fig:PvsQ} we are scanning the initial state's parameters $p\equiv |\psi_1|^2\in[0,1]$ and $q\equiv{\rm\arg}\psi_2-{\rm\arg}\psi_1\in[0,2\pi]$ and depict how the condensate fraction $\lambda_0/n_{tot}$ behaves at different time instants. We have chosen weak particle losses, $\gamma_1=0.1J$, since the depletion of the
condensate mode is expected to be more intense. As one can see in only three cases we have a pure BEC at all times: (i) around $p=0.5$ and $q=0$, (ii) for $p=0$ and (iii) for $p=1$. The latter two cases are trivial since in case (ii) there are no particles in the first well, where the loss occurs, so the particles remain in the second due to self trapping \cite{Selftrap}, and in case (iii) all the particles are in the first well so all of them will leave the trap. In case (i) we have initial states around the symmetric state, $(\psi_1,\psi_2) = (1/\sqrt{2},1/\sqrt{2})$, where we have equal populations in the two wells and zero phase difference between them. The symmetric state corresponds to an elliptic fixed point in the non-dissipative classical phase space on the Bloch sphere, that is why the BEC remains almost pure in the whole evolution. For the antisymmetric state, $p=0.5$, $q=\pi$, and along the separatrix we observe that we have a fast depletion of the BEC, which is expected since this state corresponds to the saddle point on the Bloch sphere making the dynamics unstable. Thus, we can conclude that the BBR method will give good predictions even for weak dissipation as long as we start from a BEC state which corresponds to a stable fixed point of the MF dynamics. It is now easy to generalize the above results for an extended lattice. We expect that the method will break down if the initial state is dynamically unstable, e.g., when it starts at the edge of the Brillouin zone~\cite{wu03}.
\subsection{Engineering coherent dark solitons}
\label{subsec:3.1}
It was first proposed in the framework of Gross-Pitaevskii equation~\cite{carr01} that one can create a standing dark soliton by properly engineering the phase and the density of the BEC. We will use phase imprinting in order to engineer the phase and localized particle losses in order to achieve the required density profile, in a lattice with eleven sites filled with a pure BEC. The BBR method will help us to find out for how long the standing dark soliton will survive.

Phase imprinting is a well-known, both theoretical~\cite{carr01,wu02,frantz10} and experimental~\cite{burg99,Densc00}, method to create dark solitons. The BEC is exposed to a pulsed, off-resonant laser light, such that the atoms experience a spatially varying light-shift potential of the form $V(\ell) = V_0\Theta(\ell-k)$ for a time $T$. Here $\Theta$ is the Heaviside function and $k$ is the lattice site where a phase jump will occur. We choose the exposition time $T$ such that $V_0T=\pi$. If $T$ is small enough, we can neglect the tunneling during this time, and just the phase of the condensate wavefunction changes as ${\rm e}^{-iV_0T}$ for $\ell>k$. In this way we have imprinted a sudden change of the
condensate phase at $\ell=k$. However, a phase jump is possible only if the density at $\ell=k$ vanishes. In this way a coherent dark soliton is created at the lattice site $k$ which is however travelling in the lattice~\cite{witt11}.

\begin{figure}
%\resizebox{0.4\hsize}{!}{
\centering
\resizebox{0.9\columnwidth}{!}{
\includegraphics{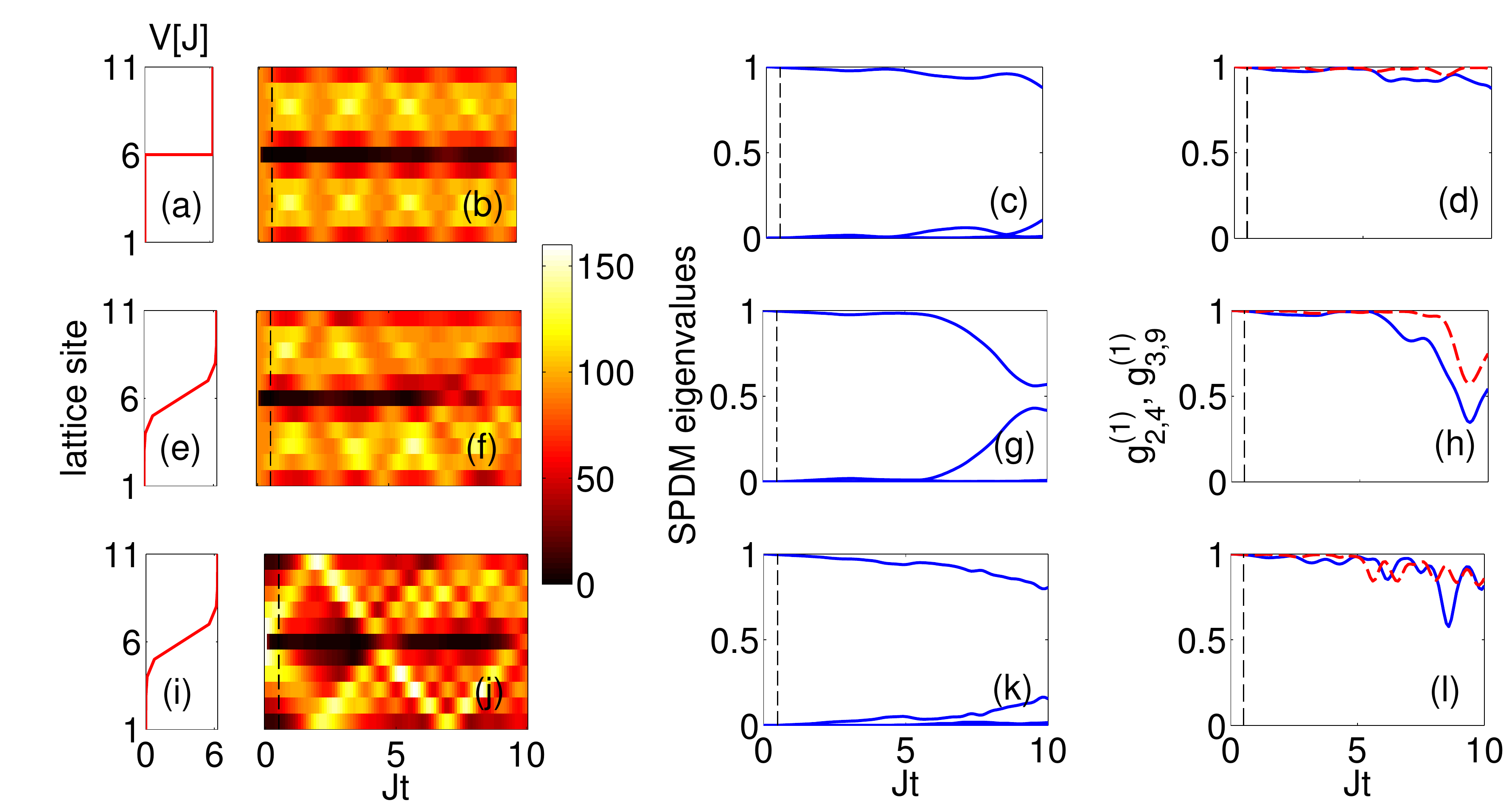}}
\caption{\label{fig:DS}
Generation of standing a dark soliton using phase imprinting and localized particle losses both for times $t<0.5J^{-1}$. The first column depicts the local potential, $V$, which is applied during the phase imprinting stage, the second, third and forth columns show the evolution of particle density, of the eigenvalues of the SPDM and of the first-order coherence between sites 2 and 4, and sites 3 and 9, respectively.
First row: the phase imprint function is a Heaviside step function and the initial state is a pure homogeneous BEC, second row: the phase imprinting function is the smooth function Eq. (\ref{eq:Vfun}), with $C=2$ and the initial state is again a pure homogeneous BEC, and (c) third row: the phase imprinting function as before but now the initial state is the ground state of the DNLS equation.
The parameters are $U=0.02J$, $\gamma_6=20J$, $\kappa=0$ and $n(0)=1000$.
}
\end{figure}

To create a standing dark soliton, in addition to the local potential, which is applied on the upper half of the lattice for $t<0.5J^{-1}$, we use particle loss in the central site for the same period of time, see the first row of Fig.~\ref{fig:DS}. Here the initial state is a pure homogeneous BEC, which is an excited state of the system with our dissipation. The result is a standing dark soliton in the central site. The condensate remains pure over a long time, as it is depicted in Figs.~\ref{fig:DS} (c,d). A careful observation of Figs.~\ref{fig:DS} (b,c) reveals that the small deviations from the pure BEC state in Fig.~\ref{fig:DS} (c) happens when waves in the condensate collide with the dark soliton. These waves are produced from the boundaries of the lattice and from the initial generation of the dark soliton. The collisions will finally lead to the depletion of the dark soliton on a time scale depending on the system size.

In real life experiments the local potential, used for the phase imprinting, is never sharp enough to be accurately described by a step function, but there is a finite width at the edge~\cite{burg99}. In order to make our simulations more realistic we use a potential of the form
\begin{equation}
 V(\ell) = \frac{V_0}{1+{\rm e}^{-C(\ell - k)}},
\label{eq:Vfun}
\end{equation}
where $k$ is the central site of the lattice. For $C\rightarrow\infty$ this function is the Heaviside step function. As previously, we use this potential for $t<0.5J^{-1}$, in combination with particle loss in the central site $k=6$ for the same period of time, for an initially pure homogeneous BEC. In the second row of Fig.~\ref{fig:DS} we have the resulting dynamics for $C=2$. We observe the formation of a dark soliton which survives for some time. However, for $t\approx 5.5J^{-1}$ the particles start to fill the vacancy and the soliton is destroyed, see Fig.~\ref{fig:DS} (f). Moreover, the condensate fragments as a consequence of the fact that a second macroscopic eigenvalue of the SPDM emerges, see Fig.~\ref{fig:DS} (g). So, a smooth potential tends to destroy the dark soliton.

Since the collisions of the waves in the BEC with the dark solitons tend to destroy the coherence one can start from a state with vanishing particle density on the boundaries, for example the ground state of the DNLS equation, see Figs.~\ref{fig:DS} (j-l). In this case the waves come only from the generation of the dark soliton, and then the collision happens only half of the times compared to the previous cases. Figs.~\ref{fig:DS} (j-l) show that the dark soliton survives for longer times and it also keeps its coherence.

\section{The Gutzwiller method}
\label{sec4}
The Gutzwiller ansatz consists in assuming that a product of the form 
\begin{equation}
\label{GA0}
|\Psi \rangle = \bigotimes_{j=1}^M |\psi_j \rangle,~|\psi_j \rangle = \sum_{n=0}^\infty C_n^j |n\rangle,
\end{equation}
provides a good description of the state of the (isolated) system described by the BH Hamiltonian (\ref{eq:BH}). The dynamics of the factors in Eq. (\ref{GA0})  is obtained by requiring that the Schr\"odinger equation associated to Hamiltonian (\ref{eq:BH}) is satisfied ``on average", $\langle \Psi| (i \partial_t -{\hat H}_{\rm BH})| \Psi \rangle = 0$ \cite{Jaksch_PRL_89_040402}. A variational procedure results in a set of equations coupling the factors in Eq. (\ref{GA}) \cite{Buonsante_JPA_41_175301}. 
In compact form
\begin{equation}
\label{GA}
i |\dot{\psi}_j \rangle = -i {\hat{\cal H} }_j\,|{ \psi}_j\rangle, \quad {\hat{\cal H} }_j = \varepsilon_j {\hat \alpha}_j^\dag {\hat \alpha}_j^\dag + \frac{U}{2} {\hat \alpha}_j^\dag {\hat \alpha}_j^\dag{\hat \alpha}_j {\hat \alpha}_j - J \left(\Phi_j {\hat \alpha}_j^\dag+ \Phi_j {\hat \alpha}_j^*\right).
\end{equation}
The coupling between neighboring sites characterizing Hamiltonian (\ref{eq:BH}) survives in a MF form in the on-site quantities
\begin{equation}
\label{Phi}
\Phi_j = \sum_{\ell \in j} \langle\psi_\ell| {\hat \alpha}_\ell |\psi_\ell\rangle
\end{equation}
where the sum over $\ell$ is extended to the nearest neighbors of site $j$. 

Despite their MF character, the Gutzwiller equations (\ref{GA}) ---specifically, their lowest-energy stationary state--- allow a qualitative description of the quantum phases characterizing the Bose-Hubbard model (\ref{eq:BH}) in many situations. These include the ``pure case'', where the on-site energies are uniform,  \cite{Krauth_PRB_45_3137,Sheshadri_EPL_22_257,VanOosten_PRA_63_053601},  the trapped case typical of experiments  \cite{jaks98,Zakrzewski_PRA_71_043601} and the disordered case, where the on-site energies are extracted from a random distribution \cite{Sheshadri_PRL_75_4075,Buonsante_PRA_76_011602R,Buonsante_PRA_79_013623,Niederle_NJP_15_075029}.  
 
Likewise, the time-dependent equations (\ref{GA}) proved to be useful in describing dynamic phenomena  in systems governed by Hamiltonian (\ref{eq:BH}), such as the creation of a molecular condensate \cite{Jaksch_PRL_89_040402}, dipole oscillations in a trapped system \cite{Snoek_PRA_76_051603}, the excitation of ``Higgs'' amplitude modes at the two-dimensional superfluid-insulator transition~\cite{enders12}, the creation of negative-temperature states \cite{Rapp_PRA_87_043611} and the disorder-induced decay of superfluid currents \cite{Buonsante_PRA_91_031601}.
 \begin{figure}
\centering
\resizebox{0.9\columnwidth}{!}{
\includegraphics{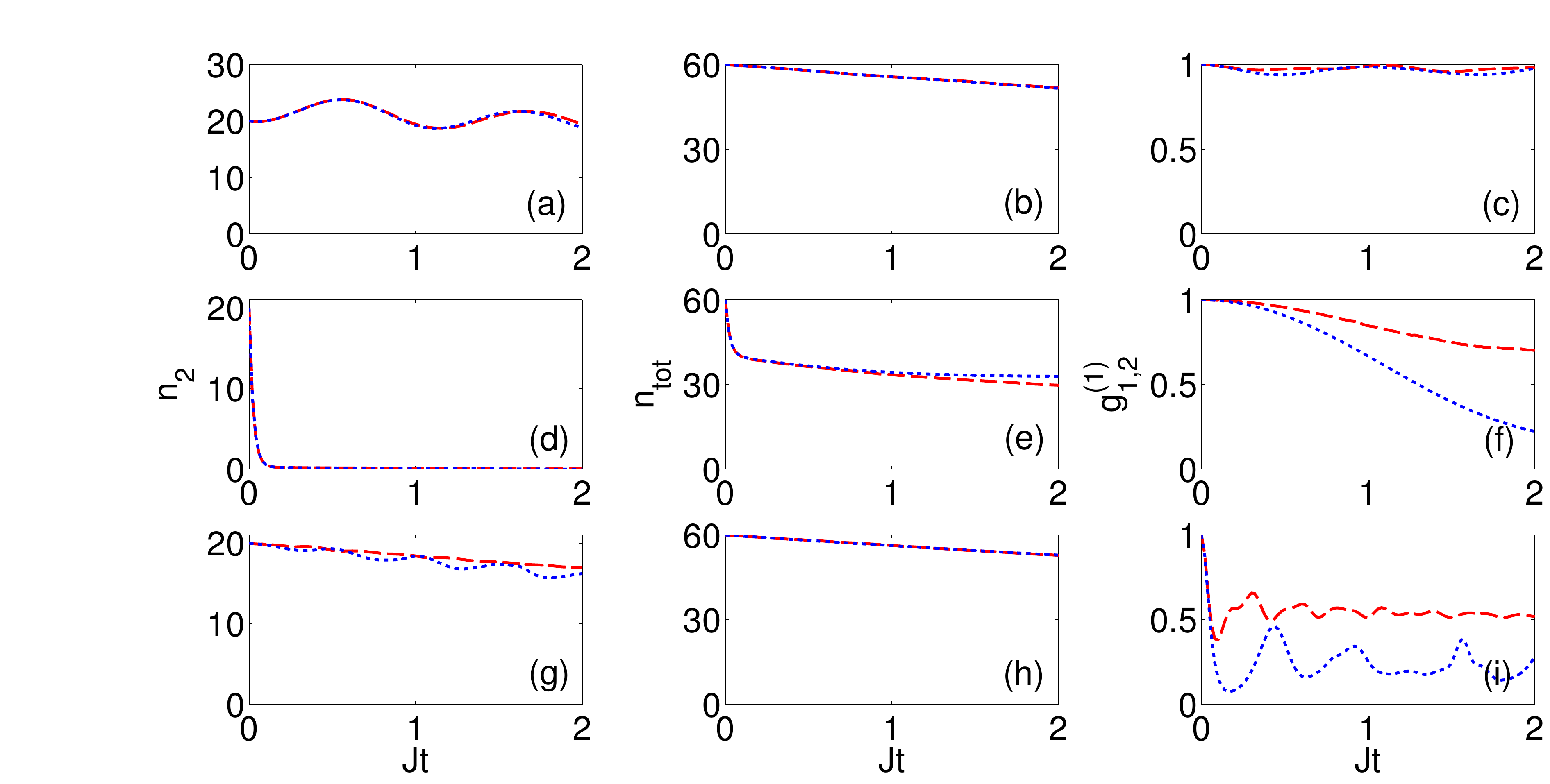}}
\caption{\label{fig:GWvsQJ}
Comparison between the Gutzwiller approximation (blue dotted line) and the quantum jump method (red dashed line). The evolution of the particle number in the second site (first column), of the total particle number (second column) and of the phase coherence between the first and second site (third column). First row, $U=0.2J$, $\gamma_2=0.2J$, second row $U=0.2J$, $\gamma_2=40J$ and third row $U=4J$, $\gamma_2=0.2J$. In all calculation the initial state is a pure BEC with $n_1(0)=n_2(0)=n_3(0)=20$.
}
\end{figure}
 
The dynamical Gutzwiller approach was generalized to dissipative systems as well, see refs.\cite{Diehl_PRL_105_015702,Pizorn_PRA_88_043635,Vida14}.  Eq.~(\ref{eq:DBH}) becomes a set of coupled equations for the on-site density operators 
\begin{equation}
\label{GAL}
\frac{d {\hat \rho}_j}{dt} = -i [\hat {\cal H}_j,{\hat \rho}_j] - \frac{1}{2} \left({\hat L}_j^\dag {\hat L}_j {\hat \rho}_j+{\hat \rho}_j {\hat L}_j^\dag {\hat L}_j -2 {\hat L}_j {\hat \rho}_j {\hat L}_j^\dag\right),
\end{equation}  
where the ``local'' Hamiltonians, $\hat {\cal H}_j$, are the same as in Eq. (\ref{GA}), and the relevant MF parameters assume the more general form
\begin{equation}
\label{PhiL}
 \Phi_j = \sum_{\ell \in j} {\rm Tr}({\hat a}_j {\hat \rho}_j)
\end{equation}  
Note that  Eqs. (\ref{GAL}) and (\ref{PhiL}) are entirely equivalent to Eqs. (\ref{GA}) and (\ref{Phi}) in the absence of the Lindblad terms, when $\rho_j = |\psi_j\rangle\langle\psi_j|$. Even in the presence of a Lindblad term for some $j$, Eq. (\ref{GA}) can be used for sites where no such term exist. Indeed, any site  is influenced by its neighbors in a MF way, only through the c-number in Eq. (\ref{PhiL}), irrespective of the presence of a Lindblad term. In other words, an initially pure Gutzwiller factor, $\rho_j = |\psi_j\rangle\langle\psi_j|$, can become mixed only through the action of a Lindblad term on the relevant site.

In Fig.~\ref{fig:GWvsQJ} we compare the Gutzwiller approximation with the results of the quantum jump method. The system is a flat triple-well with hard-wall boundary conditions and particle loss only in the middle site.
For this system the Gutzwiller equations of motion are written in the following simple form
\begin{eqnarray}
\nonumber\frac{d}{dt}|\psi_1\rangle &=& -i\hat{\mathcal{H}}_1 |\psi_1\rangle, \\
\frac{d}{dt}\hat{\rho}_2 &=& -i[\hat{\mathcal{H}}_2,\hat{\rho}_2] - \frac{\gamma_2}{2} (\hat{\alpha}_2^\dag \hat{\alpha}_2 \hat{\rho}_2 + \hat{\rho}_2 \hat{\alpha}_2^\dag \hat{\alpha}_2 \hat{\rho}_2 - 2\hat{\alpha}_2 \hat{\rho}_2\hat{\alpha}_2^\dag),\\
\nonumber \frac{d}{dt}|\psi_3\rangle &=& -i\hat{\mathcal{H}}_3 |\psi_3\rangle,
\end{eqnarray}
where $\hat{\mathcal{H}}_j$ is given by the Hamiltonian Eq. (\ref{GA}) with $\Phi_1 = \Phi_3 = {\rm Tr}\{ \hat{\alpha}_2 \hat{\rho}_2\}$ and $\Phi_2 = \langle\psi_1|\hat{\alpha}_1|\psi_1\rangle + \langle\psi_3|\hat{\alpha}_3|\psi_3\rangle$. In all calculations the initial condition is a pure homogeneous BEC. In the first row of Fig.~\ref{fig:GWvsQJ} we have weak interactions, $U=0.2J$, and weak dissipation, $\gamma_2=0,2J$. The two methods show a very good agreement. This is a consequence of the fact that for these values of the parameters we have a coherent decay of the BEC~\cite{witt11,trim11}, that is a state of the form (\ref{GA0}), $|\Psi\rangle = |\alpha_1\rangle|\alpha_2\rangle|\alpha_3\rangle$, where $|\alpha_j\rangle$ is eigenstate of $\hat{\alpha}_j$. In the second row of Fig.~\ref{fig:GWvsQJ}, the interactions remain weak, but now the dissipation rate is large, $\gamma_2=40J$. The Gutzwiller method reproduces very accurately the population in the leaky site Fig.~\ref{fig:GWvsQJ} (d), but one observes small deviations in the evolution of the total particle number for $Jt\gtrsim 1.5$, Fig.~\ref{fig:GWvsQJ} (e). Furthermore, we have significant deviations in the phase coherence between the first and second sites, Fig.~\ref{fig:GWvsQJ} (f), overestimating the loss of coherence. The failure of the method to predict the right behaviour for the loss of coherence is expected, since the equations include only a MF contribution for the tunneling between the sites. Finally, in the third row of Fig.~\ref{fig:GWvsQJ}, we have strong interaction, $U=4J$, but weak losses, $\gamma_2=0.2$. In this case the agreement is very good and the only significant deviations are observed in the evolution of the phase coherence Fig.~\ref{fig:GWvsQJ} (i), which again are overestimated. However, it faithfully reproduces the drop of coherence for short times, which is due to the strong interactions. 

In \cite{Vida14}, the Gutzwiller ansatz was used to describe a two-dimensional BH system with a local leaky site. Generally, the method is expected to work the better the higher the coordination number in the lattice, i.e. the larger the spatial dimension and the more the sites are coupled with each other \cite{BV2007}.

\subsection{The quantum Zeno effect}
\label{subsec:4.1}
\begin{figure}%\sidecaption
%\resizebox{0.4\hsize}{!}{
\centering
\resizebox{0.85\columnwidth}{!}{
\includegraphics{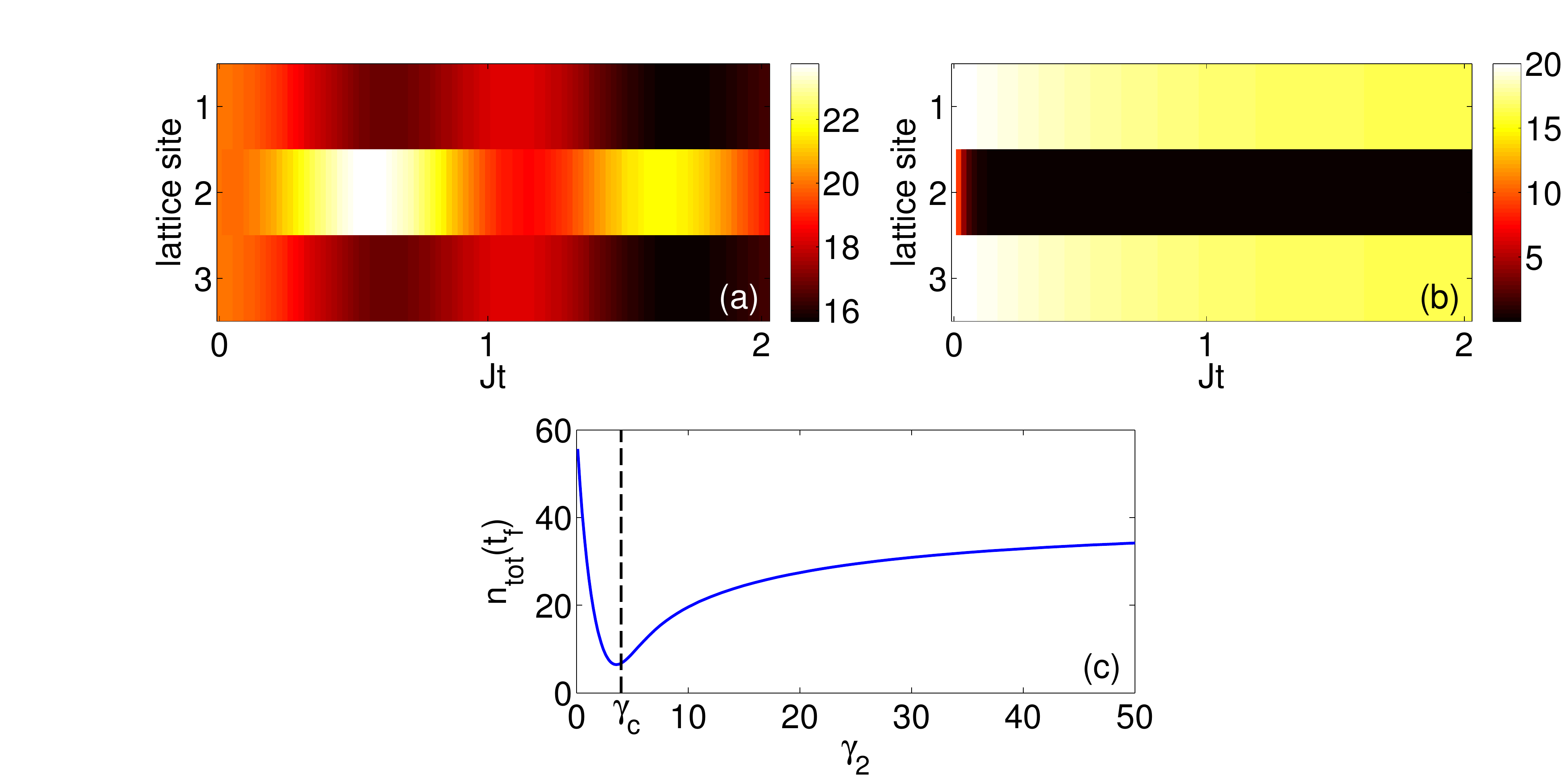}}
\caption{\label{fig:zeno}
(a,b) Evolution of the particle density, $n_j$, for weak, $\gamma_2=0.2J$, and strong, $\gamma_2=40J$, dissipation, respectively. (c) Final value of the total particle number after a fixed propagation time, $Jt_{\rm final} = 2$, as a function of the loss rate in the second site, $\gamma_2$. In all calculations $U=0.2J$ and the initial state is a pure homogeneous BEC with $n_1(0)=n_2(0)=n_3(0)=20$.
}
\end{figure}
Also in our context of bosonic lattice models, strong particle loss can inhibit quantum tunneling towards the leaky lattice site. For single-particle systems, Caldeira and Leggett provided the theoretical framework of such an inhibition of tunneling in the presence of dissipation and decoherence \cite{CL1983}. This suppression of tunneling can be understood by simple analogy to wave optics: a large mismatch of the index of refraction leads to an almost complete reflection of a wave from a surface. This is true for a real index of refraction just as well as for an imaginary index describing an absorbing material such as metal. Similarly, a large difference of the on-site potential, real or imaginary, prevents tunneling of the atoms to the respective lattice site. Another interpretation has been discussed in~\cite{trim11,Cira94,zezy12,Shchesnovich10_1} in terms of the quantum Zeno effect. The particle losses can be viewed as a continuous measurement of the quantum state of the leaky lattice site. This measurement causes a Zeno effect which prevents the tunneling to the observed site. This interpretation is natural in the Ott's group experiments~\cite{gerike08,gerik10,santra15} where the lost atoms (in the form ions) are actually measured by the ion detector.

In Fig.~\ref{fig:zeno} we once again study the time evolution in a three-well trap with hard-wall boundary conditions and with losses in the middle well only. As an initial condition we use a pure homogeneous BEC. In Fig.~\ref{fig:zeno} (a,b) we have plotted the evolution of the atomic density, $\langle \hat{n}_j\rangle$, for weak, $\gamma_2=0.2J$, and strong, $\gamma_2=40J$, dissipation, respectively. For weak dissipation the atoms tunnel to the middle well where they are dissipated with the rate, $\gamma_2$. On the other hand, strong losses lead to the formation of a stable vacancy. But here we can observe something interesting if we compare the two figures. In Fig.~\ref{fig:zeno} (a) the particles tunnel back and forth to the leaky site, the well-known oscillations of the BEC. On the contrary, in Fig.~\ref{fig:zeno} (b) we can see clearly that the particles find it difficult to tunnel, they tend to remain in their original site. So the tunneling rate slows down and the particles that tunnel to the leaky site leave the system. The cause of this effect is dissipation.

The situation becomes clearer in Fig.~\ref{fig:zeno} (c),
where we have plotted the total particle number after fixed propagation time, $t_{{\rm final}}=2J^{-1}$, as
a function of the loss rate, $\gamma_2$. 
As one can see, the residual atom number assumes a minimum for a finite loss rate $\gamma_{{\rm c}}$.
After this critical value, one faces the paradoxical situation that an increase of the loss rate
can suppress the decay of the BEC. We can estimate the critical loss rate, $\gamma_{{\rm c}}$, by matching
the time scale of dissipation $\tau_D=2/\gamma_2$ and tunneling $\tau_J=1/(2J)$, where in the later the factor
$1/2$ accounts for atoms tunneling from two sites to the leaky site. Hence, the critical loss rate is estimated
as $\gamma_{{\rm c}}=4J$. We find a good agreement of our qualitative estimate for $\gamma_{{\rm c}}$ [dashed black line
in Fig.~\ref{fig:zeno} (c)] with the dip in the total particle number.
\section{The truncated Wigner approximation}
\label{sec:5}
Powerful approaches for the numerical simulation of dissipative quantum systems were developed in the field of quantum optics \cite{Gard04,Wern97,Sina02}. One such a method, also known as stochastic phase-space method for obvious reasons, is known by now as truncated Wigner approximation (TWA). Indeed, an equivalent approach to the operator master equation (\ref{eq:DBH}) is that of the c-number equation of motion for the Wigner function, $\mathcal{W}(\{\alpha_j,\alpha_j^* \};t)$, where $\alpha_j\in \mathbb{C}$ are the eigenvalues of the destruction operator:
\begin{equation}
   \hat{\alpha}_j|\alpha_j\rangle = \alpha_j |\alpha_j\rangle, \phantom{0} 
       \langle\alpha_j|\hat{\alpha}_j^\dag =\alpha^*_j \langle\alpha_j|.
\end{equation}
The Wigner function is a quasi-probability distribution, which is defined by the relation
\begin{equation}
\mathcal{W}(\{\alpha_j,\alpha_j^* \};t) = \prod_j \int d\lambda_j \int d\lambda_j^* \exp\{-\lambda_j \alpha_j^* + \lambda_j^*\alpha_j\} \chi_{\rm w}(\{\lambda_j,\lambda_j^* \};t),
\end{equation}
that is the Fourier transform of the characteristic function
\begin{equation}
\chi_{\rm w}(\{\lambda_j,\lambda_j^* \};t) = {\rm Tr} \left\{\hat{\rho}(t) \prod_j \exp\{\lambda_j \hat{\alpha}_j^\dag - \lambda_j^*\hat{\alpha}_j\} \right\},
\end{equation}
with $\lambda_j\in \mathbb{C}$.

In order to map the operator master equation (\ref{eq:DBH}) onto an equation for the Wigner function one uses the following operator correspondences~\cite{Gard04}:
\begin{eqnarray}
   \hat{\alpha}_{j}\hat{\rho} &\leftrightarrow& 
      \left(\alpha_{j}+\frac{1}{2}\frac{\partial}{\partial\alpha_{j}^*}\right)\mathcal{W}, \label{eq:wig1}\\
   \hat{\rho}\hat{\alpha}_{j} &\leftrightarrow &
       \left(\alpha_{j}-\frac{1}{2}\frac{\partial}{\partial\alpha_{j}^*}\right)\mathcal{W}, \\
   \hat{\alpha}_{j}^\dagger\hat{\rho} &\leftrightarrow &
      \left(\alpha_{j}^*-\frac{1}{2}\frac{\partial}{\partial\alpha_{j}}\right)\mathcal{W}, \\
  \hat{\rho}\hat{\alpha}_{j}^\dagger &\leftrightarrow &
     \left(\alpha_{j}^*+\frac{1}{2}\frac{\partial}{\partial\alpha_{j}}\right)\mathcal{W}.\label{eq:wig4}
\end{eqnarray}
For example, let us prove the relation (\ref{eq:wig1}). We have
\begin{eqnarray}
\nonumber {\rm Tr} \left\{ \hat{\alpha}_j\hat{\rho}(t) \prod_k e^{\lambda_k \hat{\alpha}_k^\dag - \lambda_k^* \hat{\alpha}_k} \right\} &=& 
{\rm Tr} \left\{\hat{\rho}(t) \prod_k  e^{\lambda_k \hat{\alpha}_k^\dag}  e^{-\lambda_k^* \hat{\alpha}_k} e^{-\frac{|\lambda_k|^2}{2}} \hat{\alpha}_j\right\} \\
\nonumber &=& {\rm Tr} \left\{\hat{\rho}(t) \prod_k  e^{\lambda_k \hat{\alpha}_k^\dag} e^{-\frac{|\lambda_k|^2}{2}} \left(-\frac{\partial}{\partial \lambda_j^*}e^{-\lambda_k^* \hat{\alpha}_k}\right)\right\}\\
\nonumber &=& {\rm Tr} \left\{\hat{\rho}(t) \prod_k  e^{\lambda_k \hat{\alpha}_k^\dag} \left(\frac{\lambda_j}{2} - \frac{\partial}{\partial \lambda_j^*} \right) e^{-\lambda_k^* \hat{\alpha}_k} e^{-\frac{|\lambda_k|^2}{2}}\right\}\\
\nonumber &=& \left(\frac{\lambda_j}{2} - \frac{\partial}{\partial \lambda_j^*} \right) {\rm Tr} \left\{\hat{\rho}(t) \prod_k  e^{\lambda_k \hat{\alpha}_k^\dag} e^{-\lambda_k^* \hat{\alpha}_k} e^{-\frac{|\lambda_k|^2}{2}}\right\}\\
&=& \left(\frac{\lambda_j}{2} - \frac{\partial}{\partial \lambda_j^*} \right) \chi_{\rm w}(\{\lambda_k,\lambda_k^* \};t).
\end{eqnarray}
The Fourier transform of the last expression confirms (\ref{eq:wig1}).

In order to demonstrate the method we once again use the localized single particle loss mechanism. So, substituting the correspondences (\ref{eq:wig1})-(\ref{eq:wig4}) into the master equation (\ref{eq:DBH}) for $\hat{L}_\ell = \hat{\alpha}_\ell$, we obtain the following evolution equation for the Wigner function:
\begin{eqnarray}\nonumber
  \frac{\partial\mathcal{W}}{\partial t} &=& 2J\sum_{j=1}^{M-1}\Im \bigg[
    \left(\alpha_{j}-\frac{1}{2}\frac{\partial}{\partial\alpha_{j}^*}\right)
\left(\alpha_{j+1}^*+\frac{1}{2}\frac{\partial}{\partial\alpha_{j+1}}\right)\\
   && - \left(\alpha^*_{j+1}-\frac{1}{2}\frac{\partial}{\partial\alpha_{j+1}}\right)
    \left(\alpha_{j}+\frac{1}{2}\frac{\partial}{\partial\alpha_{j}^*}\right)\bigg]\mathcal{W} 
    \nn \\
   &&  +U \sum_{j=1}^{M} \Im\left(\alpha_{j}-\frac{1}{2}\frac{\partial}{\partial\alpha_{j}^*}\right)^2
         \left(\alpha_{j}^*+\frac{1}{2}\frac{\partial}{\partial\alpha_{j}}\right)^2\mathcal{W} \label{eqn-wig}\\
        && - \sum_{j=1}^{M}\frac{\gamma_j}{2}\bigg[  
              \left(\alpha^*_{j}-\frac{1}{2}\frac{\partial}{\partial\alpha_{j}}\right)
              \left(\alpha_{j}+\frac{1}{2}\frac{\partial}{\partial\alpha_{j}^*}\right)  
            \nn\\
   && +\left(\alpha_{j}-\frac{1}{2}\frac{\partial}{\partial\alpha_{j}^*}\right)
           \left(\alpha_{j}^*+\frac{1}{2}\frac{\partial}{\partial\alpha_{j}}\right) 
            -2\left(\alpha_{j}+\frac{1}{2}\frac{\partial}{\partial\alpha_{j}^*}\right)
           \left(\alpha_{j}^*+\frac{1}{2}\frac{\partial}{\partial\alpha_{j}}\right)\bigg]\mathcal{W}.\nn
\end{eqnarray}
The above equation includes third order derivatives arising from the interaction term (the $U$-dependent term in the third line of equation (\ref{eqn-wig})), which make the equation quickly numerically unstable. An approximation that is widely used in optical systems is the truncated Wigner method~\cite{Wern97,Sina02}, which is good as far as the mode occupation numbers are large.
In this approximation one neglects all the terms that include third order derivatives, thus we have the equation
\begin{eqnarray}
   \label{epn-fok}\nonumber
   \frac{\partial\mathcal{W}}{\partial t} &=& \sum_j\frac{\partial}{\partial x_j}\Bigg[J(y_{j+1}+y_{j-1}) 
    +U(y_j - x_j^2y_j - y_j^3) + \frac{\gamma_j}{2}x_j\Bigg] \mathcal{W}  \\
    \nonumber && + \sum_j\frac{\partial}{\partial y_j}\bigg[-J(x_{j+1} +x_{j-1}) - U(x_j - x_j y_j^2 - x_j^3) 
       + \frac{\gamma_j}{2}y_j \bigg]  \mathcal{W} \\
    && + \frac{1}{2} \sum_j \frac{\gamma_j}{4} \left(\frac{\partial^2}{\partial x_j^2}
       + \frac{\partial^2}{\partial y_j^2}\right)\mathcal{W},
\end{eqnarray}
where $x_j,y_j$ are the real and imaginary part of $\alpha_j$ respectively.

Equation (\ref{epn-fok}) is a Fokker-Planck equation, thus it can be rewritten in the language of stochastic differential or Langevin equations. 
To be more precise, consider the Fokker-Planck equation of the form~\cite{Gard09}:
\begin{equation}
 \label{eqn-stoch}
    \frac{\partial\mathcal{W}}{\partial t} = - \sum_j \frac{\partial}{\partial z_j} A_j(\textbf{z},t)\mathcal{W}
+\frac{1}{2}\sum_{j,k}\frac{\partial}{\partial z_j}\frac{\partial}{\partial z_k}\left[\textbf{B}(\textbf{z},t)\textbf{B}^T(\textbf{z},t)\right]_{jk}\mathcal{W}
\end{equation}
where the diffusion matrix $D=\textbf{B}\textbf{B}^T$ is positive definite. 
Now, we can write equation (\ref{eqn-stoch}) as a system of stochastic equations:
\begin{equation}
\frac{d\textbf{z}}{dt}=A(\textbf{z},t)+\textbf{B}(\textbf{z},t)\textbf{E}(t),
\end{equation}
where the real noise sources $E_j(t)$ have zero mean and satisfy 
$\langle E_j(t) E_k(t')\rangle=\delta_{jk}\delta(t-t')$. 
In our case, equation (\ref{epn-fok}) can be rewritten:
\begin{eqnarray}
   && \frac{dx_j}{dt}=-J(y_{j+1}+y_{j-1}) - U(y_j - x_j^2y_j - y_j^3) 
   -\frac{\gamma_j}{2}x_j + \frac{\sqrt{\gamma_j}}{2}\xi_j(t), \label{stoch1} \\
   && \frac{dy_j}{dt}=J(x_{j+1}+x_{j-1}) + U(x_j - x_j y_j^2 - x_j^3)
     -\frac{\gamma_j}{2}y_j + \frac{\sqrt{\gamma_j}}{2}\eta_j(t), \label{stoch2}
\end{eqnarray}
where $\xi_j(t),\eta_j(t)$ for $j=1,...,M$ are $\delta$-correlated in time with zero mean. 
Here it must be noted that $\xi_j(t),\eta_j(t)$ are not real noise sources, but are included 
only to recapture the commutation relations of the operators.

\begin{figure}
\centering
\resizebox{0.9\columnwidth}{!}{%
 \includegraphics{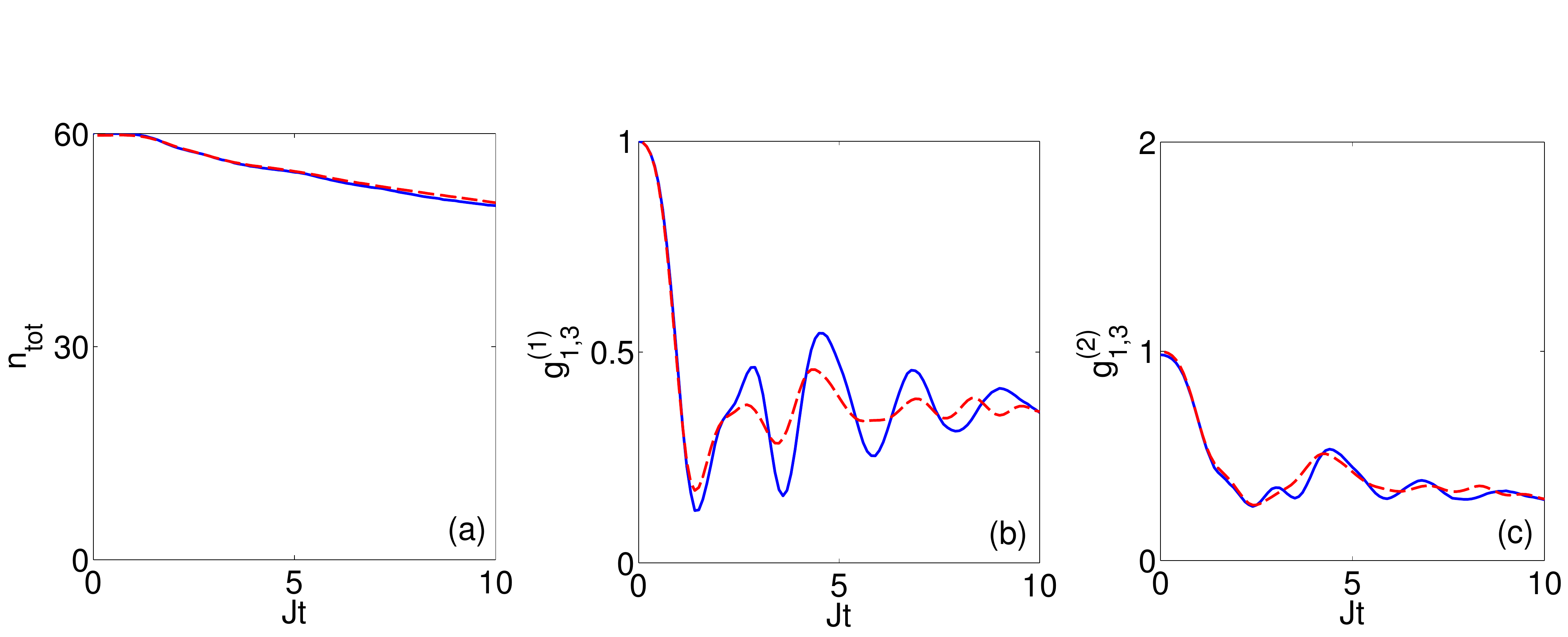} }
\caption{\label{fig:comp}
A comparison of a truncated Wigner simulation (dashed red lines)
and a quantum jump simulation (solid blue lines) shows a very good 
agreement. Shown is the time evolution of (a) the total atom number $\langle \hat n_{\rm tot} \rangle$, (b) the phase coherence $g^{(1)}_{1,3}$ and (c) the density-density correlations $g^{(2)}_{1,3}$. The parameters are: $U=0.1J$, $\gamma_2=0.2J$. The initial state is a pure BEC with an antisymmetric wavefunction, see the discussion around Eq. \eqref{init3}, with $n_1(0)=n_3(0)=30$ and $n_2(0)=0$.}
\end{figure}

As initial state one uses a product state of the form
\begin{equation}
  |\Psi(t=0)\rangle = \bigotimes_{j=1}^M |\psi_j\rangle,
  \label{init1}
\end{equation}
where $|\psi_j\rangle$ is a Glauber coherent state in the $j$th well. This state represents a pure BEC in a grand-canonical framework.
The Wigner function of a Glauber coherent state $|\psi_j\rangle$ is a Gaussian, so the Wigner function that corresponds to the initial state (\ref{init1}) is given by
\begin{equation}
  \mathcal{W}(\{\alpha_j,\alpha_j^*\};t=0) = \prod_{j=1}^M \frac{2}{\pi}e^{-2|\alpha_j-\psi_j|^2} \, .
  \label{init2}
\end{equation}
Thus we can take the initial values for $\alpha_j=x_j+iy_j$ to be Gaussian  random numbers with mean $\psi_j$.
For a BEC in a Bloch state with quasi momentum $k$, we have
\begin{equation}
 \psi_j=e^{ikj}\sqrt{\frac{N}{M}} \, .
  \label{init3}
\end{equation}
For example, if we consider a pure BEC accelerated to the edge of the Brillouin zone then $k=\pi$.

The truncated Wigner method is used to calculate the evolution of expectation values of symmetrized observables as follows. The Wigner function is treated  as a probability distribution in phase space. An ensemble of trajectories is sampled according to the Wigner function of the initial state and then propagated according to Eqs.~(\ref{stoch1}) and (\ref{stoch2}). Then one takes the stochastic average over this ensemble:
\begin{equation}
\langle \hat{O}_j(t)...\hat{O}_k(t)\rangle_{\rm sym} = \int \prod_{i=1}^M d^2 \alpha_i \, O_j...O_k \, \mathcal{W}(\{\alpha_i,\alpha_i^*\};t)
=\frac{1}{N_{\rm t}}\sum_{\ell=1}^{N_{\rm t}}O_j(t)...O_k(t),
\end{equation}
where $O_j$ stands for $\alpha_j$ or $\alpha_j^*$, $N_{\rm t}$ is the number of trajectories and the subscript ``sym'' reminds us that only expectation values of symmetrized observables can be calculated. Furthermore, the truncated Wigner approximation can be used to calculate non-equal time correlation functions~\cite{polk10,berg09}.

In Fig.~\ref{fig:comp} we compare the results of the truncated Wigner 
approximation with the results of the exact quantum jump method
for a triple-well trap. The initial state is the anti-symmetric (\ref{eq:antiS}), in which we expect significant deviations from the pure BEC state, see subsection~\ref{subsec:1.1}, for strong interparticles interactions.
The simulations show a very good agreement and we only observe small deviations in the oscillations of the correlation functions, which are slightly less pronounced. As typical for any method going beyond MF, the truncated phase space approximations become more accurate with increasing filling factors~\cite{trim08,trim09}.

Finally, we must note that the author in~\cite{polk03} developed a consistent perturbation theory in which the classical Gross-Pitaevskii equation is the zero-order approximation and the truncated Wigner is the next order.
From this approach, it becomes apparent that the evolution, in the TWA, is still classical but the initial conditions are distributed according to the Wigner transform of the initial density matrix. Additional corrections can be viewed as quantum scattering events, which appear in the form of a nonlinear response of the observable to an infinitesimal displacement of the field along its classical evolution.
We conclude this section by noting that the TWA has recently been adopted to transport problems with bosons in a quasi-continuous space \cite{TWAcont}. 

\subsection{Non-equilibrium transport}
\label{subsec:5.1}
\begin{figure}
\centering
\resizebox{0.95\columnwidth}{!}{%
 \includegraphics{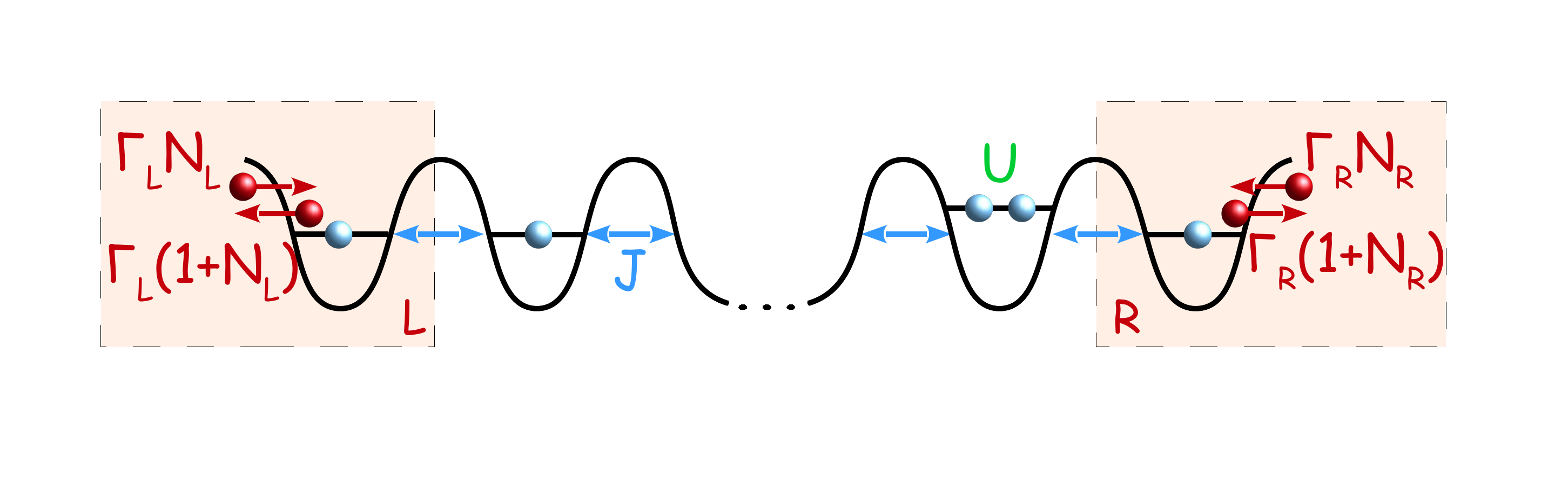} }
\caption{\label{fig:tran0}
Schematic of the non-equilibrium transport scenario, described by the master equation (\ref{eqn:TMA}).}
\end{figure}
In order to study a non-equilibrium scenario with ultracold atoms \cite{Pepino2010,Mirlin2012,Belzig2012,Reimann2013,Prosen2014,Brandes2014}, an appropriate combination of destruction and creation of single particles in the outer sites of a BH chain can be chosen, such as to keep a fixed drop of chemical potential across the system \cite{Ivan13,kordas15}.
Let us look at a chain of quantum dots for bosons, with $M+2$ sites, in which in the outer two sites, $L,R$, we destroy and create single particles with rates $\Gamma_i(1+N_i)$ and $\Gamma_i N_i$, respectively, see Fig.~\ref{fig:tran0}. The master equation that describes the system is given by the equation
\begin{eqnarray}\nonumber
\frac{\partial}{\partial t}\hat{\rho}(t)&=&-i[\hat{H}_{\rm BH},\hat{\rho}(t)]-\\
\nonumber &&-\sum_{k=R,L}\Bigg\{ \frac{\Gamma_k(1+N_k)}{2}[\hat{\alpha}_k^\dagger \hat{\alpha}_k \hat{\rho}(t) + \hat{\rho}(t)\hat{\alpha}_k^\dagger \hat{\alpha}_k - 2 \hat{\alpha}_k\hat{\rho}(t)\hat{\alpha}_k^\dagger]+\\
&&+\frac{\Gamma_k N_k}{2}[\hat{\alpha}_k \hat{\alpha}_k^\dagger \hat{\rho}(t) + \hat{\rho}(t)\hat{\alpha}_k \hat{\alpha}_k^\dagger - 2 \hat{\alpha}_k^\dagger\hat{\rho}(t)\hat{\alpha}_k]\Bigg\}.\label{eqn:TMA}
\end{eqnarray}
As it was shown in~\cite{kordas15}, this simultaneous destruction and creation of particles with specifically chosen rates can be used to control and to keep constant the particle number in the outer sites during the entire time evolution. Thus we can create the equivalent of a ``voltage" drop in a controlled manner between the outer sites of the chain.
\begin{figure}
\centering
\resizebox{0.9\columnwidth}{!}{%
 \includegraphics{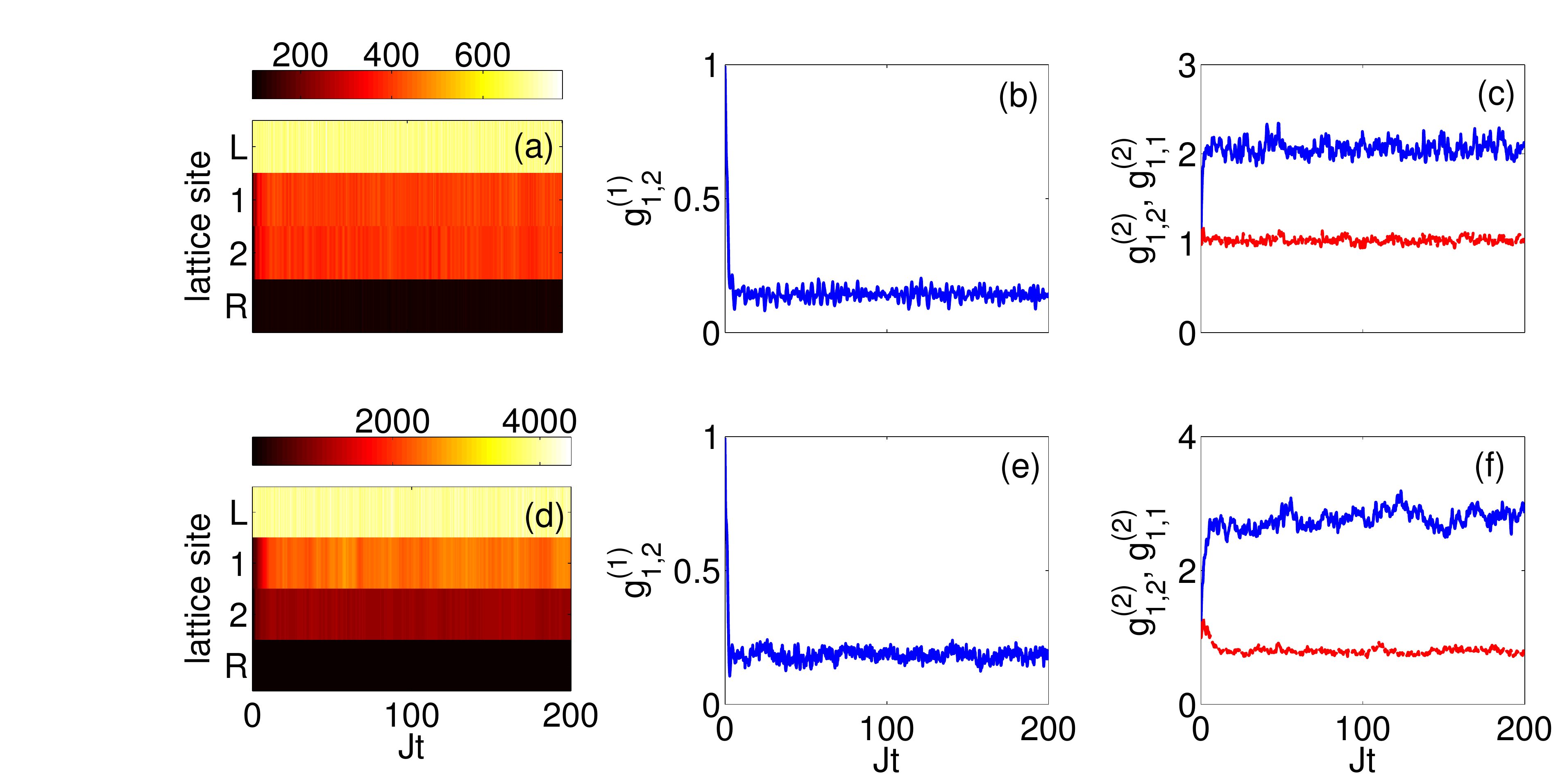} }
\caption{\label{fig:tran}
The evolution of (a,d) the particle density in each lattice site, (b,e) the phase coherence, $g^{(1)}_{1,2}$, and (c,f) the density-density correlations (red lines), $g^{(2)}_{1,2}$, and the density fluctuations (blue lines), $g^{(2)}_{1,1}$. The parameters are $U=10^{-3}$, $\eta_L=\eta_R=J$, $\Gamma_L=\Gamma_R=10J$. For the first row we have $N_L=700$, $N_R=100$ and for the second row $N_L=4000$, $N_R=100$, and the initial state is a pure BEC with $n_1(0)=n_2(0)=n_R(0)=100$ with $n_L(0)=700$ and $n_L=4000$, respectively.}
\end{figure}

In the following we will briefly discuss the role of interparticle interactions is such a system, with the help of the truncated Wigner method. This method is most appropriate in this case since, as we are going to see, the system reaches a steady state in which the coherence is largely lost.
We are interested in transport, so it is useful to introduce the following current operators
\begin{equation}
 \hat{j}_L=i\eta_L(\hat{\alpha}_1^\dagger \hat{\alpha}_L - \hat{\alpha}_L^\dagger \hat{\alpha}_1),\phantom{0}
\hat{j}_R=i\eta_R(\hat{\alpha}_R^\dagger \hat{\alpha}_M - \hat{\alpha}_M^\dagger \hat{\alpha}_R) \,.
\label{cur}
\end{equation}
Thus the current from the left reservoir to the chain and the current from the chain to the right reservoir are given by the expressions
\begin{equation}
 j_L\equiv \langle\hat{j}_L\rangle=-2\eta_L\Im\langle \hat{\alpha}_1^\dagger \hat{\alpha}_L\rangle,\phantom{0}
 j_R\equiv \langle\hat{j}_R\rangle=-2\eta_R\Im\langle \hat{\alpha}_R^\dagger \hat{\alpha}_M\rangle \,.
\label{flux1}
\end{equation}
We have defined the currents in such a way that they will be both positive if the particles flow from the left reservoir to the right one.
For the current operators the following continuity equation holds
\begin{equation}
 \hat{j}_L - \hat{j}_R=\partial_t \hat{n}_{tot} \equiv \partial_t (\hat{n}_1+...+\hat{n}_M).
\label{contin}
\end{equation}

As we discussed above, the dissipation mechanism we use allows us to control the particle number in the reservoirs. We will now answer the question what happens if we change the ``voltage'', that is the particle difference of the reservoirs: $V=n_L-n_R\approx N_L-N_R$. Our example system consists of four wells in which the outer two are the reservoirs. In Fig.~\ref{fig:tran} we have plotted the evolution of the system for small and large voltage, upper and lower row respectively, in the presence of interactions $U=10^{-3}J$. In both cases, a steady state is reached after a short time, however, the resulting steady states have different characteristics. In the small voltage case at the steady state, the particle number in the first and second well are the same, as one can see in Fig.~\ref{fig:tran} (a). Moreover, the correlation functions have the behavior of a thermal state~\cite{Glau99}: phase coherence $g^{(1)}_{1,2}\approx0$, density-density correlations $g^{(2)}_{1,2}\approx1$ (see Fig.~\ref{fig:tran} (b)) and density fluctuations $g^{(2)}_{1,1}\approx2$ (see Fig.~\ref{fig:tran} (c)). On the other hand, in the large voltage case, we observe a large particle imbalance between the two wells (see Fig.~\ref{fig:tran} (d)) density-density correlations smaller than one (see Fig.~\ref{fig:tran} (e)) and density fluctuations greater than two (see Fig.~\ref{fig:tran} (f)). As expected, also in this case the phase coherence tends to zero. The fact that the density-density correlations are smaller than one means that the particle densities in the two wells are correlated and we have anti-bunching: we will have many particles in the first well and fewer in the second one.
\begin{figure}
\centering
\resizebox{0.9\columnwidth}{!}{%
 \includegraphics{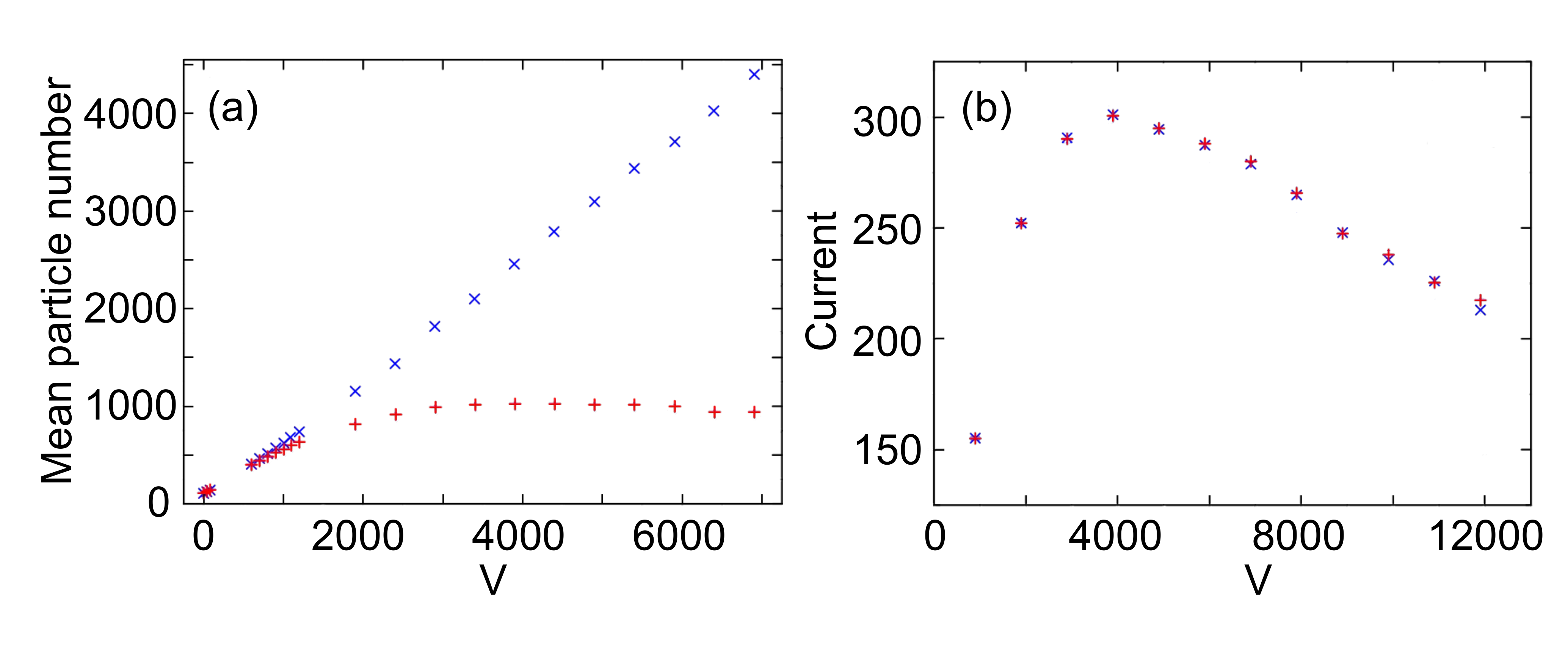} }
\caption{\label{fig:vol}
The steady state (a) mean particle number in the first (blue x) and second (red crosses) site, and (b) current as a function of the ``voltage" $V=N_L-N_R$. The parameters are $U=10^{-3}J$ and $\Gamma_L=\Gamma_R=10J$. Data and figure adapted from~\cite{kordas15}.}
\end{figure}

The above observations become clearer if we look at  Fig.~\ref{fig:vol}, where we have the mean particle number in the two wells and the currents, at the steady state, as a function of the voltage. In these examples we increase the particle number in the left reservoir, while we keep constant the particle number in the right one. In Fig.~\ref{fig:vol} (a) we have the particle number in the two wells as a function of voltage in the interacting case. In this figure we observe two regimes. In the first one, for $V\lesssim 1000$, the particle number in the two wells are the same and increase linearly with the voltage. In the second regime, for $V\gtrsim 1000$, the particle number in the first well (the well that is connected with the reservoir with the larger particle number) increases linearly with the voltage, while in the second well the particle number becomes almost constant. This is consequence of the self trapping effect \cite{Selftrap}: the change of behaviour appears when the macroscopic interaction strength is greater than the tunneling strength, $Un_{\rm tot}(0)>4J$. 
%On the contrary, in the non-interacting case, Fig.~\ref{fig:vol} (c), the particle number in the two wells are exactly the same and they increase linearly with the voltage. This linear scaling is expected for ohmic conduction across a periodic lattice~\cite{caroli71}.

Finally, let us discuss the behaviour of the steady state current as a function of the voltage. 
%In the non-interacting case, Fig.~\ref{fig:vol} (d), the current increase linearly with the voltage. 
For the interacting case, Fig.~\ref{fig:vol} (b), the current has a maximum after which the current drops. This means that for large voltages the transport of the particles through the lattice is blocked. The same qualitative behaviour was observed in~\cite{Ivan13} by approximating the interaction contribution to the self-energy by the tabpole diagram. This one-loop contribution in a systematic diagrammatic expansion takes into account weak interaction induced effects in the non-equilibrium Green's function framework. Physically, we can explain the atom blockade by noting that the first site fills very quickly with particles reaching a chemical potential that is much higher than the one of the second site to be filled. This difference in chemical potential blocks the hopping transport of particles across the chain, see also \cite{Blockade2010} for a similar context. The observed blockade effect is analogous to the dipole blockade in Rydberg atoms excited out of a cloud of cold atoms \cite{ates12,dipoleB} and to the Coulomb blockade in electronic transport across quantum dots \cite{IN1992}.
\section{The coherent-state path integral formalism for open systems}
\label{sec:7}
Since the time when they were first introduced by Feynman~\cite{Feyn1948} path integrals have been proven a powerful tool for understanding and handling quantum mechanics, statistical physics and quantum field theory~\cite{kleinert06}. With the help of the overcomplete basis of coherent states, one can expand the concept of path integration into a complexified phase space, making possible applications in many areas of physics. Indeed the coherent-state path integral has been successfully used,  mainly, as a tool for semiclassicals approximations~\cite{kochetov31,baranger01,ademir96,stone00}. However, the coherent-state path integral, as it was presented, had an important drawback. The definition and the calculation of the coherent-state path integral were based on lattice regularization and the continuum limit was taken after the relevant calculations had been performed. Calculations directly in the continuum limit show quantitative differences with well-known results~\cite{wilson11}. Only recently, a simple recipe has been presented to define and calculate the coherent-state path integral in a continuous form~\cite{Kor14}. The advance to define and calculate the coherent-state path integral in a continuous form~\cite{Kor14} allows us to use the method as a basis for systematic approximations, possibly well-known in the quantum-field-theory community. Here we will generalize this formalism to open quantum systems with the help of the so-called Feynman-Vernon influence functional~\cite{fenma63}.

In order to present the method we will use a BH chain which is coupled to a thermal reservoir. So, the Hamiltonian of the composite system we are dealing with reads as follows
\begin{equation}
\hat{H} = \hat{H}_{\rm S} + \hat{H}_{\rm R} + \hat{H}_{\rm I},
\end{equation}
where the system is described by the $M$-site BH Hamiltonian (\ref{eq:BH}), $\hat{H}_{\rm S}=\hat{H}_{\rm BH}$. Here we must note that the formalism we present can be used for any second-quantized bosonic Hamiltonian. For the environment we follow the usual convention that it can be simulated by an infinite collection of harmonic oscillators~\cite{cald83,raja15}
\begin{equation}
\hat{H}_R = \sum_k E_k \hat{R}_k^\dag \hat{R}_k
\end{equation}
where $\hat{R}_k$ and $\hat{R}_k^\dag$ are the annihilation and creation bosonic operators of the $k$-th oscillator, respectively. Furthermore, the system-environment interaction is considered to be linear
\begin{equation}
\hat{H}_{\rm I} = \sum_{j,k} \gamma_{j,k} \hat{R}_k \hat{\alpha}_j^\dag + \gamma_{j,k}^* \hat{R}_k^\dag \hat{\alpha}_j.
\end{equation}

The crucial quantity for the description of the system's dynamics is the reduced density matrix
\begin{equation}\label{eq:pi5}
\hat{\rho}_{\rm S}(t) = {\rm Tr}_{\rm R} \left\{e^{-i\hat{H}t} \hat{\rho}_{\rm S}(0)\otimes \hat{\rho}_{\rm R}(0) e^{i\hat{H}t}\right\}.
\end{equation}
In order to derive the dynamics of the above quantity we will use the Feynman-Vernon influence functional technique. This task, which is usually implemented in configuration space, can also be accomplished in the framework of the coherent-state path integrals~\cite{an07,tu08,zhang12}. To begin with, we introduce the basis states $|\vec{z}_{\rm S}\rangle = |z_{1 \rm S},...,z_{M \rm S}\rangle$ for the system and $|\vec{z}_{\rm R}\rangle = |z_{1 \rm R},...,z_{k \rm R},...\rangle$ for the environment. For the composite system we use the notation $|\vec{Z}\rangle = |\vec{z}_{\rm S},\vec{z}_{\rm R}\rangle$, while the completeness relation can be casted in the abbreviated form
\begin{equation}\label{eq:pi6}
\int d^2 \vec{Z}|\vec{Z}\rangle\langle\vec{Z}| \equiv \prod_{j\in{\rm S}}\int \frac{dz_{j\rm S}dz_{j\rm S}^*}{2\pi i} \prod_{k\in{\rm R}}\int \frac{dz_{k\rm R}dz_{k\rm R}^*}{2\pi i} = \hat{I}.
\end{equation}
In  this basis the reduced density operator is represented by the matrix
\begin{equation}
\rho_{ba,\rm S}(t) \equiv \langle \vec{z}_{{\rm S},b}|\hat{\rho}_{\rm S}(t)|\vec{z}_{{\rm S},a}\rangle = \int d^2 \vec{z}_{\rm R} \langle \vec{z}_{{\rm S},b},\vec{z}_{{\rm R}}|\hat{\rho}(t)|\vec{z}_{{\rm S},a},\vec{z}_{{\rm R}}\rangle.
\end{equation}
Inserting the identity (\ref{eq:pi6}) between each term in Eq.~(\ref{eq:pi5}), the system's density matrix reads
\begin{equation}
\rho_{ba,\rm S}(t) = \int d^2 \vec{z}'_{\rm S} \int d^2 \vec{z}''_{\rm S} \langle \vec{z}'_{\rm S}|\hat{\rho}_{\rm S}(0)|\vec{z}''_{\rm S}\rangle \mathcal{J}(\vec{z}_{{\rm S},b}^*,\vec{z}_{{\rm S},a};\vec{z}'_{\rm S},\vec{z}''_{\rm S}).
\end{equation}
Following the terminology inherited from the configuration space we introduced the quantity $\mathcal{J}$ as the propagating functional in the coherent-state language
\begin{eqnarray}
\nonumber \mathcal{J}(\vec{z}_{{\rm S},b}^*,\vec{z}_{{\rm S},a};\vec{z}'_{\rm S},\vec{z}''_{\rm S}) &\equiv & \int d^2\vec{z}_{\rm R} \int d^2\vec{z}'_{\rm R} \int d^2\vec{z}''_{\rm R} K(\vec{z}_{{\rm S},b}^*,\vec{z}_{\rm R}^*;\vec{Z}';t)\times \\
&& \times K^*(\vec{z}_{{\rm S},a}^*,\vec{z}_{\rm R}^*;\vec{Z}'';t) \langle \vec{z}'_{\rm R}|\hat{\rho}_{\rm R}(0)|\vec{z}''_{\rm R}\rangle,
\end{eqnarray}
where the dynamical factors in this expression are the correlation functions
\begin{equation}\label{eq:pi10}
 K(\vec{Z}_1^*,\vec{Z}_2;t) = \langle \vec{Z}_1|e^{-i\hat{H}t}|\vec{Z}_2\rangle.
\end{equation}
The proper definition of the path integration in the complexified phase space requires the introduction of the ``classical" Hamiltonian $H^F$ which is obtained from the quantum one via a simple route~\cite{Kor14}
\begin{equation}\label{eq:pi11}
\hat{H}(\hat{\alpha}^\dag,\hat{\alpha})\rightarrow \hat{H}(\hat{p},\hat{q})\rightarrow H^F(p,q) \rightarrow H^F(z^*,z).
\end{equation}
The first step in this chain is the replacement of the creation and annihilation operators by
the corresponding ``momentum" and ``position" operators. Next, one passes to the classical Hamiltonian appearing in the Feynman phase space integral and eventually performs a canonical change of variables: $q=(z^*+z)/\sqrt{2}$ and $q=i(z^*-z)/\sqrt{2}$.

With these clarifications the correlator (\ref{eq:pi10}) assumes the form
\begin{eqnarray}\label{eq:pi12}
\nonumber K(\vec{Z}_1^*,\vec{Z}_2;t) &=& \mathop{\int \mathcal{D}^2 \vec{z}_{\rm S}}\limits_{\substack{\vec{z}^*_{\rm S}(t)=\vec{z}^*_{\rm S1} \\ \vec{z}_{\rm S}(0)=\vec{z}_{\rm S2}}} \exp \left\{ -\Gamma_{\rm S}(\vec{z}_{\rm S}^*,\vec{z}_{\rm S}) + iS_{\rm S}[\vec{z}_{\rm S}^*,\vec{z}_{\rm S}]\right\}\times\\
\nonumber && \times\mathop{\int \mathcal{D}^2 \vec{z}_{\rm R}}\limits_{\substack{\vec{z}^*_{\rm R}(t)=\vec{z}^*_{\rm R1} \\ \vec{z}_{\rm R}(0)=\vec{z}_{\rm R2}}} \exp \left\{ -\Gamma_{\rm R}(\vec{z}_{\rm R}^*,\vec{z}_{\rm R}) + iS_{\rm R}[\vec{z}_{\rm R}^*,\vec{z}_{\rm R}]\right\}\times\\
&& \times\prod_{j\in {\rm S},k\in {\rm R}} \exp\left\{-i\int_0^t d\tau (\gamma_{kj}z_{k\rm R} z^*_{j\rm S} + c.c.) \right\}.
\end{eqnarray}
The structure of this amplitude is more or less, obvious: The first factor refers to the system, the second one contains the environmental degrees of freedom and the last one arises from their coupling. The action functionals entering Eq. (\ref{eq:pi12}) are defined as follows
\begin{equation}\label{eq:pi13}
S_{\rm S} = \int_0^t dt \left[\frac{i}{2} (\dot{\vec{z}}_{\rm S}\vec{z}_{\rm S}^* - \vec{z}_{\rm S}\dot{\vec{z}}_{\rm S}^*) - H^F_{\rm S}\right],~S_{\rm R} =\int_0^t dt \left[\frac{i}{2} (\dot{\vec{z}}_{\rm R}\vec{z}_{\rm R}^* - \vec{z}_{\rm R}\dot{\vec{z}}_{\rm R}^*) - H^F_{\rm R}\right],
\end{equation}
and the $\Gamma$ factors have the form
\begin{eqnarray}
\nonumber \Gamma_{\rm S(R)}\left(\vec{z}_{\rm S(R)}^*,\vec{z}_{\rm S(R)}\right) &=& \frac{1}{2}\left( |\vec{z}_{\rm S(R)1}|^2 + |\vec{z}_{\rm S(R)2}|^2\right)\\
&& - \frac{1}{2}\left( \vec{z}_{\rm S(R)1}^* \cdot \vec{z}_{\rm S(R)}(t) + \vec{z}_{\rm S(R)2} \cdot \vec{z}_{\rm S(R)}^*(0)\right).
\end{eqnarray}
The classical Hamiltonians in Eq.~(\ref{eq:pi13}) are constructed following the rule (\ref{eq:pi11}) and they are read as follows~\cite{Kor14}
\begin{equation}
H^F_{\rm S} = \sum_j(\varepsilon_j + U)|z_{j\rm S}|^2 - J\sum_{<i,j>}(z_{i\rm S}^*z_{j\rm S} + c.c.) + \frac{U}{2} |z_{j\rm S}|^4 + \sum_j \left(\varepsilon_j + \frac{3U}{8} \right)
\end{equation}
and
\begin{equation}
H^F_{\rm R} = \sum_k E_k |z_{k\rm R}|^2 - \frac{1}{2}\sum_k E_k.
\end{equation}

Due to the fact that the environment is just a collection of harmonic oscillators and the interaction with the system is linear, the integration of the environmental degrees can be easily performed. One needs only a change of variables in order to get rid of the boundary conditions: $\vec{z}_{\rm R} = \vec{z}_{\rm R}^{cl.} + \vec{\eta}_{\rm R}$, $\vec{z}_{\rm R}^* = \vec{z}_{\rm R}^{cl.*} + \vec{\eta}_{\rm R}^*$ with $\vec{z}_{\rm R}^{cl.*}(t) = \vec{z}_{\rm R1}^*$ and $\vec{z}_{\rm R}^{cl.}(0) = \vec{z}_{\rm R2}^*$. The functions
\begin{eqnarray}
z_{k\rm R}^{cl.}(\tau) &=& e^{-iE_k\tau}z_{k\rm R2} - i\sum_j\gamma_{kj}^* \int_0^\tau d\tau' e^{-iE_k(\tau-\tau')}z_{j\rm S}(\tau'),\label{eq:pi17a}\\
z_{k\rm R}^{cl.*}(\tau) &=& e^{-iE_k(t-\tau)}z_{k\rm R1}^* - i\sum_j\gamma_{kj} \int_\tau^t d\tau' e^{-iE_k(\tau-\tau')}z_{j\rm S}^*(\tau')\label{eq:pi17b}
\end{eqnarray}
have been chosen to enforce stationarity, with respect to the environmental degrees, of the exponent in Eq. (\ref{eq:pi12}). The rest of the calculation is just a quadratic fluctuation integral that can be evaluated by standard means. Its contribution yields an exponential factor $e^{-it\sum_k E_k/2}$ that it is exactly cancelled by the constant term appearing in $H^F_{\rm R}$. Thus the integration of the environment is encapsulated in the classical solutions Eqs. (\ref{eq:pi17a}) and (\ref{eq:pi17b}), and through them taking into account that $S_{\rm R}^{cl.}=0$, in the $\Gamma_{\rm R}$ factor. After these explanations one can easily confirm that the correlator (\ref{eq:pi12}) takes the form
\begin{equation}
K(\vec{Z}_1^*,\vec{Z}_2;t)=e^{\frac{1}{2}(|\vec{z}_{\rm R1}|^2 + |\vec{z}_{\rm R2}|^2)} \mathop{\int \mathcal{D}^2 \vec{z}_{\rm S}}\limits_{\substack{\vec{z}^*_{\rm S}(t)=\vec{z}^*_{\rm S1} \\ \vec{z}_{\rm S}(0)=\vec{z}_{\rm S2}}}  e^{ -\Gamma_{\rm S}(\vec{z}_{\rm S}^*,\vec{z}_{\rm S}) + iS_{\rm S}[\vec{z}_{\rm S}^*,\vec{z}_{\rm S}]- i\sum_{k\in\rm R} I_k},
\end{equation}
with
\begin{eqnarray}\label{eq:pi19}
\nonumber I_k &=& i z_{k\rm R1}^* z_{k\rm R2} e^{-iE_k t} + z_{k\rm R1}\sum_{j\in \rm S} \gamma_{kj}^* \int_0^t d\tau e^{-E_k(t-\tau)} z_{j\rm S}(\tau) \\
\nonumber &&+ z_{k\rm R2}\sum_{j\in \rm S} \gamma_{kj} \int_0^t d\tau e^{-E_k\tau} z_{j\rm S}^*(\tau)\\
&& - \sum_{i,j\in \rm S} \gamma_{kj}\gamma_{ki}^* \int_0^t d\tau \int_0^\tau d\tau' e^{-iE_k(\tau-\tau')} z_{j\rm S}^*(\tau) z_{i\rm S}(\tau').
\end{eqnarray}

Besides the correlation functions the other ingredient needed for the propagation kernel is the initial density matrix of the reservoir. Assuming that the environment is in thermal equilibrium at some temperature $T$ we write
\begin{equation}\label{eq:pi20}
\langle \vec{z}_{\rm R}'|\hat{\rho}_{\rm R}(0)|\vec{z}_{\rm R}''\rangle = \frac{1}{Q} \vec{z}_{\rm R}'|e^{-\beta(\hat{H}_{\rm R} -\mu \hat{N}_{\rm R})} |\vec{z}_{\rm R}''\rangle,
\end{equation}
where $Q$ is the grand partition function
\begin{equation}
Q = \prod_{k\in \rm R}\frac{1}{1-e^{-\beta(E_k - \mu)}}.
\end{equation}
For determining the density matrix (\ref{eq:pi20}) one has to face a simple quadratic coherent state path integral that can be easily calculated by standard techniques
\begin{equation}\label{eq:pi22}
\langle \vec{z}_{\rm R}'|\hat{\rho}_{\rm R}(0)|\vec{z}_{\rm R}''\rangle = \frac{1}{Q} \exp \left\{ \frac{1}{2} (|\vec{z}_{\rm R}'|^2 + |\vec{z}_{\rm R}''|^2) + \sum_{k\in \rm R} z_{k\rm R}'^* z_{k\rm R}'' e^{-\beta(E_k - \mu)}\right\}.
\end{equation}
Our next step is to insert the results (\ref{eq:pi19}) and (\ref{eq:pi22}) into the propagating functional
\begin{eqnarray}\label{eq:pi23}
\nonumber \mathcal{J} &=& \mathop{\int \mathcal{D}^2 \vec{z}_{\rm S}}\limits_{\substack{\vec{z}^*_{\rm S}(t)=\vec{z}^*_{{\rm S}b} \\ \vec{z}_{\rm S}(0)=\vec{z}_{\rm S}'}}
\mathop{\int \mathcal{D}^2 \vec{w}_{\rm S}}\limits_{\substack{\vec{w}^*_{\rm S}(0)=\vec{z}''^*_{\rm S} \\ \vec{w}_{\rm S}(t)=\vec{z}_{{\rm S}a}}} \exp \left\{ -\Gamma_{\rm S}(\vec{z}_{\rm S}^*,\vec{z}_{\rm S}) + iS_{\rm S}[\vec{z}_{\rm S}^*,\vec{z}_{\rm S}]\right\}\times\\
&& \times \exp \left\{ -\Gamma_{\rm S}^*(\vec{w}_{\rm S}^*,\vec{w}_{\rm S}) - iS_{\rm S}^*[\vec{w}_{\rm S}^*,\vec{w}_{\rm S}]\right\} \mathcal{F}[\vec{z}_{\rm S}^*,\vec{z}_{\rm S};\vec{w}_{\rm S}^*,\vec{w}_{\rm S}].
\end{eqnarray}
The functional $\mathcal{F}$ appearing in the last expression contains the influence of the environment onto the dynamics of the system and defines the influence functional in the coherent state representation. Due to the fact that all the factors in the influence functional are at the most quadratic, the integrals can be easily performed and the result reads~\cite{an07,tu08,zhang12}
\begin{eqnarray}\label{eq:pi24}
\nonumber \mathcal{F} &=& \exp\left\{\sum_{k\in \rm R} \sum_{i,j\in \rm S} \gamma_{kj}^* \gamma_{ki} \int_0^t d\tau \int_0^\tau d\tau' e^{iE_k(\tau-\tau')} (z_{j \rm S}(\tau) - w_{j \rm S}(\tau) ) w_{i\rm S}^*(\tau') \right\}\times \\
\nonumber &&\exp\left\{\sum_{k\in \rm R} \sum_{i,j\in \rm S} \gamma_{kj}^* \gamma_{ki} \int_0^t d\tau \int_0^\tau d\tau' e^{-iE_k(\tau-\tau')} (z_{j \rm S}(\tau) - w_{j \rm S}(\tau) ) z_{i\rm S}^*(\tau') \right\}\times \\
&& \exp\left\{ -\sum_{k\in \rm R}\frac{1}{e^{\beta(E_k-\mu)}-1} \left| \sum_{j\in \rm R} \gamma_{kj} \int_0^t d\tau e^{-iE_k \tau} (z_{j\rm S}^* (\tau) - (w_{j\rm S}^* (\tau)) \right|^2 \right\}.
\end{eqnarray}
In the above result, the first two exponentials correspond to the purely quantum influence of the environment, which appears also for zero temperature~\cite{an07,tu08,zhang12}. The third factor contains the temperature dependence.
We note that until now the only assumption we have made is that the initial state can be written as a product state. The obvious advantage of the expression of Eq. (\ref{eq:pi24}) is that one can handle non-Markovian environments~\cite{an07,tu08,zhang12}. Even in the simpler case of the Markovian environments one can, of course, also use the Feynman-Vernon influence functional as a basis of systematic approximations. In the next subsection we adopt a Markovian approximation for the environment and derive the expression for the influence functional for a general $M$-site BH chain.

\subsection{The Markovian approximation}
At this point we assume that $\gamma_{kj} = |\gamma_k|e^{i\theta_j}$, $\varepsilon_j = \varepsilon,~\forall~j\in \rm S$ and we adopt the Markovian approximation. Then it is enough to work out the first factor in the influence functional (\ref{eq:pi24}) only. By redefining $z_{j\rm S} =\bar{z}_{j\rm S} e^{-i(\varepsilon + U)\tau}$ and $w_{j\rm S} =\bar{w}_{j\rm S} e^{-i(\varepsilon + U)\tau}$ this term is written as
\begin{equation}\label{eq:pi25}
\sum_{i,j\in \rm S} e^{i\theta_{ij}} \int_0^t d\tau (\bar{z}_{j\rm S}(\tau) - \bar{w}_{j\rm S}(\tau)) \int_0^\tau d\tau' \left[ \sum_{k\in \rm R} |\gamma_k|^2 e^{i(E_k-\varepsilon - U)(\tau-\tau')} \right]\bar{w}_{i\rm S}^*(\tau'),
\end{equation}
where $\theta_{ij} = \theta_i - \theta_j$. In the last integral we assume that the integrand is a very fast decaying function of the time difference and we expand $\bar{w}_{i \rm S}^*(\tau') \approx \bar{w}_{i \rm S}^*(\tau) + \mathcal{O}(\tau' - \tau)$. In this way we are left with the integral
\begin{eqnarray}\label{eq:pi26}
\nonumber \int_0^\tau d\tau'\sum_{k\in \rm R} |\gamma_k|^2 e^{i(E_k-\varepsilon - U)(\tau-\tau')} &\approx & i {\rm P} \int_0^\infty dE \frac{g(E)}{E-(\varepsilon +U)} + \pi g(\varepsilon+U)\\
&\equiv& i\delta \varepsilon + \frac{\Gamma}{2},
\end{eqnarray}
where
\begin{equation}\label{eq:pi27}
g(E) \equiv \sum_{k\in \rm R} |\gamma_k|^2 \delta(E-E_k).
\end{equation}
Note that if we had not assumed that $\varepsilon_j = \varepsilon$, the value of the energy entering (\ref{eq:pi26}) would be the $\max\{ \varepsilon_j,~j\in\rm S\}$.

Inserting this result into expression (\ref{eq:pi25}) and restoring the original fields we are led to the following form for the first factor in Eq. (\ref{eq:pi24})
\begin{equation}
i\left( \delta\varepsilon - i\frac{\Gamma}{2}\right) \sum_{i,j\in \rm S} e^{i\theta_{ij}} \int_0^t d\tau (z_{j\rm S}(\tau) - w_{j\rm S}(\tau))w_{i\rm S}^*(\tau).
\end{equation}
Working in the same spirit the adoption of the Markov approximation yields the following time-local form of the influence functional
\begin{eqnarray}\label{eq:pi29}
\nonumber \mathcal{F} &=& \exp \left\{ -i(\delta\varepsilon + i \tilde{\Gamma})\sum_{i,j\in \rm S} e^{i\theta_{ij}} \int_0^t d\tau z_{j\rm S}^*(\tau) z_{i\rm S}(\tau) \right\}\times\\
\nonumber && \times \exp \left\{ i(\delta\varepsilon - i \tilde{\Gamma})\sum_{i,j\in \rm S} e^{i\theta_{ij}} \int_0^t d\tau w_{j\rm S}^*(\tau) w_{i\rm S}(\tau) \right\}\times\\
\nonumber && \times \exp \left\{\Gamma\langle n \rangle\sum_{i,j\in \rm S} e^{i+\theta_{ij}} \int_0^t d\tau z_{i\rm S}^*(\tau) w_{j\rm S}(\tau) \right.\\ 
 &&\left. + \Gamma(\langle n \rangle+1) \sum_{i,j\in \rm S} e^{i\theta_{ij}} \int_0^t d\tau w_{i\rm S}^*(\tau) z_{j\rm S}(\tau) \right\},
\end{eqnarray}
where in the last factor we used the abbreviations
\begin{equation}\label{eq:pi30}
\tilde{\Gamma} = \Gamma \left(\langle n \rangle + \frac{1}{2}\right),~\langle n \rangle = \frac{1}{e^{\beta(\varepsilon +U - \mu)}-1}.
\end{equation}
Inserting the result (\ref{eq:pi29}) into Eq. (\ref{eq:pi23}) we get for the propagating functional
\begin{equation}\label{eq:pi31}
\mathcal{J} = \mathop{\int \mathcal{D}^2 \vec{z}_{\rm S}}\limits_{\substack{\vec{z}^*_{\rm S}(t)=\vec{z}^*_{{\rm S}b} \\ \vec{z}_{\rm S}(0)=\vec{z}_{\rm S}'}}
\mathop{\int \mathcal{D}^2 \vec{w}_{\rm S}}\limits_{\substack{\vec{w}^*_{\rm S}(0)=\vec{z}''^*_{\rm S} \\ \vec{w}_{\rm S}(t)=\vec{z}_{{\rm S}a}}} e^{-\Gamma_{\rm S}(\vec{z}_{\rm S}^*,\vec{z}_{\rm S}) -\Gamma_{\rm S}^*(\vec{w}_{\rm S}^*,\vec{w}_{\rm S}) + i\tilde{S}_{\rm S}[\vec{z}_{\rm S}^*,\vec{z}_{\rm S}] - i\tilde{S}_{\rm S}^*[\vec{w}_{\rm S}^*,\vec{w}_{\rm S}] + I_{zw}[\vec{z}_{\rm S}^*,\vec{z}_{\rm S};\vec{w}_{\rm S}^*,\vec{w}_{\rm S}]}.
\end{equation}
In $\tilde{S}_{\rm S}$ appearing in the last expression we have done the following changes in the defining Hamiltonian
\begin{eqnarray}\label{eq:pi32}
\nonumber \varepsilon + U &\rightarrow& \Omega = \varepsilon + U - \delta\varepsilon - i\tilde{\Gamma} \\
-J \sum_{<i,j>}(z_{i\rm S}^* z_{j\rm S} + c.c.) &\rightarrow& -J(\delta\varepsilon + i\tilde{\Gamma}) \sum_{<i,j>}(e^{i\theta_{ij}}z_{i\rm S}^* z_{j\rm S} + c.c.).
\end{eqnarray}
We have also included an $z-w$ interaction term that can be read from Eq. (\ref{eq:pi29})
\begin{equation}
I_{zw} = \Gamma \langle n \rangle \sum_{i,j\in\rm S} e^{i\theta_{ij}} \int_o^t d\tau z_{i\rm S}^*(\tau) w_{j\rm S}(\tau) + \Gamma (\langle n \rangle + 1) \int_o^t d\tau w_{j\rm S}^*(\tau) z_{i\rm S}(\tau).
\end{equation}
The functional integrations in Eq. (\ref{eq:pi31}) can be analytically performed only when the non-linear term in the system's Hamiltonian is absent. So the presence of interparticle interactions calls for the application of some kind of approximation depending on the parameters of the system. In the framework of the coherent-state path integrals, semiclassical calculations can be systematically performed~\cite{kochetov31,baranger01}. From this point of view, the expression (\ref{eq:pi31}) sets the stage for approximations that go beyond the ones we have presented previously in this review. In the following subsection we use a very simple system in which an analytical solution is possible and we compare the result with that of the corresponding master equation.
\subsection{A bosonic quantum dot coupled to a thermal environment}
\label{subsec:7.1}
Here we will compute, analytically, the propagating functional $\mathcal{J}$ for the trivial case of single quantum dot without interactions
\begin{equation}
\hat{H}_{\rm S} = \varepsilon\hat{\alpha}^\dag \hat{\alpha} \rightarrow H_{\rm S}^F = \varepsilon |z_{\rm S}|^2 - \frac{\varepsilon}{2},
\end{equation}
which will give the time evolution of the reduced density matrix.

The strategy goes as follows: First, one finds the ``classical'' functions $z_{\rm S}^{cl.}$ and $w_{\rm S}^{cl.}$ that extremise the exponent in (\ref{eq:pi31}) and makes the replacements $z_{\rm S} = z_{\rm S}^{cl.}+\eta$ and $w_{\rm S} = w_{\rm S}^{cl.}+\eta'$. Since the exponent is quadratic and the time derivatives are of first order, the propagation functional is just
\begin{equation}
\mathcal{J} = e^{-\Gamma_{\rm S}^{cl.}(\vec{z}_{\rm S}^*,\vec{z}_{\rm S})-\Gamma_{\rm S}^{cl.*}(\vec{w}_{\rm S}^*,\vec{w}_{\rm S})}.
\end{equation}
The reason for this simple result is that the two fluctuation integrals cancel each other exactly. The coupled classical solutions corresponding to the dot case are easily obtained
\begin{eqnarray}
\dot{z}_{\rm S}^{cl.} + i\Omega z_{\rm S}^{cl.} = \Gamma\langle n\rangle  w_{\rm S}^{cl.}&,&~\dot{w}_{\rm S}^{cl.} + i\Omega^* w_{\rm S}^{cl.} = -\Gamma(\langle n\rangle + 1) z_{\rm S}^{cl.}\label{eq:pi36}\\
\dot{z}_{\rm S}^{cl.*} - i\Omega z_{\rm S}^{cl.*} = -\Gamma(\langle n\rangle + 1)  w_{\rm S}^{cl.*} &,&~\dot{w}_{\rm S}^{cl.*} - i\Omega^* w_{\rm S}^{cl.*} = \Gamma\langle n\rangle z_{\rm S}^{cl.*},\label{eq:pi37}
\end{eqnarray}
with boundary conditions
\begin{eqnarray}
\nonumber z_{\rm S}^{cl.*}(t) = z_{b\rm S}^* &,&~z_{\rm S}^{cl.}(0) = z_{\rm S}'\\
\nonumber w_{\rm S}^{cl.}(t) = z_{a\rm S} &,&~w_{\rm S}^{cl.*}(0) = {z_{\rm S}''}^*.
\end{eqnarray}
The constants $\Omega$, $\Gamma$ and $\langle n\rangle$ in (\ref{eq:pi36}) and (\ref{eq:pi37}), can be deduced from Eqs. (\ref{eq:pi26}), (\ref{eq:pi27}), (\ref{eq:pi30}) and (\ref{eq:pi32}) by setting $U=0$. Decoupling the equations (\ref{eq:pi36}) and (\ref{eq:pi37}) we get
\begin{eqnarray}
\nonumber \ddot{z}_{\rm S} + 2i\epsilon \dot{z}_{\rm S} - (\epsilon^2 + \Gamma^2/4)z_{\rm S} = 0 &,&~ \ddot{w}_{\rm S} + 2i\epsilon \dot{w}_{\rm S} - (\epsilon^2 + \Gamma^2/4)w_{\rm S} = 0, \\
\nonumber \ddot{z}_{\rm S}^* - 2i\epsilon \dot{z}_{\rm S}^* - (\epsilon^2 + \Gamma^2/4)z_{\rm S}^* = 0 &,&~ \ddot{w}_{\rm S}^* - 2i\epsilon \dot{w}^*_{\rm S} - (\epsilon^2 + \Gamma^2/4)w_{\rm S}^* = 0,
\end{eqnarray}
where we have denoted $\epsilon=\varepsilon-\delta\varepsilon$. The boundary conditions accompanying these equations can be deduced by using the coupled equations (\ref{eq:pi36}) and (\ref{eq:pi37})
\begin{eqnarray}
\nonumber z_{\rm S}^{cl.}(0) = z_{\rm S}' &,&~ \dot{z}_{\rm S}^{cl.}(t) + i\Omega z_{\rm S}^{cl.}(t) = \Gamma\langle n\rangle z_{a\rm S},\\
\nonumber w_{\rm S}^{cl.}(t) = z_{a\rm S} &,&~ \dot{w}_{\rm S}^{cl.*}(0) + i\Omega^* w_{\rm S}^{cl.*}(0) = -\Gamma(\langle n\rangle + 1) z_{a\rm S}',\\
\nonumber z_{\rm S}^{cl.*}(t) = z_{b\rm S}^* &,&~ \dot{z}_{\rm S}^{cl.*}(0) - i\Omega z_{\rm S}^{cl.*}(0) = -\Gamma(\langle n\rangle + 1) {z_{a\rm S}''}^*,\\
\nonumber w_{\rm S}^{cl.}(0) = {z_{\rm S}''}^* &,&~ \dot{w}_{\rm S}^{cl.*}(t) - i\Omega^* w_{\rm S}^{cl.*}(t) = \Gamma\langle n\rangle z_{b\rm S}^*.
\end{eqnarray}
The ``classical'' solutions needed for the propagating kernel are easily determined to be
\begin{eqnarray}
\nonumber z_{\rm S}^{cl.}(t) &=& [z_{\rm S}'e^{-i\Omega t} + z_{a\rm S}\langle n\rangle(1-e^{-\Gamma t})]\varphi(t),\\
\nonumber w_{\rm S}^{cl.}(0) &=& [z_{a\rm S}'e^{i\Omega^* t} + z_{\rm S}'(\langle n\rangle +1) (1-e^{-\Gamma t})]\varphi(t),\\
\nonumber w_{\rm S}^{cl.*}(0) &=& [z_{b\rm S}^* e^{-i\Omega t} + {z_{\rm S}''}^* (\langle n\rangle +1) (1-e^{-\Gamma t})]\varphi(t),\\
\nonumber z_{\rm S}^{cl.*}(t) &=& [{z_{\rm S}''}e^{i\Omega^* t} + z_{b\rm S}^*\langle n\rangle(1-e^{-\Gamma t})]\varphi(t),
\end{eqnarray}
where
\begin{equation}
\varphi(t) = \frac{1}{1 + \langle n\rangle (1-e^{-\Gamma t})}.
\end{equation}

All the ingredients for the determination of the reduced density matrix are now in our hands. We only need the system's density matrix at $t =­ 0$. Assuming that initially was at the state $|\psi_{\rm S}\rangle = |N\rangle$ we find that
\begin{equation}
\langle z_{\rm S}'|\hat{\rho}_{\rm S}(0)|z_{\rm S}''\rangle = \frac{1}{N!}({z_{\rm S}'}^* z_{\rm S}'')^N e^{-\frac{1}{2}(|z_{\rm S}'|^2 + |z_{\rm S}''|^2)}.
\end{equation}
Now the calculation of the reduced density matrix, in the coherent state representation, proceeds with no difficulties. The result reads as follows
\begin{eqnarray}
 \rho_{ba,\rm S}(t) &=& e^{-\frac{1}{2}(|z_{b \rm S}|^2 + |z_{a\rm S}|^2) + z_{b\rm S}^* z_{a\rm S}\langle n\rangle (1-e^{-i\Gamma t})\varphi(t)}\times\\
\nonumber && \times\sum_{\ell=0}^N 
\left( \begin{array}{c}
N \\
\ell \end{array} \right)
\frac{1}{(N-\ell)!} [z_{b\rm S}^* z_{a\rm S} e^{-\Gamma t}(1-e^{-\Gamma t})^2 \varphi^2(t)]^{N-\ell} [(\langle n\rangle+1)\varphi(t)]^{\ell}.
\end{eqnarray}
One can immediately check that
\begin{equation}
{\rm Tr_S}\hat{\rho}_{\rm S}(t) = \int d^2 z_{b\rm S}~\rho_{bb,\rm S}(t) =1.
\end{equation}

A quantity of interest that can be easily calculated is the mean population of the site as a function of time
\begin{eqnarray}
\nonumber \langle \hat{\alpha}^\dag \hat{\alpha}\rangle_t &=& {\rm Tr_S} \{\hat{\alpha}^\dag \hat{\alpha} \hat{\rho}_{\rm S}(t)\}\\
&=& \int d^2 z_{b \rm S} \int d^2 z_{a \rm S}~z_{b \rm S} z_{a \rm S}^* e^{-\frac{1}{2}(|z_{b\rm S}|^2 + |z_{a\rm S}|^2) + z^*_{a\rm S} z_{b\rm S}} \rho_{ba,\rm S}(t).
\end{eqnarray}
The last integral can be trivially performed yielding the result
\begin{equation}\label{eq:pi46}
n_{\rm dot} \equiv \langle \hat{\alpha}^\dag \hat{\alpha}\rangle_t = (N-\langle n\rangle)e^{-\Gamma t} + \langle n\rangle.
\end{equation}

The same very simple result can be easily obtained using the master equation controlling the time evolution of the reduced density matrix. Using the well-known Born-Markov elimination method~\cite{bre06} one can arrive to the master equation
\begin{eqnarray}
\nonumber \frac{d}{d t}\hat{\rho}_{\rm S}(t) &=& -i\epsilon [\hat{\alpha}^\dagger\hat{\alpha},\hat{\rho}_{\rm S}(t)]
-\frac{\Gamma(1+\langle n\rangle)}{2}[\hat{\alpha}^\dagger\hat{\alpha}\hat{\rho}_{\rm S}(t) + \hat{\rho}_{\rm S}(t)\hat{\alpha}^\dagger\hat{\alpha} -2\hat{\alpha}\hat{\rho}_{\rm S}(t)\hat{\alpha}^\dagger]\\
&&-\frac{\Gamma\langle n\rangle}{2}[\hat{\alpha}\hat{\alpha}^\dagger\hat{\rho}_{\rm S}(t) + \hat{\rho}_{\rm S}(t)\hat{\alpha}\hat{\alpha}^\dagger -2\hat{\alpha}^\dagger\hat{\rho}_{\rm S}(t)\hat{\alpha}],
\label{master5}
\end{eqnarray}
where $\epsilon = \varepsilon - \delta\varepsilon$, with $\delta\varepsilon$ and $\Gamma$ given again by the relation (\ref{eq:pi26}) for this system. The evolution equation for the dot's population can be easily obtained
\begin{equation}
 \frac{d}{dt}\langle \hat{\alpha}^\dagger \hat{\alpha}\rangle\equiv {\rm Tr}_{\rm S}\{\hat{\alpha}^\dagger \hat{\alpha}\dot{\hat{\rho}}_{\rm S}\}=-\Gamma\langle \hat{\alpha}^\dagger \hat{\alpha}\rangle + \Gamma\langle n\rangle,
\end{equation}
and using the same initial condition for the subsystem, as previously, we finally arrive again at the result (\ref{eq:pi46}).
\begin{figure}%\sidecaption
\centering
\resizebox{0.5\hsize}{!}{%
 \includegraphics{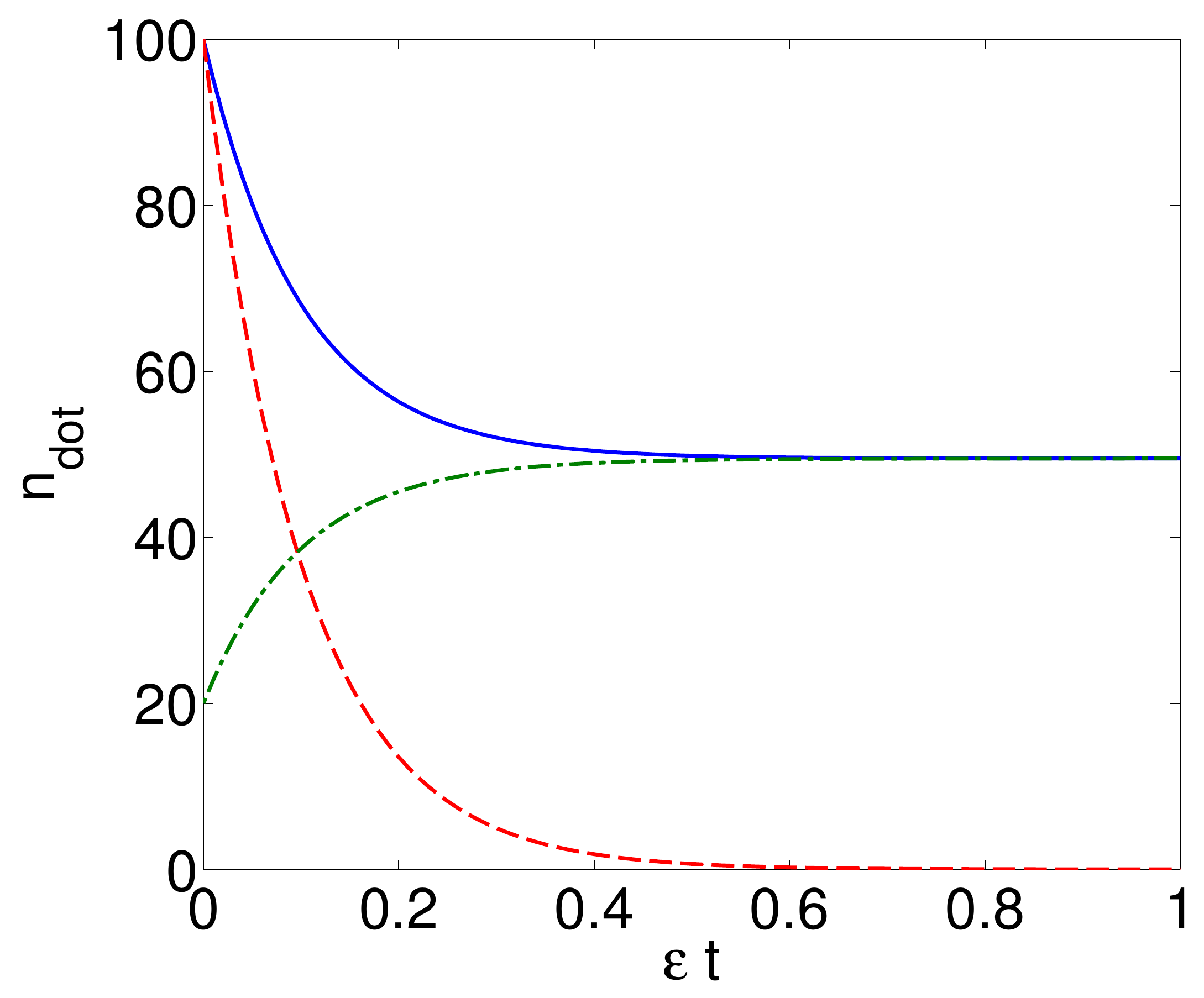} }
\caption{\label{fig:ndot} The evolution of dot's population for a reservoir in the canonical ensemble. The parameters are: $1/\beta=50$ and $N=100$ (blue solid line), $1/\beta=0$ and $N=100$ (red dashed line) and $1/\beta=50$ and $N=20$ (green dashed dotted line). In all cases $\Gamma=10\varepsilon$.}
\end{figure}
The evolution of the dot's population for different initial populations and reservoir's temperatures is depicted in Fig.~\ref{fig:ndot}. At zero temperature we have $\langle n\rangle = 0$, so the master equation (\ref{master5}) is the same as in the single particle loss case with rate $\Gamma$. Thus at zero temperature the population will go to zero.

Analytical solutions are possible not only in this simple case but also for larger non-interacting systems. In the case of interacting BH chains analytical solutions are not possible. However, the formalism we presented and which extends the Feynamn-Vernon technique into the realm of coherent states,  can be used for approximations that can go, in a systematic way, beyond the ones presented in this review. We finally note that the above presented self-consisted formulation of path integration in terms of coherent states  can be used not only for BH lattices, but also for any second quantized bosonic system.
\section{Summary}
\label{sec:8}
In this article we review methods which allow the study of the dynamics of dissipative and one-dimensional Bose-Hubbard models. We focused on systems that are coupled to Markovian environments, so one can write down a master equation in Lindblad form.  

Exact numerical unravelling of the master equation (\ref{eq:DBH}) can be achieved with the quantum jump method. The density matrix is decomposed into pure states whose continuous non-Hermitian evolution is interrupted by stochastic quantum jumps. The exact density matrix of the system is recovered when we average over these trajectories. 

The first approximation one can do is the mean-field approximation. In this limit the quantum fluctuation are totally suppressed and the dynamics is given by a system of nonlinear equations (\ref{eq:MF}). This approach is valid for large numbers of particles with weak interparticle interactions at the same time. These equations can be viewed as the zeroth order of a perturbative expansion with the complement of the condensate fraction as a small parameter. The next order in this expansion is described by the BBR equations of motion. This approximation predicts the deviation of the system's state from a pure BEC state and allows the calculation of quantities, like the condensate fraction, which are not accessible by the mean-field approximation. However, the method breaks down if the system deviates significantly from a pure BEC. 

The Gutzwiller method is another mean-field-like method which assumes that the state can be always written as a product state. The advantage of the method is that it works not only for a pure BEC but also for product-type states, as for example Mott insulator. However, the method systematically overestimates the loss of coherence in the system. 

The truncated Wigner approximation is a semiclassical method (in the sense of large particle numbers), which can, in principle, calculate the evolution of any correlation function, even if the system deviates from the pure BEC state. 
It just requires a large particle number. Because of the applied truncation of higher derivatives of the Wigner function it is, unfortunately, not easy to predict a priori up to which time the TWA remains valid for a state propagation.
%The basic assumptions of the method is that the system starts from  pure BEC state with large particle number. 

Finally, we have presented the Feynman-Vernon coherent-state path integral approach to the open BH chains. This general framework allows the study of the system's evolution even in the case of non-Markovian environments. Furthermore, in the case of Markovian environments, the formalism can be the basis for systematic approximations, which can go beyond the usual ones as summarized above. Further work will be necessary to apply the new method described in sec. \ref{sec:7} for the treatment of realistic, i.e. more complex quantum systems. Moreover, other novel methods should be developed which can simultaneously deal with non-Markovian strong coupling between the system and the environment and strong particle interaction.

As a final note, we report that a semiclassical approach based on interfering paths in Fock space has been recently introduced directly for discrete systems, such as for a closed Bose-Hubbard model \cite{Richter2012,Richter2014,Strunz2014}. Also a semiclassical Gutzwiller trace formula is brought forward \cite{Bristol2015} for the Bose-Hubbard model in the chaotic regime \cite{Buchleitner2004,Andrea2007,Carlos2014,Lubasch2009}. It remains to be seen whether these approaches can be extended to open dissipative and decohering many-body quantum systems as well \cite{Strunz2013}, analogues to the general technique reported in sec. \ref{sec:7}. Another interesting aspect would be the investigation of tilted many-body Bose-Hubbard models including more than one energy band \cite{Carlos2014}, which are naturally open systems due to decay into the continuum of free states accelerated by the tilting force \cite{Pisa2007}, see also \cite{Alon2012} in the context of continuum coupling.

%_________________________________________________________________________
%
\appendix
\section{An Entanglement criterion}
\label{sec:ent}
In this appendix we review the entanglement criterion based on Eq. (\ref{eqn:ent_para}),
which is optimally suited to the breather states discussed in sec. \ref{subsec:1.1}. The criterion 
was introduced in \cite{kordas12,kordas13}. It somewhat generalizes established entanglement criteria in terms of spin squeezing \cite{Sore01} and is derived
in a similar way, as one may see in the following. In contrast to spin squeezing inequalities, it shows that
a state is entangled if the variance defined below in Eq. (\ref{eqn:ent_para2}) is \emph{larger} than a certain threshold value.

Let us assume that the many-body quantum state $\hat \rho$ is decomposed into a mixture of pure states
\begin{equation}
   \hat{\rho} = \sum_k p_k \hat \rho_k = \sum_k p_k |\psi_k\rangle \langle\psi_k|,
   \label{eqn:rhodecompose}
\end{equation}
where every pure state $\hat \rho_k = |\psi_k\rangle\langle\psi_k|$ 
has a fixed particle number $N_k$. Note that the quantum jump 
simulation, see sec. \ref{sec:1}, of the dynamics directly provides such a decomposition.
The entanglement parameter is defined by
\begin{eqnarray}
       \label{eqn:ent_para2}    
    E_{a,b} &\equiv& \langle (\hat n_a - \hat n_b)^2 \rangle
     - \langle \hat n_a - \hat n_b \rangle^2 
     - \langle \hat n_a + \hat n_b \rangle  \\
    &&  - \frac{1}{2} \sum_{k,\ell} p_k p_\ell 
     \left[ \langle (\hat n_a - \hat n_b) \rangle_k 
         - \langle (\hat n_a - \hat n_b) \rangle_\ell \right]^2 \nonumber
\end{eqnarray}
for the sites $a$ and $b$. In this expression $\langle \cdot \rangle_{k,\ell}$ denotes the expectation value in the pure state $|\psi_{k,\ell}\rangle$.  Now we can prove that $E_{a,b}<0$ for every separable state such that a value $E_{a,b}>0$ unambiguously reveals the presence of many-particle entanglement. Note that $E_{a,b}$ provides an entanglement criterion, it is not a quantitative entanglement measure in the strict sense, see e.g. \cite{HHHH} for a definition of stricter ``measures''.

To prove the above statement, we consider an arbitrary separable state and show that $E_{a,b}<0$ for this class of states. 
If a pure state $\hat \rho_k$ is separable, it may be written as a tensor product of single particle states
\begin{equation}
   \hat \rho_k =   \hat \rho_k^{(1)} \otimes   \hat \rho_k^{(2)} 
           \otimes \cdots \otimes   \hat \rho_k^{(N_k)},
   \label{eqn:psstate}
\end{equation}
We introduce the abbreviation
\begin{equation}
  \hat S_\pm := \hat n_a \pm \hat n_b.
\end{equation}
This operator is also written as a symmetrized tensor product of single-particle operators
\begin{eqnarray}
   \hat S_\pm = \sum_{i=1}^{N_k} {\bf 1} \otimes \cdots \otimes {\bf 1} 
       \otimes \hat s_\pm^{(i)} \otimes {\bf 1} \otimes \cdots \otimes {\bf 1},
\end{eqnarray}
where the superscript $(i)$ denotes that the single-particle operator $\hat s_\pm^{(i)}$ acts on the $i$th atom. The single-particle operators are given by
\begin{equation}
    \hat s_\pm = |a\rangle\langle a| \pm  |b\rangle\langle b|,
\end{equation}
where $|a\rangle$ is the quantum state at which the particle is localized in the site $a$.

For a separable pure state  $\hat \rho_k$, the expectation values of the population imbalance 
$ \langle \hat S_- \rangle_k = {\rm Tr} [\hat{\rho}_k \hat S_-]$
and its square may be written as (dropping the subscript $k$ for notational clarity)
\begin{eqnarray}
   \langle \hat S_- \rangle &=& \sum_{i=1}^N {\rm Tr}\left[ \rho^{(i)} \hat s_-^{(i)}  \right] \\
   \langle \hat S_-^2 \rangle &=& \sum_{j \neq i}^N {\rm Tr} \left[ (\rho^{(j)} \otimes \rho^{(i)} )
            (\hat s_-^{(j)} \otimes \hat s_-^{(i)}) \right] + \sum_{j=1}^N  {\rm Tr}\left[ \rho^{(j)} \hat s_-^{(j)2}  \right] \nonumber \\
     &=& \sum_{j,i=1}^N {\rm Tr}\left[ \rho^{(j)} \hat s_-^{(j)}  \right] {\rm Tr}\left[ \rho^{(i)} \hat s_-^{(i)}  \right] \nonumber \\
        && -  \sum_{j=1}^N {\rm Tr}\left[ \rho^{(j)} \hat s_-^{(j)}  \right] {\rm Tr}\left[ \rho^{(j)} \hat s_-^{(j)}  \right] 
           + \sum_{j=1}^N {\rm Tr}\left[ \rho^{(j)} \hat s_-^{(j)2}  \right] \nonumber \\
      &=& \langle \hat S_- \rangle^2 +
           \sum_{j=1}^N  {\rm Tr}\left[ \rho^{(j)} \hat s_-^{(j)2}  \right]   - \left\{ {\rm Tr}\left[ \rho^{(j)} \hat s_-^{(j)}  \right] \right\}^2.
\end{eqnarray}
Using ${\rm Tr}[ \rho^{(j)} \hat s_-^{(j)2} ] = {\rm Tr}[ \rho^{(j)} \hat s_+^{(j)}]$ we thus find that 
every pure products state $\hat \rho_a$ satisfies the following condition
\begin{equation}
   \langle \hat S_-^2 \rangle_k - \langle \hat S_- \rangle^2_k \le \langle \hat S_+ \rangle_k \, .
\end{equation}

If the total quantum state $\hat{\rho}$ is separable, such that it can be written as a mixture of
separable pure states (\ref{eqn:rhodecompose}), the expectation values are 
\begin{eqnarray}
   && \langle \hat S_-^2 \rangle = \sum_k p_k \langle \hat S_-^2 \rangle_k \le \langle \hat S_+ \rangle + \sum_k p_k  \langle \hat S_- \rangle_k^2 \\
   && \langle \hat S_- \rangle^2 = \sum_{k,\ell} p_k p_\ell \langle \hat S_- \rangle_k \langle \hat S_- \rangle_\ell
= \sum_k p_k  \langle \hat S_- \rangle_k^2
            - \frac{1}{2} \sum_{k,\ell} p_k p_\ell \left[ \langle \hat S_- \rangle_k - \langle \hat S_- \rangle_\ell \right]^2.
\end{eqnarray}
We hence find that every separable quantum state satisfies the following inequality for the variance of the population imbalance $\hat S_-$: 
\begin{equation}
   \langle \hat S_-^2 \rangle - \langle \hat S_- \rangle^2
   \le \langle \hat S_+ \rangle +
    \frac{1}{2} \sum_{k,\ell} p_k p_\ell \left[ \langle \hat S_- \rangle_k - \langle \hat S_- \rangle_\ell \right]^2.
\end{equation}
For separable quantum states, the above inequality can be rewritten in terms of the entanglement parameter from Eq. (\ref{eqn:ent_para2}) as $E_{a,b} < 0$.

\end{document}